\documentclass[12pt]{article}
\usepackage{amsmath,amsfonts,amssymb}
\usepackage{hyperref}
\usepackage{graphicx}
\usepackage{slashed}
\usepackage{epsfig}
\usepackage{psfrag}
\makeatletter\@addtoreset{equation}{section}\makeatother

\DeclareMathOperator{\Tr}{Tr}

\def\bC {\mathbb{C}}
\def\bH {\mathbb{H}}

\def\bR {\mathbb{R}}
\def\bS {\mathbb{S}}
\def\bZ {\mathbb{Z}}

\newcommand{\beq}{\begin{equation}}
\newcommand{\eeq}{\end{equation}}
\newcommand{\bal}{\begin{equation}\begin{aligned}}
\newcommand{\eal}{\end{aligned}\end{equation}}
\newcommand{\half}{\frac{1}{2}}

\newcommand{\vev}[1]{{\left< {#1} \right>}}
\newcommand{\bra}[1]{{\left< {#1} \right|}}
\newcommand{\ket}[1]{{\left| {#1} \right>}}

\newcommand{\eqn}[1]{(\ref{#1})}

\newcommand{\address}[1]{\vbox{\center\em#1}}
\renewcommand{\title}[1]{\vbox{\center\LARGE{#1}}\vspace{5mm}}

\newcommand{\cL}{{\mathcal L}}

\newcommand{\cN}{{\mathcal N}}
\newcommand{\cP}{{\mathcal P}}

\newcommand{\cS}{{\mathcal S}}
\newcommand{\cW}{{\mathcal W}}

\begin{document}
\bibliographystyle{utphys}

\begin{titlepage}
\begin{center}
\phantom{.}

\hfill {\tt CERN-PH-TH/2013-314}\\
\hfill {\tt UT-Komaba/12-13}\\

\vspace{8mm}

\title{Exact results for vortex loop operators\\ in 3d supersymmetric theories}
\vspace{5mm}

\renewcommand{\thefootnote}{$\alph{footnote}$}

Nadav Drukker\footnote{\href{mailto:nadav.drukker@gmail.com}
{\tt nadav.drukker@gmail.com}},
Takuya Okuda\footnote{\href{mailto:takuya@hep1.c.u-tokyo.ac.jp}
{\tt takuya@hep1.c.u-tokyo.ac.jp}} and
Filippo Passerini\footnote{\href{mailto:filippo.passerini@cern.ch}
{\tt filipasse@gmail.com}}
\vskip 5mm
\address{
${}^{a}$Department of Mathematics, King's College,\\
The Strand, WC2R 2LS, London, UK
}
\address{
${}^{b}$Institute of Physics, University of Tokyo, \\
Komaba, Meguro-ku, Tokyo 153-8902, Japan}
\address{
${}^{c}$PH-TH division, CERN, CH-1211 Geneva, Switzerland
}

\renewcommand{\thefootnote}{\arabic{footnote}}
\setcounter{footnote}{0}

\end{center}

\vskip5mm

\abstract{
\noindent
Three dimensional field theories admit disorder line operators, dubbed vortex loop operators.  
They are defined by the path integral in the presence of prescribed singularities along the 
defect line.  We study half-BPS vortex loop operators for $\cN = 2$ supersymmetric theories 
on $\bS^3$, its deformation $\bS^3_b$ and $\bS^1\times\bS^2$.   
We construct BPS vortex loops defined by the 
path integral with a fixed gauge or flavor holonomy for infinitesimal curves linking the loop.  
It is also possible to include a singular profile for matter fields.  For vortex loops defined by 
holonomy, we perform supersymmetric localization by calculating the fluctuation modes, or 
alternatively by applying the index  theory  for transversally elliptic operators.  We clarify 
how the latter method works in situations without fixed points of relevant isometries.   
Abelian mirror symmetry transforms Wilson and vortex loops in a specific way.   In particular an 
ordinary Wilson loop transforms into a vortex loop for a flavor symmetry.   Our localization 
results confirm the predictions of abelian mirror symmetry.
}

\end{titlepage}

\setcounter{tocdepth}{2}
\tableofcontents

\section{Introduction}
\label{sec:intro}

In this paper we initiate the study of the exact expectation value of 
supersymmetric vortex loop operators in 
$\cN=2$ gauge theories in three dimensions.%
\footnote{Preliminary versions of these results were presented by 
T.O.  at the ``Autumn Symposium on String/M Theory'', KIAS, Seoul
September 17-21 2012, at Kyoto University and at Rikkyo University 
and by F.P.  at the ``II Workshop on Geometric Correspondences of Gauge Theories,'' 
SISSA-Trieste Italy, September 17-21 2012.}
In the case of $\cN=6$ supersymmetric 
Chern-Simons-matter theories \cite{abjm} (known as ABJM theory) such operators 
were defined in \cite{vortex} and evaluated at strong coupling.  In this paper 
we will define them in more general theories.
For abelian gauge groups, we perform the exact localization calculation of their expectation value, 
reducing the infinite dimensional path integral to a finite dimensional integral.

The basic definition of a vortex loop operator is that a gauge field has 
a singularity along a curve in space.  Stated differently, it is the result of 
quantizing the theory in a background with a non-trivial singular connection.  
We start this paper by considering in great detail 
theories on a round $\bS^3$ and later 
generalize to the case of the squashed sphere $\bS^3_b$ and to the index 
calculation on $\bS^2\times\bS^1$.  In all these examples it is possible to 
introduce such singularities and with the appropriate choice of curve and 
boundary terms they preserve half of the supersymmetries.

The allowed vortex loop operators depend intimately on the choice of 
gauge and global symmetries, matter content and the action.  Particularly,
the non-trivial connection may be dynamical and for gauge symmetry, or non-dynamical and for a global symmetry.
In the next section we classify the possible types of 1/2 BPS loop operators in 
$\cN=2$ supersymmetric theories on $\bS^3$.  
Once we fix the path to be a large circle, a vortex loop operator 
is specified by some singularities of the gauge field, parameterized by a real 
diagonal matrix $H$ (or for abelian theories a number $\eta$) 
and by singularities in the matter fields, encoded by a complex 
vector $B$ (or number $\beta$).%
\footnote{There is an obstruction to having a singularity for the matter fields 
in the cases of $\bS^3_b$.}

When $B$ is completely generic, it serves effectively as a Higgs vacuum near 
the locus of the singularity.  
This can be made more precise by considering the 
gauge theory on $\bH^2\times\bS^1$, where the singular classical solution 
becomes essentially a constant Higgs VEV on hyperbolic space $\bH^2$
(with a holonomy along the $\bS^1$).  
Localization reduces the partition function of supersymmetric theories on $\bS^3$ 
to a finite dimensional integral over constant matrices, parameterizing the Coulomb 
branch.  Even though the Higgs mechanism 
breaks the gauge symmetry only at the singularity, since the remaining fields 
are constant, they are effected by this local breaking and are frozen at the origin 
of the Coulomb branch.  
This can therefore be considered as 
localization on the Higgs branch, rather than the usual Coulomb branch.  

We shall not perform the localization calculation of the  operators with 
singularities for the matter fields in this paper and restrict ourselves to the case with 
$B=0$.

Going back to the vortex loops without a singularity for the matter fields, these are studied in the following sections and supersymmetric 
localization is used to evaluate their expectation values.  
The calculation is similar to that in \cite{pestun, kwy1,Hama:2010av,Jafferis:2010un,Hama:2011ea,Imamura:2011su} and the 
final result is a very simple modification of the resulting matrix model.

In order to perform the localization calculation one should choose a localizing action, 
which for $\bS^3$ is the usual supersymmetric Yang-Mills action and the dimensional 
reduction of the 4d chiral action to 3d.  Both are known to be exact under some of 
the supercharges which preserve the vortex loops, though it also requires keeping 
track of boundary terms in the action near the singularity.  
Supersymmetric localization should allow therefore to compute the exact vacuum 
expectation value of the vortex loop operators by modifying the action with this 
exact term.  With a diverging prefactor the calculation reduces to evaluating the classical 
action and one-loop quantum corrections around it.

We proceed to study the one-loop determinant by doing the spectral analysis in the 
background of the vortex loop operator.  It breaks the supersymmetry of the vacuum, 
so the supersymmetry multiplets are short (similar to those on the deformed $\bS^3_b$ 
\cite{Hama:2011ea}).  It also effectively imposes modified non-periodic boundary conditions on the 
fields.  We therefore classify non-periodic spherical harmonics on $\bS^3$, which is 
mainly done in Appendices~\ref{sec:1-loop} and~\ref{sec:harmonics} 
and discussed in Section~\ref{sec:localize}.  
For small vorticity we expect that the spectrum does not change 
much (except for possible new almost zero modes).  Indeed the spectrum is 
continuous with the vorticity parameter.

The result of the calculation is the usual finite dimensional matrix integral with 
an imaginary shift of the Coulomb brach parameters.  The same shift appears in the 
one-loop determinant and in the classical action.  When the vortex is in a gauge 
connection, these parameters are integrated over and by contour deformation the 
result is trivial (up to a simple overall factor).  When the vortex is defined for 
a background global symmetry the result is non-trivial.

As the vorticity grows larger, some modes which were perfectly regular turn singular, 
and worse, non-normalizable.  
We will assume that the expectation value of the vortex loop is analytic in vorticity.
We thus propose a prescription for which modes to include in the spectrum 
and how to perform the integration over the Coulomb-branch parameters which 
guarantees this behavior for the gauge vortex loops and gives a prediction for  
vortex loops of global symmetries.

We also compute the partition function and the expectation value of loop operators on 
the deformed sphere $\mathbb S^3_b$ as well as $\mathbb S^1\times \mathbb S^2$.
In both cases the effect of the vortex loop operator is similar to that on 
the round $\bS^3$: vorticity in the gauge connection has no effect and vorticity 
for a flavor symmetry leads to shifts in physical parameters.
To compute the one-loop determinants on these geometries, we apply the Atiyah-Singer 
index  theory  for transversally elliptic operators by generalizing the method used in \cite{pestun}.
In particular we manage to apply the Atiyah-Singer index   theory  despite the absence 
of fixed points for relevant isometries on these manifolds.%
\footnote{%
 The paper \cite{Kim:2012qf} also uses the fixed-point formula in such a situation, based on a similar logic.  
} 

We will also provide an intuitive explanation for how abelian mirror symmetry
acts 
on vortex and Wilson loop operators, using the BF coupling between dynamical 
and non-dynamical gauge fields.

The vortex loop operators share some similarities to 't~Hooft loop operators in 
four dimensions, whose exact expectation value in $\cN=2$ supersymmetric 
gauge theories was recently calculated in \cite{Gomis:2011pf,Ito:2011ea}.  They are both 
disorder line operators.  They are also related to surface operators in 4d, 
see \cite{gw-surface, dgm} being co-dimension two defects.%
\footnote{%
When a 3d theory lives on the boundary of a 4d spacetime, a bulk surface 
operator \cite{gw-surface} 
ending on the boundary along a loop induces a vortex loop in the 3d theory.} 
Like the 
surface operators, the vortex loop operators may involve a singularity 
for the matter fields as well as a non-trivial holonomy.  We hope this work 
would be useful for an exact calculation of the expectation value of a 
BPS spherical surface operator in 4d.

Vortex loop operators with quantized vorticities are the same as Dirac strings, they 
may start and end on monopole operators.  The ones we consider, though, 
permeate all of space (or a closed curve) instead of starting at a monopole.  
While in the presence of a monopole a cycle wrapping the string 
can be deformed and contracted to zero in a regular way on the other side of the 
monopole (so a Wilson loop around this cycle has to have trivial VEV), in the absence 
of the monopole, when considering an infinite or closed vortex loop operator, the 
holonomy does not have to be trivial and the vorticity may be non-integer.

It is important to distinguish between the vortex loop operators and dynamical 
vortices, like those of Nielsen-Olesen or Abrikosov, or those in supersymmetric 
theories studied in \cite{Hanany:2003hp}.  These vortices are dynamical objects, 
solutions to the vacuum equations of motion, while the vortex loop operators 
are external probes of the theory.  If it were not for special boundary terms, the 
action of the vortex loop operators would diverge.  But there is a relation, as 
a singular limit of the smooth solitonic vortices does reproduce the semiclassical 
vortex loop operator.  The relation between the two is analogous to that between 
an 't~Hooft-Polyakov monopole and an 't~Hooft loop.

As this manuscript was being finalized the paper of Kapustin, Willett and Yaakov 
\cite{Kapustin:2012iw} appeared.%
\footnote{%
The mapping under Abelian mirror symmetry of loop operators was also obtained by Benjamin Assel, Ricardo Couso Santamaria and Jaume Gomis.  
We thank them for discussions.

}
That paper shares the same topic as ours and has a great 
deal of overlap to our discussion of vortex loop operators on the round $\bS^3$.

Note added: in the replacement on the arXiv, we elaborated on the index theory calculation of the one-loop determinants on $\mathbb S^3_b$ and $\mathbb S^1\times\mathbb S^2$.
We also made several corrections in the computation of the vortex loop expectation values.
For the analysis of gauge vortex loops, we made use of the $SL(2,\mathbb Z)$ action in the presence of loop operators considered in \cite{Kapustin:2012iw}.

\section{Half-BPS loop operators}
\label{sec:loops}

Loop operators are non-local gauge invariant operators that are supported on a closed 
one-dimensional line.  In three dimensional gauge theories there are two types 
of loop operators: Wilson loops, that are order type operators, and vortex loops 
that are disorder type operators.  In the following, we provide a definition of 
the latter in a generic Euclidean theory on $\bS^3$ with ${\cal N}\geq2$ 
supersymmetry.  Most of this is carried over to the cases of $\bS^3_b$ and 
$\bS^2\times\bS^1$ discussed in Sections~\ref{sec:S3b} and~\ref{sec:S2S1}.  
It is assumed that the field content of the theory includes at 
least an ${\cal N}=2$ vector multiplet, that is a gauge field $A_\mu$, two 
spinors $\lambda$ and $\bar{\lambda}$, and two auxiliary real scalars $D$ 
and $\sigma$.  This multiplet may be gauged or associated to a global symmetry.  
The matter vortices require of course matter fields, the 
dimensional reduction of a chiral multiplet in 4d with scalar $\phi$, spinor $\psi$ and 
auxiliary field $F$ and anti-chiral multiplet with $\bar\phi$, $\bar\psi$ and $\bar F$.  
The  parameterizations of the round $\bS^3$ are described in Appendix~\ref{sec:metrics3} and a few aspects of  supersymmetry 
on $\bS^3$ are collected in Appendix~\ref{sec:susys3}.

\subsection{Half-BPS Wilson loop}
\label{sec:wilson}

Before focusing on vortex loop operators let us recall the construction of 
the half-BPS Wilson loops in $\cN=2$ supersymmetric theories in 3d 
\cite{gaiotto-yin}.  We will then study all the singular field configurations preserving 
the same supercharges.

The ansatz for a supersymmetric Wilson loop operator is given by
\beq
\label{WL}
W_R=\frac{1}{\dim R}\Tr_{R} \cP
\left[\exp\left(\oint d\tau\left(iA_\mu\dot x^\mu+\sigma|\dot x|\right)\right)\right]
\eeq
where $x^\mu(\tau)$ parameterizes the curve on which the Wilson loop is defined, 
$\cP$ denotes path-ordering and $R$ is a representation of the gauge group.  
Applying the supersymmetry variations (\ref{susyV}) to this operator it results \cite{kwy1}
\beq
\delta W_R \propto -\frac{1}{2} \bar{\epsilon} \left( \gamma_\mu \dot x^\mu - |\dot x| \right)\lambda 
+ \frac{1}{2} \bar{\lambda} \left( \gamma_\mu \dot x^\mu - |\dot x| \right)\epsilon
\eeq
and this is zero if 
\beq
\bar{\epsilon} \left( \gamma_\mu \dot x^\mu - |\dot x| \right) = 0\,,\qquad
\left( \gamma_\mu \dot x^\mu - |\dot x| \right) \epsilon = 0 \,,
\eeq 
or equivalently, using $\epsilon\bar{\epsilon}=\bar{\epsilon}\epsilon$ and 
$\epsilon\gamma^\mu\bar{\epsilon}=-\bar{\epsilon}\gamma^\mu\epsilon$ 
for fermionic SUSY parameters
\beq
\left( \gamma_\mu \dot x^\mu + |\dot x| \right) \bar{\epsilon} = 0\,,\qquad
\left( \gamma_\mu \dot x^\mu - |\dot x| \right) \epsilon = 0 \,.
\eeq
From these equations, it follows that a Wilson operator defined on a loop 
such that
\beq
\label{WLsphere}
\dot{x}^\mu=R\, e_{3}{}^\mu
\eeq 
preserves the supersymmetry generated by $\epsilon$ and $\bar{\epsilon}$ that satisfy 
\beq
\label{susy-S3}
\left( \gamma_{3} - 1 \right) \epsilon = 0\,,
\qquad
\left(\gamma_{3} + 1\right)\bar{\epsilon} = 0\,.
\eeq
We  consider  the Hopf fibration metric (\ref{hopf})  with the left invariant vielbein (\ref{viel}), since in this vielbein basis the Killing 
spinors $\epsilon$ and $\bar\epsilon$ are constant.  Given the expression for  the inverse left invariant vielbein \eqn{inviel},  the condition (\ref{WLsphere}) implies that the Wilson loop is extended along a curve parameterized in the Hopf metric 
\eqn{hopf} as 
\beq
\theta=\text{const}\,,
\qquad
\phi=\text{const}\,,
\qquad
\psi=2\tau\,,
\quad
0\leq\tau\leq2\pi\,.
\eeq
Or in terms of the complex coordinate $(u,v)$ in \eqn{uv} as 
\beq
u=u_0\,e^{i\tau}\,,
\qquad
v=v_0\,e^{i\tau}\,,
\qquad
0\leq\tau\leq2\pi\,,
\eeq
with arbitrary $|u_0|^2+|v_0|^2=R^2$.  We will concentrate on the case of the loop at $u_0=0$ 
which is $\theta=0$ in the Hopf coordinates.  The submanifold described by $\theta=0$ is 
codimension-2, since the metric (\ref{hopf}) reduces to 
\bal
ds^2&=\frac{R^2}{4}(d\psi+d\phi)^2
\eal
and therefore, at $\theta=0$ the loop is extended along $\psi+\phi$.   In the torus fibration  coordinates  \eqn{torusf}, the loop is extended along  $\varphi_2$.

Considering the commutators of the supersymmetries generated by $\epsilon$ and 
$\bar\epsilon$ that satisfy (\ref{susy-S3}), one obtains the expression (\ref{commuvec}) where
\beq
v^\theta=0\,,
\qquad
v^\phi=0\,,
\qquad
v^\psi=\frac{2}{R}\bar\epsilon \gamma^3 \epsilon\,,
\eeq
that implies that  the commutator of the supersymmetry includes a translation along the $\psi$ 
angle that is a symmetry of the Wilson loop.

\subsection{Half-BPS vortex loop operator: Vector multiplet}
\label{sec:g-vortex}

Our purpose is not to study Wilson loops, which are electric order operators, but 
rather vortex loop operators, which are magnetic disorder operators.  That amounts to considering the theory 
where certain fields have a singularity along a curve on $\bS^3$, which we take to be the 
same curve $\theta=0$ as the aforementioned Wilson loops.  We restrict to singularities 
which preserve the supercharges with parameters $\epsilon$ and $\bar\epsilon$ satisfying 
the conditions (\ref{susy-S3}) as above and use the localization 
scheme of \cite{pestun,kwy1} to evaluate their expectation values.

We first examine which field configurations are invariant under the supercharge generated by 
$(\epsilon,\bar\epsilon)$, which will restrict the allowed form of the singularities.  
Imposing reality of all the fields, then from the SUSY variation 
$\delta_\epsilon\lambda=0$ in \eqn{susyV} we obtain 
\bal
\label{Hitchin}
-\frac{1}{2}\varepsilon_{\rho\mu\nu}F^{\mu\nu}+D_\rho \sigma=0\,,
\qquad
D+\frac{\sigma}{R}=0\,,
\eal
and from $\delta_{\bar\epsilon}\bar\lambda=0$
\bal
\label{Hitchinbar}
\frac{1}{2}\varepsilon_{\rho\mu\nu}F^{\mu\nu}+D_\rho \sigma=0\,,
\qquad
D+\frac{\sigma}{R}=0\,.
\eal
A field configuration that is invariant under the full set of supersymmetry preserved by the half-BPS Wilson loop (\ref{susy-S3}), satisfies $\delta_\epsilon\lambda=\delta_{\bar\epsilon} \bar\lambda =0$.  Combining (\ref{Hitchin}) and (\ref{Hitchinbar}) we obtain 
\beq
\label{BPS}
F_{\mu\nu}=0\,,
\qquad
D_\mu \sigma=0\,,
\qquad
D=-\frac{\sigma}{R}\,.
\eeq
These are the same as the solutions to the localizing equations considered in \cite{kwy1,Hama:2010av} 
and therefore we will be able to use the same localizing action.

In studying the $\bS^3$ partition function the only classical solution of the 
supersymmetric Yang-Mills and Chern-Simons actions satisfying these 
conditions and the equations of motion are $A_\mu=0$, $\sigma=0$ 
and $D=0$.  
BPS configurations include also a constant matrix $\sigma=\sigma_0$ 
and $D=-\sigma_0/R$.  In studying the vortex loop operators we allow in addition singularities 
for the gauge field at $\theta=0$.  It is easiest to write the solution in the torus fibration 
coordinates (\ref{torusf}) where the vortex is at $\vartheta=0$ and is extended along the 
$\varphi_2$ circle.  The curves along $\varphi_1$ at fixed $\vartheta$ are linked to the 
vortex, and therefore may have a nontrivial holonomy.  We choose a gauge where
\beq
\label{vor}
A_{\varphi_1}^{(0)}=H=
\begin{pmatrix}
\eta_1\otimes1_{N_1}&\cdots & 0 \cr
\vdots &\ddots&\vdots\cr
0&\cdots&\eta_M\otimes1_{N_M}
\end{pmatrix}.
\eeq
The allowed choices of $H$ depend on the details of the gauge theory including the matter 
content.  The requirement is that all observables are single valued when winding around the 
vortex loop $\varphi_1\to\varphi_1+2\pi$.  This includes the action, which should be well 
defined (up to integer shifts by $2\pi$, as usual for Chern-Simons theory) and any gauge 
invariant local operator.  This is automatically satisfied if all the fields of the theory are 
single valued, which happens if the eigenvalues $\eta_i$ of $H$ are all integers.  

Local observables are gauge invariant under any gauge transformation and therefore 
will not be affected when rotating around the vortex loop.  The partition function of 
Chern-Simons is not invariant under large gauge transformations, which leads to 
the usual quantization of the Chern-Simons level $k$.  The presence of a 
vortex loop operator further restricts $k$ such that $k H/2$ is a weight vector of 
a unitary representation of the gauge group \cite{ms-zoo}.  For a fixed $k$ this 
is a quantization condition on $H$.

In the Hopf coordinates (\ref{hopf}), the vortex is associated to a constant gauge 
field $A_\mu^{(0)}$ given by 
\beq
\label{vfield}
A_\theta^{(0)}=0\,,
\qquad
A_\phi^{(0)}=-\frac{1}{2}H\,,
\qquad
A_\psi^{(0)}=\frac{1}{2}H\, .
\eeq
Away from the singularity, this constant vector field configuration satisfies $F_{\mu\nu}=0$.  
Indeed, in the presence of the vortex, the topology of the $\bS^3$ is modified to $\bS^1\times \text{disk}$ 
and it is hence possible to have non-trivial 
flat connections.

An important point to notice is that 
thus far the background of the vortex does not seem to break any supersymmetry, 
as a flat connection is a solution to both the BPS and anti-BPS equations.  One has 
to examine the singularity at $\theta\to0$ to see that it indeed breaks half the 
supercharges.  This is done in Appendix~\ref{sec:boundaries}, where we write down the 
boundary terms for the Yang-Mills, Chern-Simons and Fayet-Iliopoulos actions and 
verify that the vortex loop operator breaks half the supersymmetries.

Given the singular behavior \eqn{vor} at $\theta=0$, the most general solution to the BPS equations 
\eqn{BPS} has this exact value for the gauge field as in the classical solution and in addition we can 
turn on $\sigma=\sigma_0$ and $D=-\frac{\sigma_0}{R}$ where $\sigma_0$ is covariantly constant, {\em i.e.},
\beq
\label{covcon}
D^{(0)}_\mu\sigma_0=0\,,
\eeq
and the covariant derivative $D^{(0)}_\mu$ is defined using the constant connection 
$A_\mu^{(0)}$ (\ref{vfield}).  

If we label by $(\sigma_0)^i_{\ j}$ one of the components of $\sigma_0$ in the $i,j$ block, 
then (\ref{covcon}) gives
\bal
\label{periodicity}
&\partial_\phi(\sigma_0)^i_{\ j}-\frac{i}{2}(\eta_i-\eta_j)(\sigma_0)^i_{\ j}=0\,,
\\
&\partial_\psi(\sigma_0)^i_{\ j}+\frac{i}{2}(\eta_i-\eta_j)(\sigma_0)^i_{\ j}=0\,.
\eal
For $\eta_i\not\equiv \eta_j \pmod 1$ the only regular periodic solution to these equations is 
$(\sigma_0)^i_{\ j}=0$.  
For generic $\eta_i,\eta_j$ the only nontrivial solutions are therefore for $i=j$, and this 
component $(\sigma_0)^i_{\ j}$ can be an arbitrary constant.

If the vortex loop operator is defined for a gauged vector multiplet (rather than a background 
vector field), we should define the integration measure.  Since for generic $H$ the allowed values 
of $\sigma_0$ are automatically diagonal, there is no extra Vandermonde determinant.

If there are degeneracies, and the singularity preserves a non-trivial 
Levi group $U(N_1)\times\cdots\times U(N_M)$, the resulting Vandermonde determinant 
involves only the eigenvalues within the different blocks along the diagonal
\beq
\prod_{m=1}^M\prod_{i<j=1}^{N_m} \big[(\sigma_0)_{m,i}-(\sigma_0)_{m,j}\big]^2\,,
\eeq
where we labeled $(\sigma_0)_{m,i}$ the $i^\text{th}$ element on the diagonal of 
the $m^\text{th}$ block of $\sigma_0$.

Note that the symmetry is enlarged (and the resulting Vandermonde) also for values of 
$\eta_i$ differing by integers, as can be seen by the periodic non-trivial solutions of 
\eqn{periodicity}.  Furthermore, if some $\eta_i$ form a representation of $\bZ_n$ for 
some $n\leq N$, so $\eta_j\equiv j/n\pmod 1$ for $j=1,\cdots n$, then an $S_n$ subgroup 
is preserved allowing for twisted solutions which are periodic only up to $S_n$ transformations.  
Such solutions are important in ABJM theory \cite{vortex} and exist also for surface operators 
in $\cN=4$ SYM in four dimensions in \cite{Koh:2008kt}.  This mimics the construction of 
``long strings'' in M(atrix) theory \cite{Motl:1997th,Dijkgraaf:1997vv}.

\subsection{Half-BPS vortex loop operator: Matter multiplet}
\label{sec:m-vortex}

We turn now to the matter sector, whose supersymmetry transformations are written in 
Appendix~\ref{sec:chiral-susy}.  We shall find non-trivial profiles for the scalar field 
which are invariant under the same supercharges as the Wilson loops and the 
vortex loop operators from the vector multiplet.

\subsubsection{Abelian theory}

Let us start with an abelian theory.  
Assuming the supercharges satisfy the half-BPS conditions in \eqn{susy-S3}, 
the vanishing of the variation of $\psi$ and $\bar\psi$ in \eqn{chiralsusy} give 
the equations
\bal
\label{m-BPS}
ie_3{}^\mu D_\mu\phi +i\sigma\phi-\frac{\Delta}{R}\phi&=0\,,
\qquad&
(e_1{}^\mu+ie_2{}^\mu)D_\mu\phi&=0\,,
\qquad&
F&=0\,,
\\
ie_3{}^\mu D_\mu\bar\phi-i\sigma\bar\phi+\frac{\Delta}{R}\bar\phi&=0\,,
\qquad&
(e_1{}^\mu-ie_2{}^\mu)D_\mu\bar\phi&=0\,,
\qquad&
\bar F&=0\,.
\eal
where $e_a{}^\mu$ are the inverse vielbeins in \eqn{inviel}.  This expression applies for 
a massless field charged under a single gauge group.  For a field charged under two 
groups there would be the appropriate modification to the connection $D_\mu$ and 
likewise $\sigma$ would be replaced by the difference of $\sigma^{(i)}$ of the 
two vector multiplets.  As usual a mass term is like a $\sigma$ field for a non-dynamical 
vector field.

In terms of the Hopf coordinates \eqn{hopf} the equations for $\phi$ are%
\footnote{Hopefully there will be no confusion between the field $\phi$ and coordinate $\phi$.}
\beq
\label{chiral-matter}
D_\psi\phi=-\frac{i}{2}(\Delta-iR\sigma)\phi\,,
\qquad
(\sin\theta\,D_\theta-iD_\phi+i\cos\theta\,D_\psi)\phi=0\,.
\eeq
We saw already that the supersymmetry conditions of the vector multiplet restrict 
$\sigma=\sigma_0$ a constant.  For real $\sigma_0\neq0$ the first equation does 
not have periodic solutions other than $\phi=0$.  
For $\sigma_0=0$ (or in the case with more than one gauge multiple or a mass term, the vanishing of
their sum) there are extra solutions of the form
\beq
\label{genmv}
\phi(\theta,\phi,\psi)=e^{-i\frac{\Delta+\eta}{2}\psi}\phi(\theta,\phi)\,,
\eeq
with $\phi(\theta,\phi)$ satisfying
\beq
\left(\sin\theta\,\partial_\theta-i\partial_\phi-\frac{\eta}{2}+\frac{\Delta}{2}\cos\theta\,\right)\phi(\theta,\phi)
=0\,,
\eeq
where $\eta$ is the gauge vorticity (\ref{vor}) for an abelian theory, 
{\em i.e.}, $H=\eta$.  With the ansatz $\phi(\theta,\phi)=e^{in\phi}\phi_n(\theta)$ 
we get the solution
\beq
\phi_n(\theta)
=\frac{\beta_n}{R^{\Delta}}
\sin^{-\frac{\Delta-\eta}{2}-n}\frac{\theta}{2}
\cos^{-\frac{\Delta+\eta}{2}+n}\frac{\theta}{2}\,.
\eeq
The values of $\beta_n$ are determined by specifying the singularity of the field.  This ansatz 
allows for singularities and zeros at $\theta=0$ and $\theta=\pi$, but by taking linear combinations 
of these functions one can get singularities at any point on the base parameterized by 
$(\theta,\phi)$.

In terms of the torus coordinates \eqn{torusf} the solution is
\beq
\phi(\vartheta,\varphi_1,\varphi_2)
=\frac{\beta_n}{R^{\Delta}}\left(\sin\vartheta\,e^{i\varphi_1}\right)^{-\frac{\Delta-\eta}{2}-n}
\left(\cos\vartheta\,e^{i\varphi_2}\right)^{-\frac{\Delta+\eta}{2}+n}
e^{-i\eta\varphi_1}\,.
\eeq
Requiring periodicity in the $\varphi_2$ direction enforces $n-(\Delta+\eta)/2$ to be an integer.  
Furthermore, if we want singularities only at $\vartheta=0$, then this integer cannot be negative.  
The simplest and least singular case is when it is zero, which gives
\beq
\label{chiral-vortex}
\phi(\vartheta,\varphi_1,\varphi_2)
=\frac{\beta\,e^{-i\eta\varphi_1}}{\left(R \sin\vartheta\,e^{i\varphi_1}\right)^\Delta}\,.
\eeq
The behavior of the scalar field near the singularity is determined by its dimension 
$\Delta$ (and in addition the holonomy $\eta$).  For a scalar of canonical dimension 
$\Delta=1/2$ this is
\beq
\label{chiral-vortex-1/2}
\phi(\vartheta,\varphi_1,\varphi_2)
=\frac{\beta\,e^{-i\eta\varphi_1}}{\sqrt{R\sin\vartheta\,e^{i\varphi_1}}}\,.
\eeq
The field $\phi$ is complex, but the parameter $\beta$ can, without loss of 
generality, be taken real.  Its phase is unphysical as it is modified by taking 
$\varphi_1\to\varphi_1+2\pi$ and can be changed by a gauge transformation 
with a constant gauge parameter.

One can also formulate the vortices in flat space (and on $\bH_2\times\bS^1$), 
as was done in \cite{vortex}.  
The flat space vortex arises in the large $R$ limit after replacing 
$R\sin\vartheta\to r$, $\cos\vartheta\to1$ and $R\varphi_2\to x_3$.  After rescaling 
we get the solution
\beq
\phi(r,\varphi_1,x_3)
=\beta_n\left(r\,e^{i\varphi_1}\right)^{-n-\Delta/2}\left(r\,e^{-i\varphi_1}\right)^{\eta/2}\,.
\eeq
In the special case of $n=\frac{\Delta+\eta}{2}$ we get from \eqn{chiral-vortex} 
\beq
\phi(r,\varphi_1,x_3)
=\frac{\beta\,e^{-i\eta\varphi_1}}{\left(r\,e^{i\varphi_1}\right)^\Delta}\,.
\eeq
This indeed matches with the vortex in ABJM theory \cite{vortex} once we set 
$\Delta=1/2$ and $\eta=0$ (in \cite{vortex} there was a gauge vortex, but it was in the 
diagonal sum of the two gauge groups which the matter fields are not charged under).

So far we discussed only the field $\phi$.  The same analysis applies also for the field 
$\bar\phi$, once we require the invariance under the $\bar\epsilon$ variation.

\subsubsection{Non-abelian theory}
\label{sec:nonabel}

Turning to the general non-abelian theory, the matter 
fields are in some representation $R$ of the gauge group.  
We denote the generators of the algebra in the $R$ representation as 
$(X_\alpha^R,K_i^R)$, where $K_i^R$'s span the Cartan subset.  
The normalization of the generators is such that 
$\Tr(X_\alpha^R,X_\beta^R)=\delta_{\alpha+\beta,0}$ and 
$(X_\alpha^R)^\dagger=X_{-\alpha}^{R}$.

The weights of $R$ are denoted as $\rho$ and the associated state is $\ket{\rho}$ 
such that for the Cartan generators $K_{i}^{R}\ket{\rho}=\rho_i\ket{\rho}$.  
The scalar field of the chiral multiplet $\phi$ is expressed as
\beq
\phi=\sum_{\rho}\phi^\rho\ket{\rho}
\eeq
and likewise the other members of the multiplet.  
For the anti-chiral scalar we take 
bra states
\beq
\bar\phi=\sum_{\rho}\bar\phi^\rho\bra{\rho}
\eeq
where $\vev{\rho|\rho'}=\delta_{\rho,\rho'}$.  The fields of the vector multiplet, which are in the adjoint representation, appear in 
the chiral Lagrangian and the supersymmetry transformations accompanied by the 
generators of the algebra in the representation $R$, so for example
\beq
\sigma\to \sigma^iK_i^R+\sigma^\alpha X^R_\alpha\,.
\eeq
Then the first equation in \eqn{m-BPS} becomes
\bal
ie_3{}^\mu D_\mu\phi +i\sigma\phi-\frac{\Delta}{R}\phi
&=\sum_\rho\left(ie_3{}^\mu (\nabla_\mu\phi^\rho+iA^i_{\mu}\rho_i\phi^\rho)
+i\sigma^i\rho_i\phi^\rho
-\frac{\Delta}{R}\phi^\rho\right)\ket{\rho}
\\&\qquad\quad
+\sum_{\rho,\rho'}\left(ie_3^\mu A^\alpha_{\mu}(X^R_\alpha)^\rho_{\rho'}\phi^{\rho'}
+i\sigma^\alpha(X^R_\alpha)^\rho_{\rho'}\phi^{\rho'}\right)\ket{\rho'}
=0\,.
\eal
For $\sigma=0$ and $A_\mu$ as in \eqn{vor}, the second line of this equation vanishes 
and we find $\dim(R)$ copies of the scalar equations in \eqn{m-BPS}.  The solution 
is then as in \eqn{chiral-vortex}
\beq
\label{na-chiral-vortex}
\phi(\vartheta,\varphi_1,\varphi_2)
=\sum_\rho
\frac{\beta^\rho\,e^{-i\rho(H)\varphi_1}}{\left(R \sin\vartheta\,e^{i\varphi_1}\right)^\Delta}\ket\rho\,,
\eeq
which can also be written as
\beq
\label{na-chiral-vortex1}
\phi(\vartheta,\varphi_1,\varphi_2)
=\frac{e^{-iH\varphi_1}}{\left(R \sin\vartheta\,e^{i\varphi_1}\right)^\Delta}B\,,
\qquad
B=\sum_\rho \beta^\rho\ket\rho\,.
\eeq

The simplest solutions to the BPS equations are when either $\sigma=0$, which 
allows for arbitrary $\beta^\rho$, or when $\phi=0$ which allows for arbitrary 
constant $\sigma=\sigma_0$.  More generally one can turn on just some 
components of $\sigma$ and then there will be a restriction on which 
$\beta^\rho$ may be nonzero.  Viewed the other way, choosing non-generic $\beta^\rho$ 
will leave some residual symmetry and the components of $\sigma$ in the directions of 
the generators of this preserved symmetry will not be frozen to zero and will have to 
be integrated over after localization.  For example, if $R$ is the fundamental representation 
of $U(N)$ and $\beta^i=0$ for $i=1,\cdots,n$, then there will be a residual $U(n)$ symmetry 
and after diagonalization, $n$ elements of $\sigma_0$ to integrate over.  
If $n$ of the $\beta^i$'s are equal to each-other but non zero, the symmetry will be $SU(n)$, 
and so on.

This analysis is nothing different from the usual breaking of gauge symmetry by 
the Higgs mechanism, only that here the scalar fields get a non-trivial profile, instead of 
a constant VEV.  As mentioned in the introduction, this profile becomes constant upon 
conformal transformation to $\bH_2\times\bS^1$.

\subsection{Vortex and Wilson loop operators for flavor symmetries}
\label{sec:flavor}

Before plunging into the localization calculations we want to explore the relation 
between Wilson loop operators and vortex loop operators in different $\mathcal N=2$ 
theories.  A useful generalization of the loop operators discussed above is to 
consider operators defined with respect to global rather than gauge groups.

Given a global symmetry that commutes with SUSY (hence not an 
$R$-symmetry), it is natural to couple its current $j^\mu$ to a background, 
non-dynamical abelian gauge field $\boldsymbol A_\mu$ through the coupling in the action%
\footnote{%
If the global symmetry is non-abelian, we can use its Cartan subalgebra 
to define the flavor vortex loop.}
\begin{equation}
\label{aj}
  \int \boldsymbol A_\mu j^\mu\,,
\end{equation}
which can be supersymmetrized.
This procedure is sometimes called ``gauging'', but we reserve the term for the case 
when the gauge field is dynamical.  The vortex loop operator for this global symmetry 
is defined by letting the non-dynamical gauge field have the singularity (\ref{vor})
\begin{equation}
\label{a-vor}
\boldsymbol A\sim \eta\,d\varphi_1\,,
\end{equation}
where $\varphi_1$ is the angular variable in the locally defined polar coordinates 
on the plane orthogonal to and centered at the loop.  Whether this definition gives 
a BPS vortex loop depends on the space-time geometry since we need a globally 
defined supersymmetric profile of $\boldsymbol A_\mu$ with the singularity (\ref{a-vor}).  In 
the case of $\bS^3$ discussed so far (and $\bS^3_b$ and $\bS^1\times\bS^2$ 
studied later) this is indeed the case.

Consider a vortex loop for the topological symmetry $U(1)_J$ generated by 
the current $J^\mu=\frac 12 \epsilon^{\mu\nu\rho}F_{\nu\rho}$, the Hodge dual of the 
field strength of a dynamical abelian gauge field.  This vortex loop is simply 
a rewriting of the usual Wilson loop.  The singularity (\ref{a-vor}) 
means that the non-dynamical field strength $d\boldsymbol A$ has a delta function, and the 
coupling (\ref{aj}), which is precisely the BF coupling \eqn{SUSY-BF}, becomes
\begin{equation}\label{Wilson-BF}
  \int \boldsymbol A\wedge F=\int d\boldsymbol A\wedge A=\eta \oint A+ \int d\underline{\boldsymbol A}\wedge A\,,
\end{equation}
where $A$ is the dynamical gauge field and the underline indicates the smooth 
part of the field.  Thus the singularity induces a Wilson loop (of charge $\eta$).  
This argument can be supersymmetrized.  See Section~\ref{sec:localize}.

Though somewhat trivial, it is also natural to define Wilson loop operators associated with 
a global symmetry as the insertion of the function
\begin{equation}
  \label{flavor-WL}
e^{\oint (i\boldsymbol A+\ldots) }\,,
\end{equation}
where for supersymmetry one needs an appropriate curve for integration and 
certain terms in the ellipses.  Since $\boldsymbol A$ is non-dynamical, this term factors out of the 
path integral and is given by its background value.

In the case of the topological symmetry $U(1)_J$, the associated Wilson loop is in fact 
a gauge vortex loop, defined by the singularity (\ref{vor}) in a dynamical gauge field $A$.  This can be seen through a manipulation similar to (\ref{Wilson-BF}):
\begin{equation}\label{BF-gauge-bosonic}
\int \mathcal D\underline{A}\ldots e^{i\int \boldsymbol A\wedge d\underline{A}}
e^{-S[A,\ldots]}\ldots
=
\int \mathcal D A\ldots
e^{-i \eta \oint \boldsymbol A}
e^{i\int \boldsymbol A\wedge dA} 
e^{-S[A,\ldots]}\ldots\,.
\end{equation}
On the left side $A$ contains a singular part and $\underline{A}$ is the smooth fluctuation, while on the right side $A$ is regarded as a smooth gauge field by a change of the integration contour.%
\footnote{%
This manipulation becomes more natural when loop operators are smeared as in \cite{Kapustin:2012iw}.
}

This analysis applies to all vortex loop operators for dynamical gauge fields, and 
we can therefore conclude that they have a trivial expectation value, apart for 
a possible simple multiplicative factor.  This will be verified in the rest of the 
paper by explicit localization calculations.

To prevent the impression that the following is an exercise in futility, we should 
point out that not all vortex loop operators are trivial.  We saw above that vortex loop 
operators for the topological symmetry are the same as Wilson loops, which are 
not trivial.  That still is not so exciting, as we can use the standard definition of 
the Wilson loop and do not require to define it via the vortex loop.  If there is a 
global symmetry under which some of the chiral fields are charged 
({\em i.e.}, a flavor symmetry) 
then the flavor vortex loop for that group will not be trivial nor trivially 
related to a Wilson loop operator.%
\footnote{Likewise, it is not clear whether the matter 
vortex loop operators of Section~\ref{sec:m-vortex} are trivial or not.}
Indeed as we explain in Section~\ref{sec:discuss}, under abelian mirror symmetry 
flavor and topological symmetry are exchanged, so the flavor vortex loop operator 
gets mapped to the gauge Wilson loop operator.

This statement may seem surprising, since we are accustomed to continuous holonomies 
and discrete electric charges.  It is therefore important 
to analyze which values of charges are allowed for the BPS loop operators.  The answer 
seems to depend on the topology of the space.

As discussed after \eqn{vor}, in the case of $\bS^3$ the holonomies can be continuous, which is 
true also for the squashed sphere $\bS^3_b$ discussed in Section~\ref{sec:S3b}.  
The situation on $\bS^2\times\bS^1$ discussed in Section~\ref{sec:S2S1} 
is slightly different.  In that case introducing a non-integer vortex at the north pole of 
$\bS^2$ (wrapping the $\bS^1$) would automatically induce also a singularity at the south pole.  
The total vorticity will cancel, unless we introduce a nontrivial transition function at the equator, 
in which case the total vorticity is integral.  The conclusion is therefore that each vortex can 
have a continuous parameter, but the total vorticity has to be an integer.

Normally Wilson loops are defined only for integer electric charges, which is due to the fact that 
the gauge group (in the abelian case) is $U(1)$ rather than $\bR$.  But on $\bS^3$, which 
is simply connected there is no obstruction of using $\bR$, with continuous electric charges, 
as the gauge group.  The mirror of the flavor vortex would be such a Wilson loop.  On 
$\bS^2\times\bS^1$ there is a non-contractable cycle and large gauge transformations can wind 
around it leading to a quantization condition.  
Again, we can locally break the abelian Wilson loop into two which are not integer, say one 
at the north and one at the south poles of $\bS^2$, 
but the total charge is quantized, which matches the mirror picture of the 
vortex loops.

\section{Localization on $\bS^3$ and harmonic analysis}
\label{sec:localize}

In this section we describe the localization of ${\cal N}=2$ theories on the round $\bS^3$ 
in the presence of a vortex operator defined in (\ref{vor}).  
We use the conventions of \cite{Marino:2011nm}.

The gauge vortex loop operators are given by a choice of a real diagonal matrix $H$ \eqn{vor} 
breaking the gauge symmetry near the singularity to a subgroup.  In the most general 
case, where all eigenvalues of $H$ are distinct, the gauge symmetry is broken to the 
Cartan subalgebra.  For degenerate $H$ there will be larger residual gauge symmetry.

In the proceeding we will study the partition function of generic supersymmetric theories 
in the presence of a gauge vortex loop operator.  The calculation is done using localization 
techniques.

\subsection{Classical factor}
\label{sec:clas}

The localization calculation reduces the path integral on $\bS^3$ to a finite dimensional integral 
over BPS configurations.  This is achieved (see Appendix~\ref{sec:1-loop}) by adding $Q$-exact terms to the action, 
whose bulk part is proportional to the SYM and/or the Chiral actions.  
The modified action determines the 
localization locus and the 1-loop determinant about it.  This locus turn out to be given by 
$\delta\lambda=0$, which are just the BPS equations \eqn{BPS}.  We thus have the 
classical vortex configuration and in addition should integrate off-shell over the 
covariantly constant $\sigma_0$ matrix.

The original action does not necessarily vanish on the BPS configurations.  
We calculate this contribution first.

In the gauge sector there may be a supersymmetric Chern-Simons term with level $k$. 
The action on $\bS^3$ is
\beq
\label{CS}
\cS_\text{SCS}=\frac{k}{4\pi}\int d^3 x \sqrt{g}\, \Tr \left[
\epsilon^{\mu\nu\rho}\left(A_\mu\partial_\nu A_\rho+\frac{2i}{3}A_\mu A_\nu A_\rho\right)
-\bar\lambda\lambda+2D\sigma\right]\,.
\eeq
Including the boundary term \eqn{boundary}, which is required for supersymmetry and 
gauge invariance, and evaluated on the BPS vortex configuration we find
\bal
\label{CS-classical}
i\cS_\text{SCS}^\text{BPS}+i\cS_\text{B}^\text{BPS}
&= \frac{ik}{4\pi}\int d^3 x \sqrt{g}\, \Tr \left[
-2\frac{\sigma_0^2}{R}\right]+kR\int d\varphi_2\,\Tr[H\sigma_0]
\\&
=-\pi i k\Tr\left[(R\sigma_0+iH)^2+H^2\right].
\eal\

It is also possible to include the supersymmetric Yang-Mills action on $\bS^3$ 
\bal
\label{SYMact}
\cS_\text{SYM}=\frac{1}{g_\text{YM}^2}\int d^3 x \sqrt{g}\,\Tr \bigg[
\frac{1}{4} F_{\mu\nu}F^{\mu\nu}+\frac{1}{2} D_\mu\sigma D^\mu\sigma
+\frac{1}{2} \left(D+\frac{\sigma}{R}\right)^2
\\
+\frac{i }{2} \bar\lambda\gamma^\mu D_\mu\lambda
+\frac{i }{2} \bar\lambda[\sigma,\lambda]
-\frac{1}{4R} \bar\lambda\lambda
\bigg]\,.
\eal
Both the bulk and boundary terms \eqn{S-SYM-B} of the SYM action vanish on the 
BPS vortex configurations.

Lastly, another possible supersymmetric term for an abelian vector multiplet is the 
Fayet-Iliopoulos action
\bal
\label{FI}
\cS_\text{FI}=-\frac{i\zeta}{2\pi R}\int d^3x\sqrt g\,\left(D-\frac{\sigma}{R}\right).
\eal
Together with the boundary term \eqn{FI-B} we find
\beq
\label{FI-classical}
\cS_\text{FI}^\text{BPS}+\cS_\text{B}^\text{BPS}=2\pi i\zeta(R\sigma_0+i\eta)\,.
\eeq

The matter fields are described by a 
chiral multiplet in a generic representation $R$ of the gauge group, possibly reducible.  
The supersymmetric action for a chiral multiplet with fields with arbitrary dimension 
$\Delta$ is given by 
\bal
\label{chi}
\cL_\text{chiral} =&\,
D_\mu\bar\phi D^\mu\phi
+\bar\phi\sigma^2\phi
+\frac{i(2\Delta-1)}{R} \bar\phi\sigma\phi
+\frac{\Delta (2-\Delta)}{R^2} \bar\phi\phi
+i\bar\phi D\phi
+\bar FF 
\\ &\,
-i\bar\psi\gamma^\mu D_\mu\psi
+i\bar\psi\sigma\psi
- \frac{2\Delta-1}{2R}\bar\psi\psi
+i\bar\psi\lambda\phi
-i\bar\phi\bar\lambda\psi \,.
\eal
We perform the localization calculation here only for the vortex loops with vanishing $\phi$, 
so this term in the action vanishes on these BPS configurations.

In addition to the vortex we may have a Wilson loop which links it and does not break any further 
supersymmetry.  From \eqn{WL} we see that we will get a term
\beq
\label{V+WL}
W^\text{cl.}=\frac{1}{\dim R} \Tr_{R} \left[\exp\left(2\pi (R\sigma_0+iH)\right)\right].
\eeq

Examining the classical pieces in the different actions as well as in the Wilson loop, we see that 
the inclusion of the vortex loop amounts to the simple replacement $R\sigma_0\to R\sigma_0+iH$.  
The only exception is in the case of Chern-Simons, which has an extra $-\pi ik\Tr(H^2)$, which 
is a simple constant multiplicative factor.

\subsection{Fluctuation determinant}
\label{sec:det}

The localizing action on $\bS^3$ is written in Appendix~\ref{sec:1-loop} and is the sum 
of the SYM action and in the presence of matter fields also the chiral action, both multiplied 
by an arbitrary constant $t$.  For large $t$ the path integral reduces to the saddle points 
of the action and the one-loop determinant about it.

In Appendix~\ref{sec:1-loop} the resulting kinetic operators are written down and 
diagonalized.  For the vector multiplet they are
\beq
\nabla_\mu^{(0)}=\nabla_\mu+i\alpha(A^{(0)}_\mu)\,,
\eeq
which is the usual covariant derivative in the presence of the background gauge field.  The background 
field can be removed by a singular gauge transformation \eqn{redef0}, which makes all the 
fluctuation fields $\tilde \Phi$ non-periodic.  Rather, they satisfy
\bal
\tilde\Phi^\alpha(\vartheta,\varphi_1+2\pi,\varphi_2)
&=e^{2\pi i \alpha (H)}\tilde\Phi^\alpha(\vartheta,\varphi_1,\varphi_2)\,,\\
\tilde\Phi^\alpha(\vartheta,\varphi_1,\varphi_2+2\pi)&=\tilde\Phi^\alpha(\vartheta,\varphi_1,\varphi_2)\,.
\eal

The spherical harmonics with non-standard periodicity conditions are studied in 
Appendix~\ref{sec:harmonics} and give for the vector multiplet the product representation of the 
determinant as \eqn{ts} 
\beq
Z^{\text{vector}}_{\text{1-loop}}(\sigma_0)
=\prod_{\alpha>0}\prod_{n}^\infty\left(n^2+\alpha(R\sigma_0+iH)^2\right)\,.
\eeq
For $\alpha(H)=0$ the product over $n$ starts at $n=1$, but for $\alpha(H)\neq0$ we expect there 
to be extra fermionic (almost-)zero modes and the product starts at $n=0$.  In that case we find
after regularizing the infinite product \eqn{app-vec-result}
\beq
\label{vecdet}
Z^{\text{vector}}_{\text{1-loop}}(\sigma_0)
=\prod_{\alpha>0}\frac{1}{\pi^2}\sinh^2(\pi\alpha(R\sigma_0+iH))\,.
\eeq
For $\alpha(H)=0$ this is multiplied by an extra factor of $1/\alpha(R\sigma_0+iH)^2$.  
This extra factor exactly cancels the Vandermonde determinant, which as discussed at the 
end of Section~\ref{sec:g-vortex}, appears only in the case of degenerate $H$.

A similar analysis for the chiral multiplet in Appendix~\ref{sec:chiral} leads to \eqn{chiraldet}
\beq
\label{chiraldet1}
Z_{\text{1-loop}}^{\text{chiral}}(\sigma_0)
=\prod_{n=1}^\infty\prod_\rho\left(\frac{n+1-\Delta+i\rho(R\sigma_0+iH)}
{n-1+\Delta-i\rho(R\sigma_0+iH)}\right)^{n}
=\prod_\rho s_{b=1}(i-i\Delta -\rho(R\sigma_0 +iH))\,,
\eeq
where $\rho$ are the weights of the representation of the matter fields and 
$s_b(x)$ is the double sine function.

\subsection{Spectral analysis}
\label{sec:spectral}

\begin{figure}[t]
\begin{center}
\raisebox{20mm}{\epsfig{file=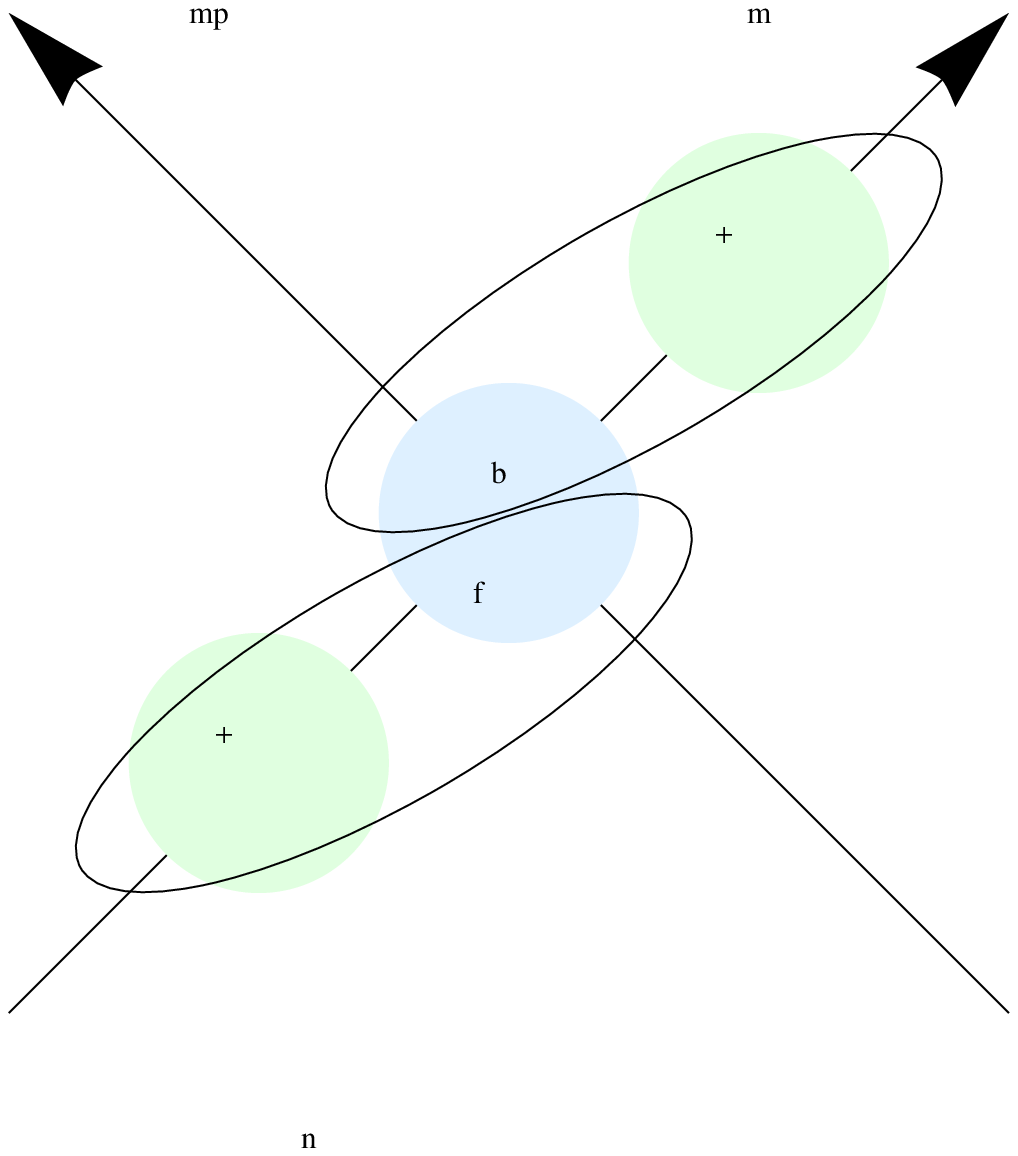,width=25mm
\psfrag{b}{$\scriptstyle{\phi}$}
\psfrag{f}{\footnotesize$F$}
\psfrag{+}{\footnotesize$\psi^+$}
\psfrag{-}{\footnotesize$\psi^-$}
\psfrag{m}{\footnotesize$m$}
\psfrag{mp}{\footnotesize$m'$}
\psfrag{n}{\footnotesize$n=0$}
}}~~~~~
\raisebox{10mm}{\epsfig{file=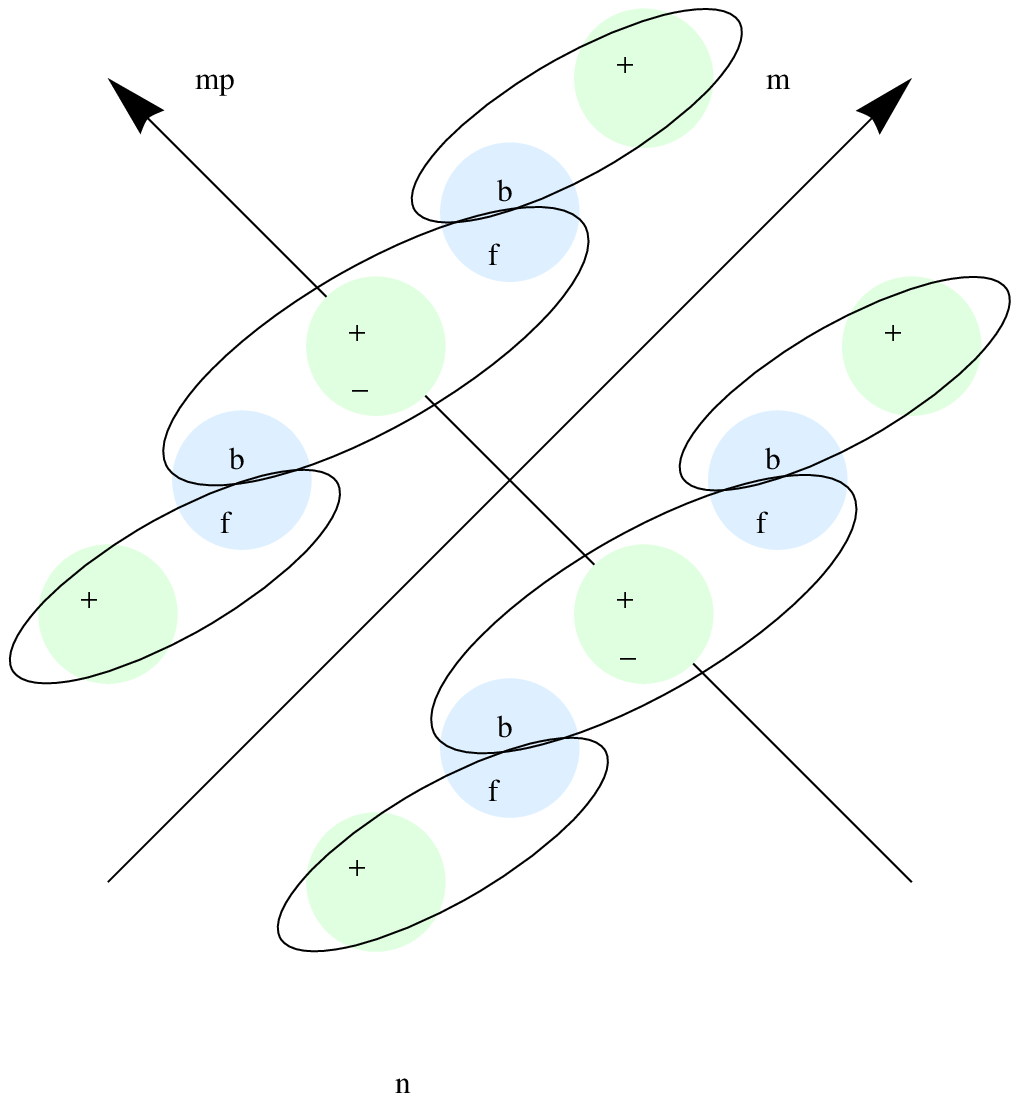,width=45mm
\psfrag{b}{\footnotesize$\phi$}
\psfrag{f}{\footnotesize$F$}
\psfrag{+}{\footnotesize$\psi^+$}
\psfrag{-}{\footnotesize$\psi^-$}
\psfrag{m}{\footnotesize$m$}
\psfrag{mp}{\footnotesize$m'$}
\psfrag{n}{\footnotesize$n=1$}
}}~~~~~
\epsfig{file=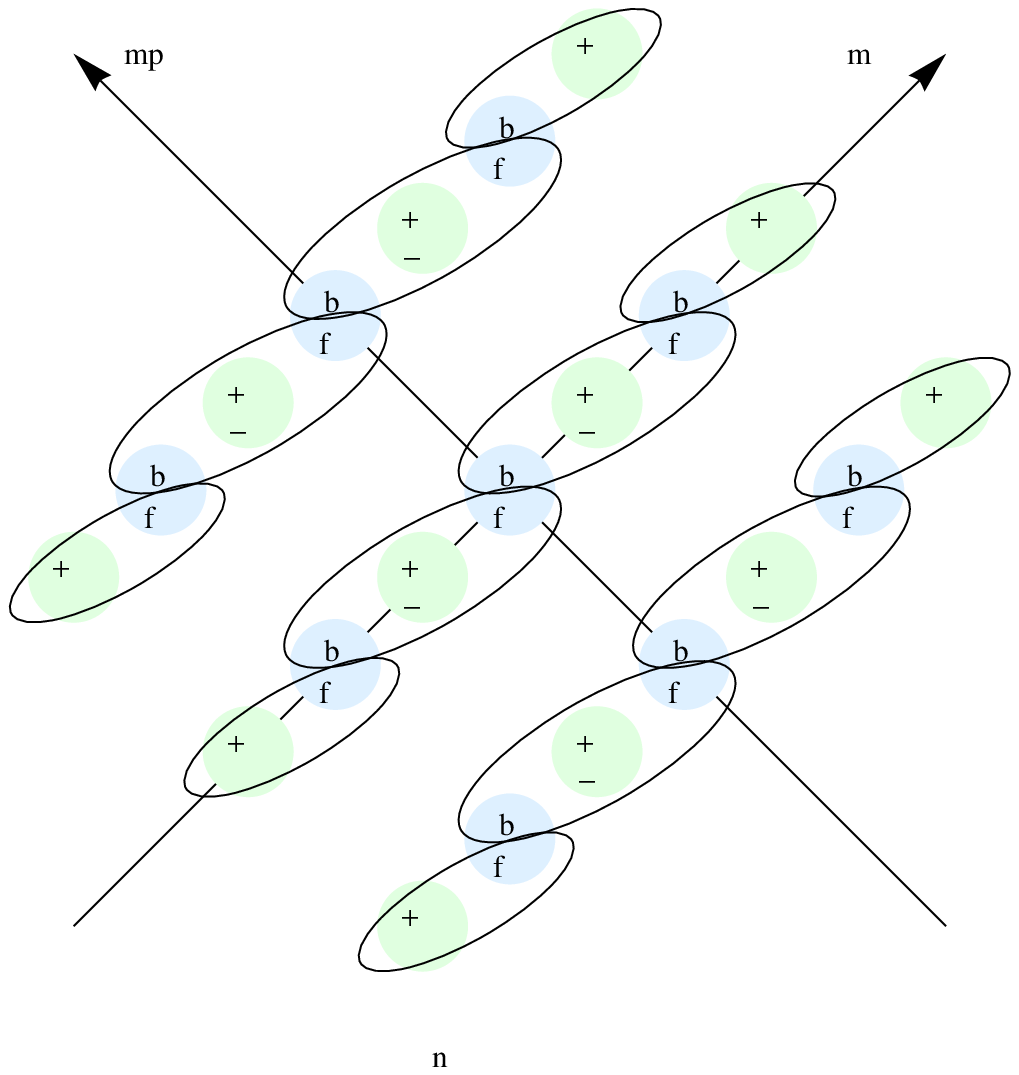,width=65mm
\psfrag{b}{\footnotesize$\phi$}
\psfrag{f}{\footnotesize$F$}
\psfrag{+}{\footnotesize$\psi^+$}
\psfrag{-}{\footnotesize$\psi^-$}
\psfrag{m}{\footnotesize$m$}
\psfrag{mp}{\footnotesize$m'$}
\psfrag{n}{\footnotesize$n=2$}
}
\parbox{15cm}{
\caption{Fluctuation modes of a chiral multiplet.   
The lattices represent states with principle quantum numbers $n=0,1,2$ and the allowed values of 
$m$ and $m'$.  In the presence of the vortex loop these multiplets are broken to smaller 
ones encapsulated by the ovals.  Only the short representations 
(with two modes) contribute to the determinant.
\label{fig:lattice}}}
\end{center}
\end{figure}

\begin{figure}[t]
\begin{center}
\raisebox{19.6mm}{\epsfig{file=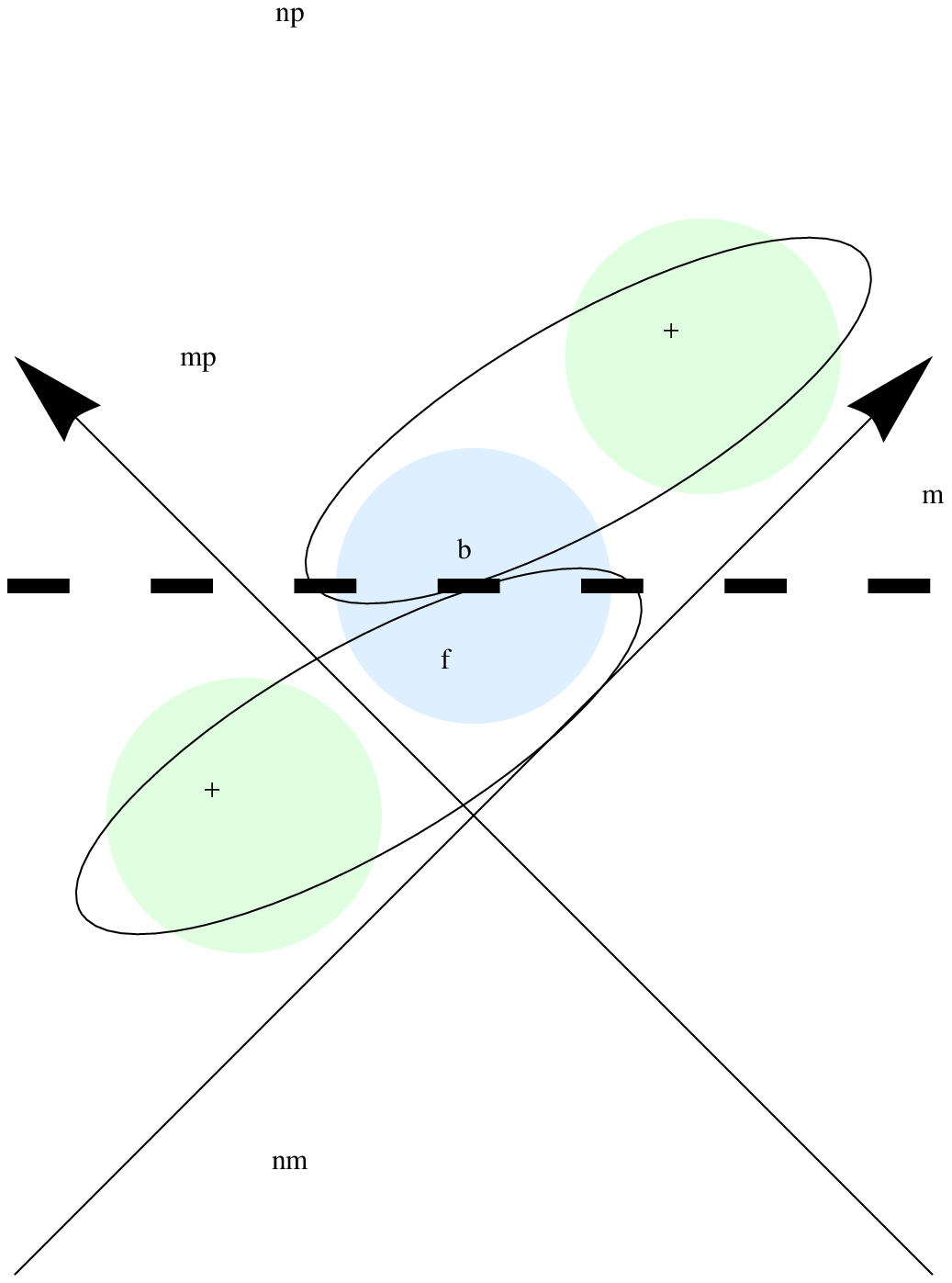,width=25mm
\psfrag{b}{\footnotesize$\phi$}
\psfrag{f}{\footnotesize$F$}
\psfrag{+}{\footnotesize$\psi^+$}
\psfrag{-}{\footnotesize$\psi^-$}
\psfrag{m}{\footnotesize$m$}
\psfrag{mp}{\footnotesize$m'$}
\psfrag{np}{\footnotesize$n=\eta$}
\psfrag{nm}{\footnotesize$n=-\eta$}
}}~~~~~
\raisebox{10mm}{\epsfig{file=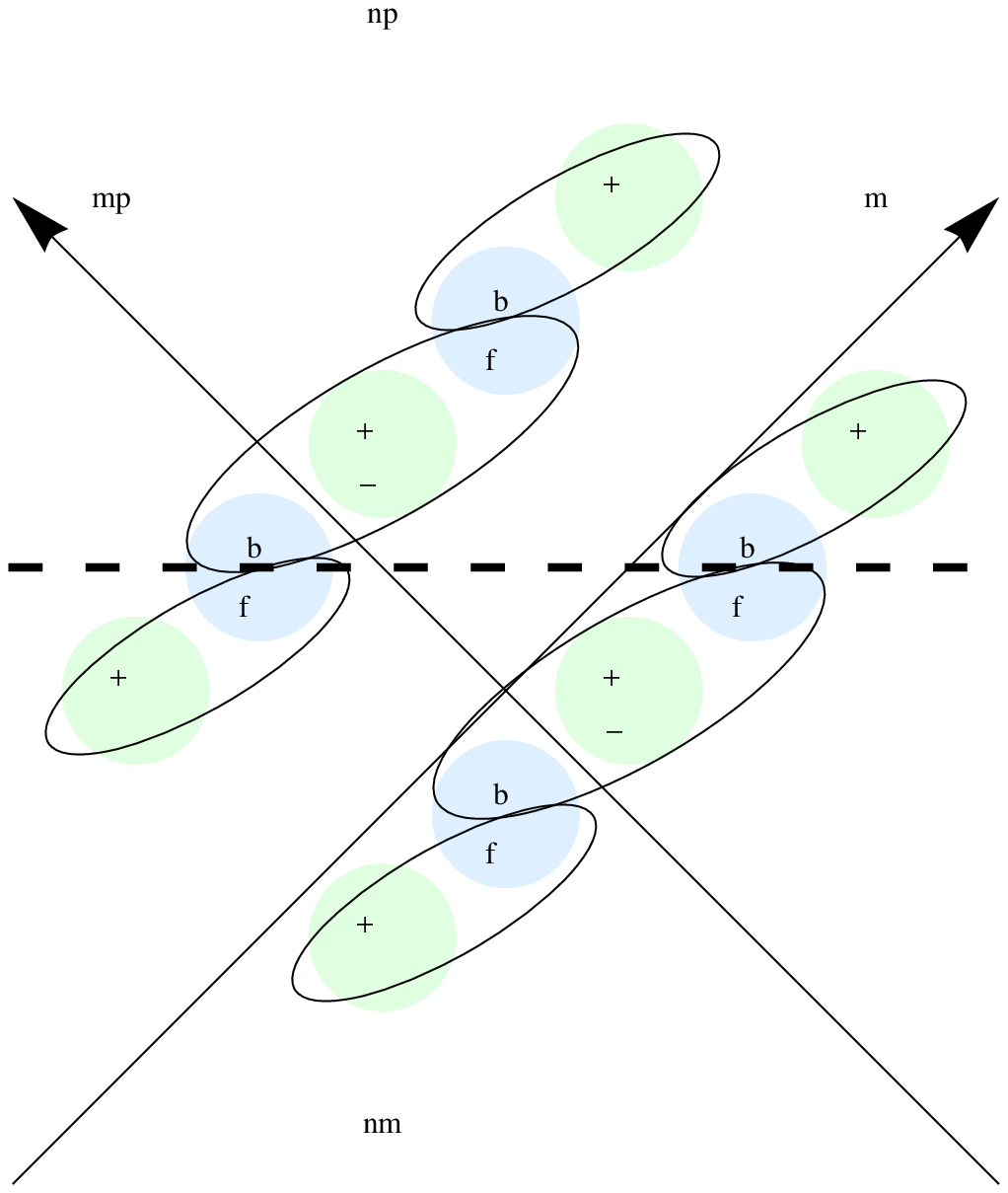,width=45mm
\psfrag{b}{\footnotesize$\phi$}
\psfrag{f}{\footnotesize$F$}
\psfrag{+}{\footnotesize$\psi^+$}
\psfrag{-}{\footnotesize$\psi^-$}
\psfrag{m}{\footnotesize$m$}
\psfrag{mp}{\footnotesize$m'$}
\psfrag{np}{\footnotesize$n=1+\eta$}
\psfrag{nm}{\footnotesize$n=1-\eta$}
}}~~~~~
\epsfig{file=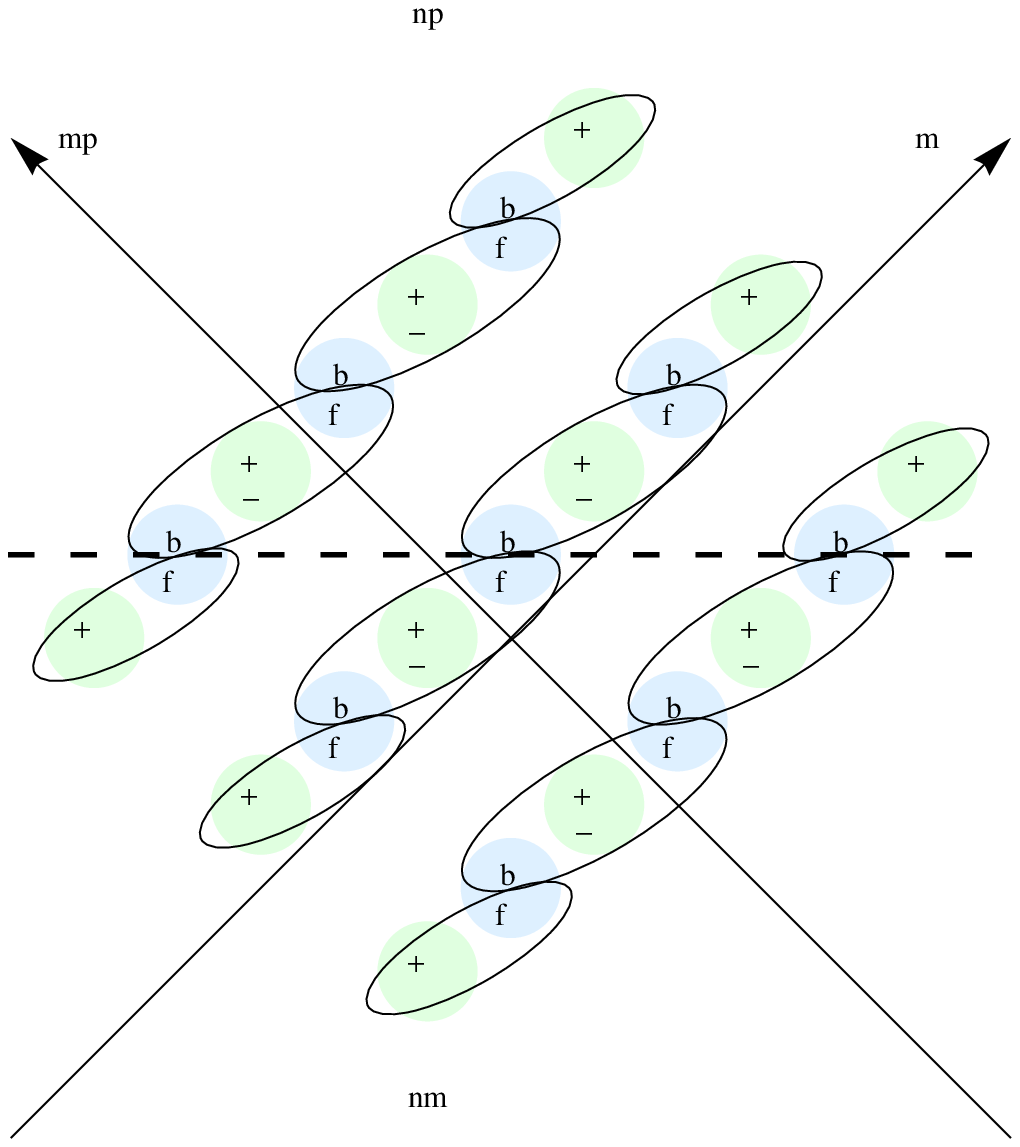,width=65mm
\psfrag{b}{\footnotesize$\phi$}
\psfrag{f}{\footnotesize$F$}
\psfrag{+}{\footnotesize$\psi^+$}
\psfrag{-}{\footnotesize$\psi^-$}
\psfrag{m}{\footnotesize$m$}
\psfrag{mp}{\footnotesize$m'$}
\psfrag{np}{\footnotesize$n=2+\eta$}
\psfrag{nm}{\footnotesize$n=2-\eta$}
}
\parbox{15cm}{
\caption{After introducing $\eta=1/2$ the entire spectrum in Figure~\ref{fig:lattice} 
is shifted by $m\to m+\eta/2$ and $m'\to m'+\eta/2$.  For the multiplets under 
the dashed line the principle quantum number is shifted $n\to n-\eta$ and above 
the dashed line $n\to n+\eta$, which effects the determinant.
\label{fig:spec}}}
\end{center}
\end{figure}

\begin{figure}[t]
\begin{center}
\raisebox{15.8mm}{\epsfig{file=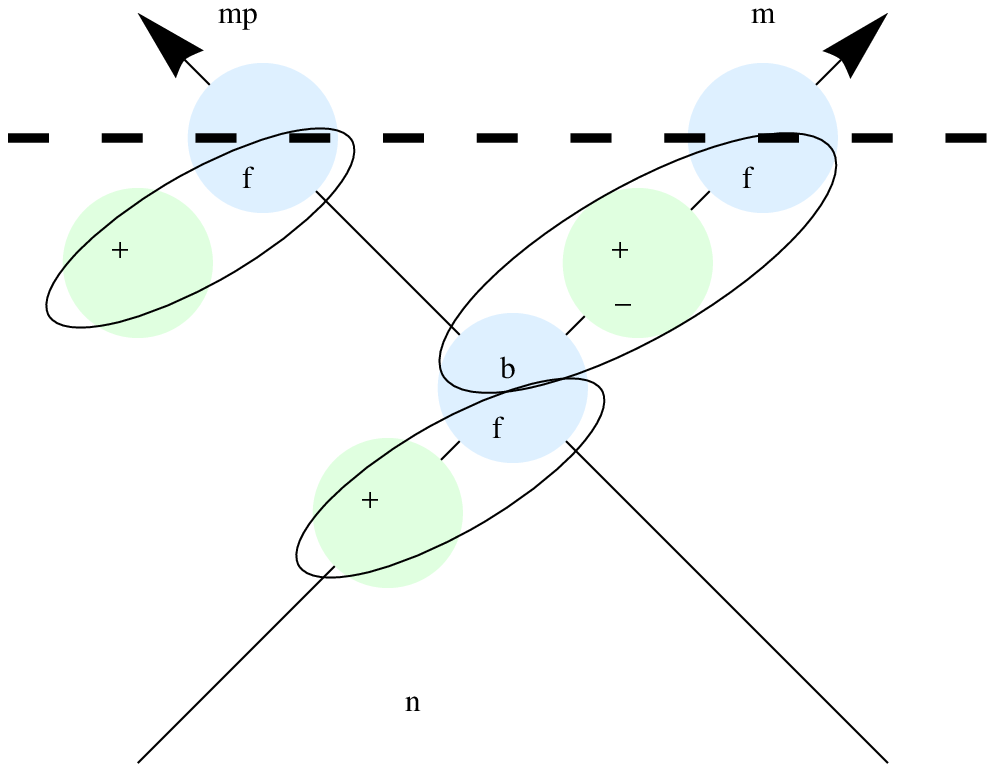,width=35mm
\psfrag{b}{\footnotesize$\phi$}
\psfrag{f}{\footnotesize$F$}
\psfrag{+}{\footnotesize$\psi^+$}
\psfrag{-}{\footnotesize$\psi^-$}
\psfrag{m}{\footnotesize$m$}
\psfrag{mp}{\footnotesize$m'$}
\psfrag{n}{\footnotesize$n=0$}
}}%
\raisebox{10mm}{\epsfig{file=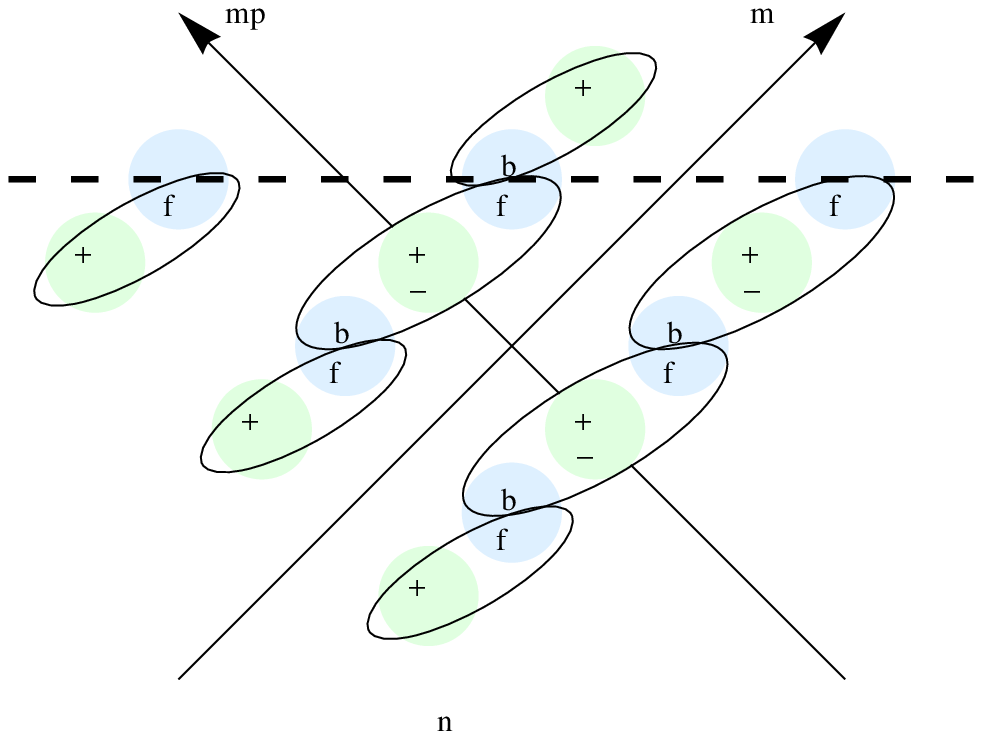,width=55mm
\psfrag{b}{\footnotesize$\phi$}
\psfrag{f}{\footnotesize$F$}
\psfrag{+}{\footnotesize$\psi^+$}
\psfrag{-}{\footnotesize$\psi^-$}
\psfrag{m}{\footnotesize$m$}
\psfrag{mp}{\footnotesize$m'$}
\psfrag{n}{\footnotesize$n=1$}
}}%
\epsfig{file=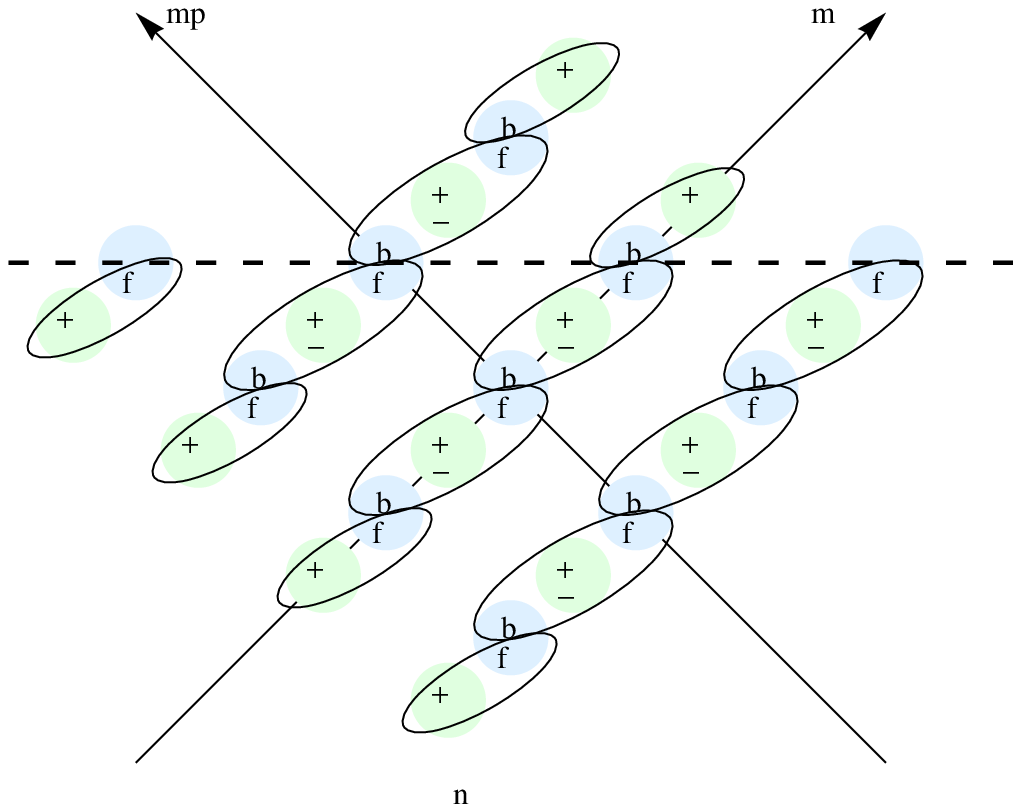,width=75mm
\psfrag{b}{\footnotesize$\phi$}
\psfrag{f}{\footnotesize$F$}
\psfrag{+}{\footnotesize$\psi^+$}
\psfrag{-}{\footnotesize$\psi^-$}
\psfrag{m}{\footnotesize$m$}
\psfrag{mp}{\footnotesize$m'$}
\psfrag{n}{\footnotesize$n=2$}
}
\parbox{15cm}{
\caption{For $\eta=1$ the spectrum is shifted by a full integer.  Here are the new states with principle 
quantum numbers 0, 1 and 2.  The states above the dashed line come from the original multiplet 
with $n-\eta$ and those below from $n+\eta$.  Compared to the spectrum in Figure~\ref{fig:lattice}, 
with the same value of $n$, there are the same number of $\phi$ modes, but an extra 
$\{\psi^+,F\}$ short multiplet, and one $\{\phi,\psi^+\}$ short multiplet gets 
enlarged by an extra $\psi^-$ and $F$ mode.
\label{fig:new}}}
\end{center}
\end{figure}

We have now found that the different ingredients making 
up the matrix model representation of the $\bS^3$ partition function of an $\cN=2$ 
supersymmetric theory in 3d are modified.  The contributions of the CS and FI actions 
evaluated on the BPS configurations are given by \eqn{CS-classical}, \eqn{FI-classical} 
respectively.  The one-loop determinants for the vector and chiral multiplets are in 
\eqn{vecdet} and \eqn{chiraldet1}.  Rather surprisingly, the change to all of them can 
be accounted for by an imaginary shift $\sigma_0\to\sigma_0+iH/R$.%
\footnote{The one exception is the CS term which has an extra $-\pi ik\Tr(H^2)$ term 
in the action, which gives an overall multiplicative factor to the partition function.}
This is also true for the expectation value of a Wilson loop in the presence of the 
vortex \eqn{V+WL}.

Since $\sigma_0$ is integrated over, the deformation of the contour of integration will 
not change the answer as long as no singularities are crossed.  The conclusion, as predicted 
in Section~\ref{sec:flavor}, is therefore 
that at least for abelian theories {\em the vorticity $H$ does not effect the partition function}.%
\footnote{Note that in pure topological Chern-Simons theory a non-supersymmetric vortex 
loop operator was defined in \cite{ms-zoo} and it was argued there that it is nontrivial 
and equal to a Wilson loop observable.}

There are some subtleties in this statement due to the fact that the double sine function 
arising in the one-loop determinant does have poles at integer values of $H$.  In this 
section we discuss these subtleties and their origin and propose a prescription to 
resolve them.

After the imaginary shift of $\sigma_0$, the determinant for a chiral field is given by 
\eqn{chiraldet}.  For $\eta=1$ this is
\beq
\label{monodromy}
s_{b=1}(i-i\Delta -(R\sigma_0 +i\eta))
=\frac{1}{\pi}\sin(\pi(i-i\Delta-R\sigma_0))\, s_{b=1}(i-i\Delta-R\sigma_0)\,.
\eeq
The situation for the vector multiplet is simpler, since the adjoint representation is self 
conjugate.  Under an integer imaginary shift 
$\sinh(\pi(R\sigma_0+in))=(-1)^n\sinh(\pi R\sigma_0)$ \eqn{vecdet}.  The other change is that the 
denominator $1/(\pi(R\sigma_0+in))$ which usually cancels the regular Vandermonde factor 
is no longer there, due to the extra goldstino zero modes.

But the transformation of the chiral multiplet \eqn{monodromy} 
is a nontrivial transformation, meaning the spectrum of fluctuations really changes 
even for integer $\eta$.  This is illustrated in Figures~\ref{fig:lattice}-\ref{fig:new}.  Starting 
with a supermultiplet of the $OSp(2|2,2)$ supersymmetry group on $\bS^3$, the 
vortex breaks the symmetry down to $SU(1|1,1)$, and the original symmetry is 
not restored at integer $\eta$, leading to a different multiplet structure.  The breaking of the 
supersymmetry multiplet of fluctuation modes is analyzed in Appendix~\ref{sec:susyrepre}.

Exactly half of the states with principle quantum number $n$ get deformed to states with 
quantum number $n-\eta$ and half to $n+\eta$.  For small $\eta>0$ there are $(n+1)(n+2)/2$ 
modes of the scalar field $\phi$ with $n+\eta$ and $n(n+1)/2$ with $n-\eta$.  For $\eta=1$ 
there are $(n+1)(n+2)/2+n(n+1)/2=(n+1)^2$ states with principle quantum number 
$n$, which is the same as the number for $\eta=0$.  The number of fermi fields does 
change, with one extra fermion of either chirality.  To keep the SUSY structure consistent, 
there are also two extra modes of the auxiliary field $F$.

In terms of the multiplets of the smaller group, for $\eta=1$ there is one extra long 
multiplet and one less short multiplet with $\phi$ and $\psi^+$ and one more short 
multiplet with $\psi^+$ and $F$.  The extra four modes are eigenstates of the Laplacian 
and Dirac operator with the relevant eigenvalues, which were not there for $\eta=0$, 
so these are modes which do not belong to the $OSp(2|2,2)$ representation.  They 
are in fact singular modes, which normally are not included in the spectral analysis.  
They are part of larger nonunitary representations of this group, which are not part 
of the unitary subrepresentation.  The explicit analysis of the spectral flow means that 
these states should be counted and they lead to the factor in \eqn{monodromy}.

Of course if the theory has only self conjugate representations, or all representations 
are paired up with their conjugates, then there are extra cancellations and for integer 
$\eta$ one finds only at most a sign factor.  This is the case for theories with $\cN=4$ 
SUSY.  But as stated, for $\cN=2$ SUSY, the effect of spectral flow is very nontrivial.

The conclusion of the above discussion would seem to imply that the partition 
function in the presence of the vortex loop, while constant for $0\leq\eta<1$, 
jumps for integer $\eta$.  
There seem to be two possible prescriptions.
The first is to use the values of the 1-loop determinant that we have found, but 
keep the integration contour such that it does not cross the poles.  The second 
possibility is to not do the spectral flow as discussed above, but as the singular 
modes show up in the spectrum, replace them with other modes which were 
singular before and now become regular.  In this way we restore the original 
spectrum for integer $\eta$.

While the second possibility seems more appealing physically, an analysis 
of flavor vortex loop operators, where $\sigma_0$ is not integrated over 
seems to prefer the first interpretation.  This allows for them to be dual to 
regular gauge Wilson loops under abelian mirror symmetry.

\section{Localization on $\mathbb S^3_b$ by index theory}
\label{sec:S3b}

In this section, we will compute by localization the expectation value of the gauge and flavor 
vortex loops on a deformation of the three-sphere, commonly denoted as  $\mathbb S^3_b$.
As in previous sections, we will consider an arbitrary ${\mathcal N}=2$ gauge theory with a chiral multiplet in representation $R$ of the gauge group.
We will first explain how to use the equivariant index theory to compute the one-loop determinant that appears in the partition function.
Then we will apply the technique to compute the expectation value of the vortex loop.
We provide many technical details in Appendix~\ref{sec:index-details}.

\subsection{Partition function}\label{sec:S3bPF}

This geometry $\mathbb S^3_b$, also known as the ellipsoid, is defined by the metric
\beq
\label{torusf-b}
ds^2=R^2\left(f(\vartheta)^2d\vartheta^2+b^2\sin^2\vartheta\,d\varphi_1^2+b^{-2}\cos^2\vartheta\,d\varphi_2^2\right)\,,
\eeq
where
\begin{equation}
  f(\vartheta)\equiv(b^{-2}\sin^2\vartheta+b^{2}\cos^2\vartheta)^{1/2}\,.
\end{equation}
To describe spinors, we will use the orthonormal frame given by%
\footnote{%
Comparison with \cite{Hama:2011ea} is simple with this definition of $e^{\hat\mu}$: $(\vartheta,\varphi_1,\varphi_2)_\text{here}=(\theta,-\chi,\varphi)_\text{there}$.
Also we have the following change of conventions: $(A_\mu,C,\bar\epsilon,\bar\lambda,\bar\psi)_\text{here}=-(A_\mu,C,\bar\epsilon,\bar\lambda,\bar\psi)_\text{there}$, 
 $(\epsilon,\bar\epsilon)_\text{here}= (-\bar \epsilon,\epsilon)_\text{there}$,
$v^\mu_\text{here}=-v^\mu_\text{there}$, $(F,\bar F)_\text{here}=(-F,-\bar F)_\text{there}$.
}
\beq
\label{S3btorusviel}
e^{\hat 1}=Rb^{-1}\cos\vartheta\,d\varphi_2\,,
\qquad
e^{\hat 2}=-Rb\sin\vartheta\,d\varphi_1\,,
\qquad
e^{\hat 3}=R\,
f(\vartheta)d\vartheta\,.
\eeq
For localization we will use the supercharge $Q=\delta_\epsilon+\delta_{\bar\epsilon}$  generated by the two spinors
\begin{equation}
\label{S3b-spinors}
\epsilon\equiv
\frac{1}{\sqrt 2}
  \begin{pmatrix}
e^{\frac{i}{2}(\varphi_1+\varphi_2+\vartheta)}
\\
e^{\frac{i}{2}(\varphi_1+\varphi_2-\vartheta)}
  \end{pmatrix}
\,,
\quad
\bar\epsilon\equiv
\frac{1}{\sqrt 2}
  \begin{pmatrix}
-e^{\frac{i}{2}(-\varphi_1-\varphi_2+\vartheta)}
\\
e^{\frac{i}{2}(-\varphi_1-\varphi_2-\vartheta)}
  \end{pmatrix}\,.
\end{equation}
As shown in \cite{Hama:2011ea}, these spinors satisfy a variant of the 
Killing spinor equations, ensuring that the algebra generated by supersymmetry 
transformations of fields on $\mathbb S^3_b$ closes.
In particular, the supercharge $Q$ squares to a sum of bosonic symmetries
\begin{equation}
\label{Q-squared}
  Q^2=i {\cal L}_v +i \sigma - v^\mu A_\mu+\frac{1}{2R}(b+b^{-1}){\cal R}\,,
\end{equation}
where ${\cal L}_v$ is the Lie derivative%
\footnote{%
Here the Lie derivative ${\mathcal L}_v$ acts as 
$(\mathcal L_v+i v^\nu A_\nu)\cdot A_\mu=v^\nu F_{\nu\mu}$, 
$\mathcal L_v w_\mu=v^\nu \nabla_\nu w_\mu + [\nabla_\mu v^\nu]w_\nu$ etc., 
and in particular includes a Lorentz rotation.
We also note that ${\cal L}_v +i v^\mu A_\mu$ is a gauge covariant Lie derivative.}
along the vector field 
\begin{equation}
\label{vS3b}
  v\equiv \bar\epsilon \gamma^{\hat \mu}\epsilon \, e_{\hat \mu}
=
R^{-1}\left(
b^{-1}\frac{\partial}{\partial \varphi_1}
+
b\frac{\partial}{\partial \varphi_2}
\right)\,.
\end{equation}
We write ${\cal R}=\mathcal R_0-\Delta \mathcal F$, where ${\cal R}_0$ generates the canonical R-symmetry, and $\mathcal F$ is the generator of flavor symmetry $U(1)_\text{F}$.
In our convention the lowest component $\phi$ has eigenvalues $\mathcal R_0=0$ and $\mathcal F =1$.

In Section~\ref{sec:localize} and Appendix~\ref{sec:1-loop}, the one-loop determinant in the presence of a 
vortex loop on the round sphere $\mathbb S^3_{b=1}$ was computed by 
expanding the fields in spherical harmonics.  With the deformation $b\neq 1$ 
turned on, the analysis of harmonics is possible but more complicated 
\cite{Hama:2011ea}, especially when the vortex loop is inserted.  We thus work 
in an alternative approach based on the equivariant index theory \cite{MR0482866}.  
First we will reproduce the known one-loop determinant 
that appears in the partition function on $\mathbb S^3_b$.

The equivariant index theory was used in \cite{pestun,Gomis:2011pf,Benini:2012ui} to compute one-loop determinants in other geometries.
In this approach, one deforms the Lagrangian by $t Q\cdot V$ for some fermionic functional $V$.
We choose
\begin{equation}
\label{Vchoice}
  V=V_\text{vec}+V_\text{chi}\,,
\end{equation}
where
\begin{equation}
\label{Vvecchi}
  V_\text{vec}=(Q\lambda)^\dagger\lambda 
+(Q\bar\lambda)^\dagger\bar\lambda\,,
\quad\quad
V_\text{chi}=(Q\psi)^\dagger\psi 
+(Q\bar\psi)^\dagger\bar\psi\,.
\end{equation}
Let $H$ and $K$ be two copies of $U(1)$ generated respectively by
\begin{equation}
  \label{S3b-HK-def}
-i(\partial_{\varphi_2} + \partial_{\varphi_1})-\mathcal R_0
\quad
\text{ and }
\quad
-i(\partial_{\varphi_2} - \partial_{\varphi_1})\,.
\end{equation}
The bosonic generator $Q^2$ in (\ref{vS3b}) specifies the action of the complexification $G_{\mathbb C}$ of the group $G$, which we define as%
\footnote{%
See also (\ref{S3b-group-para}).
}
\begin{equation}
  \label{G-def-S3b}
  G\equiv H\times K\times (\text{maximal torus of gauge group})\times U(1)_{\rm F}\,.
\end{equation}
One then computes the equivariant index 
\begin{equation}
  \label{eq-index-def}
\text{ind}_g D_{10}=\text{Tr}_{{\rm Ker}D_{10}} ( g)-\text{Tr}_{{\rm Coker}D_{10}} (g)
\end{equation}
of a differential operator $D_{10}$ that appears in $V$ as a function of $g\in G$.  
The precise definition of $D_{10}$ is given in Appendix~\ref{sec:index-details} .
We also show there that the path integral localizes to the configurations
\begin{equation}
\label{S3b-config}
A_\mu=0\,,\quad\quad
\sigma=\text{constant}\,,\quad\quad
D=-\sigma/R\,,\quad\quad
\phi=F=0\,.
\end{equation}
The one-loop determinant is obtained from the index by the rule
\begin{equation}
\label{rule}
  \text{ind}_g D_{10}|_{g=\exp(c Q^2)}=\sum_j c_j e^{ iw_j}
\quad\rightarrow\quad
Z_\text{1-loop}=\prod_j w_j^{-c_j/2}\,.
\end{equation}
Here $c_j$ is a sign $\pm 1$, $i w_j$ is the eigenvalue of $c Q^2$ for mode $j$, and $Q^2$ is evaluated at the saddle point (\ref{S3b-config}).
The constant $c$ affects only the overall normalization, and will be set to a convenient value.

In the set-ups of \cite{pestun,Benini:2012ui}, $D_{10}$ was transversally elliptic and the index $\text{ind}_gD_{10}$ received contributions from the fixed points of the vector field in $Q^2$.
In our case of $\mathbb S^3_b$, there is no fixed point with respect to the single $U(1)$ action generated by the vector field $v$.
How can we compute the equivariant index in such a situation?
We first rewrite the gauge field $v$ in terms of the Hopf fibration coordinates $\phi=\varphi_2-\varphi_1$, $\psi=\varphi_2+\varphi_1$ (See (\ref{hopf})):
\begin{equation}\label{v-decomposition-S3b}
v
=
 R^{-1}(b+b^{-1}) \partial_\psi  + R^{-1}(b-b^{-1})\partial_\phi 
\,.
\end{equation}
The vector fields $2\partial_\psi$ and $2\partial_\phi$ respectively generate the action of $H$ and $K$ above.
In particular, $H$ rotates the Hopf fibers, and thus acts on $\mathbb S^3_b$ freely.
With respect to the $H\times K$-action, $D_{10}$ fails to be elliptic
but it is transversally elliptic.
When part of the group action is free, the index of a transversally elliptic operator can be expressed in terms of the index of a transversally elliptic operator on the quotient space ($\mathbb S^2$ in our case) \cite{MR0482866}.
This is reviewed in Appendix~\ref{sec:index-details}.
By the fixed point formula then, the index receives contributions from the fixed points of the other $U(1)$ action generated by $\partial_\phi$.
In terms of the original three-dimensional geometry, these fixed points correspond to the circle fibers at the north and the south poles of the $\mathbb S^2$ ($\theta=2\vartheta$ equal to $0$ and $\pi$ respectively).

Let us set ${\rm Q}=b+b^{-1}$, $\hat \sigma\equiv R\sigma$.
We now compute the index $\text{ind}_g D_{10}$ with $g=e^{c  Q^2}$, $c= -iR$.
For the chiral multiplet, we can write $\text{ind}_g D_{10}^\text{chi}= \text{ind}_g D_{10,\mathbb C}^\text{chi}+ \text{ind}_{g^{-1}} D_{10,\mathbb C}^\text{chi}$ as we show in Appendix~\ref{sec:index-details}.
The reduction of $D_{10,\mathbb C}^\text{chi}$  to $\mathbb S^2$ near the north pole $\theta=0$
is a twisted  Dolbeault operator $D_{\bar z}$.
The local complex coordinate is given by $z\sim \theta e^{-i\phi}$.%
\footnote{%
The one-form in (\ref{hopf}) can be written as $d\psi+\cos\theta d\phi=d(\psi\pm\phi)-(\pm 1 -\cos\theta)d\phi$.   Thus at the north (south) pole $\theta=0$ $(\pi)$ the base is parameterized by $(\theta,\phi)$ and the fiber by $\psi+\phi= 2 \varphi_2$ ($\psi-\phi= 2 \varphi_1$).
}
The equivariant index for the untwisted Dolbeault operator is $(1-q^{-1})^{-1}$, where $q\in U(1)$ is the weight for the $U(1)$ action $z\mapsto t z$.
We identify $q$ with $e^{i(b-b^{ -1})}$.
As the contribution to $\text{ind}_gD_{10,\mathbb C}^\text{chi}$ from $\theta=0$ we obtain
\begin{equation}
\label{theta-zero-contribution}
  \sum_{n\in \mathbb Z} 
e^{inb}
e^{\half i\Delta {\rm Q}}
\frac{1}{1-e^{-i(b-b^{-1})}
}
\sum_w e^{w\cdot \hat{\sigma}}\,,
\end{equation}
where the sum is over the weights in the representation $R$.
Similarly, the fixed point $\theta=\pi$ on $\mathbb S^2$, where we identify $z\sim (\pi-\theta)e^{i\phi}$ and $q$ with $e^{-i(b-b^{-1})}$, contributes
\begin{equation}
\label{theta-halfpi-contribution}
    \sum_{n\in \mathbb Z} 
e^{inb^{-1}}
e^{\half i\Delta {\rm Q}}
\frac{1}{1-
e^{i(b-b^{ -1})}
}
\sum_w e^{w\cdot \hat{\sigma}}\,.
\end{equation}
As we explain in Appendix~\ref{sec:index-details}, the index theory instructs us to expand (\ref{theta-zero-contribution}) as%
\footnote{%
There are two allowed choices as explained below (\ref{Dolb-index-S2}).
}
\begin{equation}\label{theta-zero-expand}
  -
  \sum_{n\in\mathbb Z}
e^{inb}
e^{\half i\Delta {\rm Q}}
\sum_{k=1}^\infty e^{i k (b-b^{-1})}
\sum_w e^{w\cdot \hat{\sigma}}\,,
\end{equation}
and (\ref{theta-halfpi-contribution}) as 
\begin{equation}\label{theta-halfpi-expand}
  \sum_{n\in\mathbb Z}
e^{inb^{-1}}
e^{\half i \Delta {\rm Q}}
\sum_{k=0}^\infty e^{ik (b-b^{-1})}
\sum_w e^{w\cdot \hat{\sigma}}\,.
\end{equation}
By using the shift invariance $\sum_{n\in\mathbb Z} e^{in b^{\pm}} e^{\pm i k b^{\pm1 }}=\sum_{n\in\mathbb Z} e^{in b^{\pm}}$ and then splitting $\sum_{n\in\mathbb Z}$ into $\sum_{n\geq 0}+\sum_{n<0}$ in (\ref{theta-zero-expand}) and (\ref{theta-halfpi-expand}), we obtain the total contribution
\begin{equation}
   \sum_{n=0}^\infty
e^{inb^{-1}}
e^{\half i\Delta {\rm Q}}
\sum_{k=0}^\infty e^{ik b}
\sum_w e^{w\cdot \hat{\sigma}}
-
e^{-i{\rm Q}}
  \sum_{k=0}^\infty
e^{-ikb}
e^{\half i\Delta {\rm Q}}
\sum_{n=0}^\infty e^{-i n b^{-1}}
\sum_w e^{w\cdot \hat{\sigma}}
\,.
\end{equation}
The rule (\ref{rule}) applied to $\text{ind}_g D_{10}^\text{chi}= \text{ind}_g D_{10,\mathbb C}^\text{chi}+ \text{ind}_{g^{-1}} D_{10,\mathbb C}^\text{chi}$ gives, up to an overall sign, 
\begin{equation}
\label{S3b-chiral-oneloop}
Z_\text{1-loop}^\text{chi}= 
\prod_{w\in R}
  \prod_{m,n\ge0}
 \frac{mb+nb^{-1}+\frac {\rm Q}2+iw\cdot \hat{\sigma}+\frac {\rm Q}2(1-\Delta)}
      {mb+nb^{-1}+\frac {\rm Q}2-iw\cdot \hat{\sigma}-\frac {\rm Q}2(1-\Delta)}
 ~=~\prod_{w\in R} s_b\big(\tfrac{i {\rm Q}}2(1-\Delta)-w\cdot \hat{\sigma}\big).
\end{equation}
This is the well-known one-loop determinant for the chiral field \cite{Hama:2011ea}.
If we had kept the constant $c$ arbitrary, it would have canceled between the numerator and the denominator.

For a vector multiplet, the relevant differential operator is the differential in the de Rham complex twisted by the adjoint bundle (with a degree shifted by one).
Since the de Rham and Dolbeault complexes are related by complexification ($\Omega^0_{\mathbb C}=\Omega^{0,0}$, $\Omega^1_{\mathbb C}=\Omega^{1,0}\oplus \Omega^{0,1}$,  $\Omega^2_{\mathbb C}=\Omega^{1,1}$),
the index of the untwisted de Rham complex $D_\text{dR}: \Omega^0\rightarrow \Omega^1\rightarrow \Omega^2$ on $\mathbb C$ is given as
\begin{equation}
  \text{ind}\, D_\text{dR}
=
(1-t^{-1})
  \text{ind}\, \bar \partial= 1\,.
\end{equation}
Let $\alpha$ denote the roots of the gauge group.
Then the north pole $\theta=0$ contributes
\begin{equation}\label{S3b-vec-north}
  -
\sum_{n\in\mathbb Z} e^{inb}
\sum_{\alpha}  e^{\alpha\cdot\hat\sigma}
\end{equation}
to $\text{ind}\,D_{10}$ for the vector multiplet,
and the south pole $\theta=\pi$ contributes
\begin{equation}\label{S3b-vec-south}
  -
\sum_{n\in\mathbb Z} e^{inb^{-1}} 
\sum_{\alpha}  e^{\alpha\cdot\hat\sigma}
\,.
\end{equation}
The resulting one-loop determinant is
\begin{equation}
Z_\text{1-loop}^\text{vec}=  \prod_{\alpha>0} \sinh(\pi b \alpha\cdot\hat \sigma)\sinh(\pi b^{-1}\alpha\cdot\hat \sigma)\,.
\end{equation}
The product is over the positive roots.
This also agrees with the results in the literature  \cite{Hama:2011ea}.

\subsection{Vortex loop expectation values}

 On $\mathbb S^3_b$ with $b\neq 1$, BPS loop operators can only be supported along the circle fibers at $\theta=0,\pi$.
 Let us for now focus on the gauge vortex loop with vorticity $H$ along the fiber at $\theta\equiv 2\vartheta=0$.
This is characterized by the $Q$-invariant background configuration%
\footnote{%
The delta functions should be understood to be $\delta(1-\cos(\vartheta-\vartheta_0))$ with small $\vartheta_0>0$ 
}
\begin{equation}\label{S3b-config-vortex}
  \begin{aligned}
F_{\mu\nu}= \frac{H}{b^2 R^2} \delta(1-\cos\vartheta) \epsilon_{\mu\nu\rho}v^\rho\,,
\quad
\sigma={\rm const}.,
\\
D=-\frac{\sigma}{R f(\vartheta)} -i\frac{ H}{b^2 R^2} \delta(1-\cos\vartheta)\,.
  \end{aligned}
 \end{equation}
We claim that it has the effect of shifting the contributions from both the poles as $\hat{\sigma} \rightarrow \hat{\sigma}+i b^{-1}H$.

For the contribution from $\theta=\pi$, the gauge parameter in (\ref{Q-squared}) becomes $\sigma  +i(Rb)^{-1}H$ simply because the gauge field is turned on.
This induces the shift in $\hat\sigma$.

The effect on the contribution from  $\theta=0$ is more subtle.
In Section~\ref{sec:localize}, we constructed explicitly the eigenmodes of the kinetic operators on the round sphere.   We saw that when the vortex loop is inserted and the eigenvalue $\eta$ of $H$ is turned on, generically certain modes that are singular must be allowed to fluctuate, contributing to the one-loop determinant.   The analysis there was global and specific to the round metric, but the local behaviors of the allowed singular modes must be intrinsic to the vortex loop operator.   Thus in the current approach to the one-loop determinant based on the index theory, we should compute the local contributions to the index by taking into account the local modes that are singular.
This means that we should sum the $U(1)$ weights for $z^{k +\eta}$ 
($k=0,1,\ldots$) instead of $z^k$ if we work in the gauge where $A_{\varphi_1}$ is zero.%
\footnote{%
If we work in the gauge where $A_{\varphi_1}=H$, the relevant modes are $z^k |z|^\eta$ and the $H$-dependence comes from the term in $Q^2$ that involves $A_{\varphi_1}$ explicitly.
}
Then (\ref{theta-zero-contribution}) receives an extra overall factor $e^{ ib^{-1}\eta}$, which is equivalent to shifting $\hat \sigma \rightarrow \hat{\sigma} +i b^{-1}H$.

Thus the total effect of the vortex loop on the one-loop determinant is the shift $\hat{\sigma} \rightarrow \hat{\sigma} + i b^{-1}H$.
This generalizes the results (\ref{vecdet}) and (\ref{chiraldet}) for $\mathbb S^3_{b=1}$ to $\mathbb S^3_b$.

We also need to evaluate the Chern-Simons term in the presence of a vortex loop on $\mathbb S^3_b$.
For $b=1$, this was done in (\ref{CS-classical}) using the boundary term (\ref{boundary}).
We specialize to the abelian case and set $H=\eta$.
From (\ref{S3b-config-vortex}) we find that $\mathcal S_{SCS}= -\pi k (\hat \sigma +i b^{-1}\eta)^2$.%
\footnote{%
If we add a constant boundary term $\propto \Tr H^2$ in Section \ref{sec:clas}, the results here and there agree.
}
The effect of the vortex loop on the Chern-Simons action is again the shift $\hat \sigma \rightarrow \hat{\sigma} +i b^{-1} \eta$.

Repeating the same arguments above for a vortex loop with vorticity $\eta$ at $\vartheta=\pi$, we find that the effect is the shift $\hat \sigma \rightarrow \hat{\sigma} -i b \eta$.

Let us assume that the gauge group is $U(1)$ and consider the BF coupling (\ref{SUSY-BF}) that appears as $e^{-\mathcal S_{BF}}$.
It may be evaluated via the relation (\ref{BF-CS}) between the BF and Chern-Simons terms.
If we use the full gauge multiplet configuration in (\ref{S3b-config}), we find that $\mathcal S_{BF}$ equals $2\pi i (\hat\sigma+i b^{-1}\eta)\zeta$.
If this were included in the path integral, then all the contributions inside the $\hat\sigma$-integral would receive a uniform shift, so that after integration the vortex loop does not affect the partition function at all.
From the point of view of the $SL(2,\mathbb Z)$ action \cite{Witten:2003ya} on superconformal theories, however, it is more natural to not include terms proportional to $\eta$ in the BF coupling, as follows from the discussion in \cite{Kapustin:2012iw}.%
\footnote{%
See the discussion around (2.25) of \cite{Kapustin:2012iw}.
They define loop operators as an action on the partition function that depends on a background gauge field coupled to a chosen global symmetry.
The gauge vortex loop corresponds to $S D_\omega$ in their notation.
Since $(D_\omega Z)[A]=Z[A+A_\omega]$ has $A$, not $A+A_\omega$, as the argument, the $S$-action yields the BF coupling between a new background field and $A$, not $A+A_\omega$.
}
See also (\ref{BF-gauge-bosonic}).
Thus $\mathcal S_{BF}=2\pi i \hat\sigma \zeta$.
Since all other contributions uniformly receive the shifts (\ref{S1S2shifts}), the only effect of the vortex loop is to multiply the partition function by $\exp(-2\pi b^{-1} \eta\zeta)$.
For a gauge vortex loop that has a singularity in a non-abelian gauge field, we cannot rule out the existence of non-perturbative corrections.

On the other hand, if the singularity is in a non-dynamical gauge field coupled to an 
(abelian or non-abelian) flavor symmetry, $\sigma$ is replaced by a real mass and is 
not integrated over.   Then the shift has a non-trivial effect.

Let us summarize the results of localization calculation for vortex loops placed at $\vartheta=\theta/2=0$ and $\pi/2$.
The partition function $Z_{\mathbb S^3_b}(\zeta,\ldots)$ of an abelian gauge theory is a function of the FI parameter $\zeta$, and the effect of a gauge vortex loop $ V_\eta^\text{gauge}$ is the multiplication by an overall factor:
\begin{equation}
  \langle V_\eta^\text{gauge}\rangle_{\mathbb S^3_b}
=
 \left\{
   \begin{array}{lll}
e^{-2\pi b^{-1} \eta\hat\zeta}
Z_{\mathbb S^3_b}(\hat \zeta) &\text{ at }    \vartheta=0\,,
 \\
e^{2\pi b \eta\hat\zeta}
Z_{\mathbb S^3_b} (\hat\zeta) &\text{ at }  
  \vartheta=\pi/2\,.
   \end{array}
 \right.
\end{equation}
If the theory has a flavor symmetry, the partition function $Z_{\mathbb S^3_b}(\hat m=R m,\ldots)$ depends on the real mass parameters $m=\text{diag}(m_1,\ldots)$.
The expectation value of a flavor vortex loop $V_H^\text{flavor}$ is the partition function whose argument $\hat m$ is shifted in the imaginary direction:
\begin{equation}
  \langle V_H^\text{flavor}\rangle_{\mathbb S^3_b}=
 \left\{
   \begin{array}{lll}
Z_{\mathbb S^3_b}\left(
\hat m +i b^{-1}H\right) &\text{ at }    \vartheta=0\,,
 \\
Z_{\mathbb S^3_b}\left(\hat m-i bH\right) &\text{ at }    \vartheta=\pi/2\,.
   \end{array}
 \right.
\end{equation}

Later we will consider mirror symmetry.
For reference, we quote results for the gauge and flavor Wilson loop expectation values \cite{Hama:2011ea}:
\begin{equation} \label{S3b-gauge-wilson-result}
  \langle W_\eta^\text{gauge} = e^{\eta \oint (A\pm i \sigma ds)}\rangle_{\mathbb S^3_b}
=
 \left\{
   \begin{array}{lll}
Z_{\mathbb S^3_b}(\zeta + i b^{-1}\eta) &\text{ at }    \vartheta=0\,,
 \\
Z_{\mathbb S^3_b}(\zeta - i b\eta) &\text{ at }  
  \vartheta=\pi/2\,,
   \end{array}
 \right.
\end{equation}
\begin{equation} \label{S3b-flavor-wilson-result}
  \langle W_\eta^\text{flavor}\rangle_{\mathbb S^3_b}
=
 \left\{
   \begin{array}{lll}
e^{2\pi b^{-1} \boldsymbol \eta\hat m}
Z_{\mathbb S^3_b}(\hat m) &\text{ at }    \vartheta=0\,,
 \\
e^{- 2\pi b  \boldsymbol \eta \hat m }
Z_{\mathbb S^3_b}(\hat m) &\text{ at }  
  \vartheta=\pi/2\,,
   \end{array}
 \right.
\end{equation}

\section{Localization on $\mathbb S^1 \times \mathbb S^2$ by the index theory}
\label{sec:S2S1}

In this section, we compute the expectation value of a vortex loop operator on the geometry $\mathbb S^1\times \mathbb S^2$, or equivalently the (generalized) superconformal index in the presence of a vortex loop operator.
For the purpose of explaining the computation, it is enough to consider the ordinary index of a general $\mathcal N=2$ gauge theory, with a chiral multiplet of general R-charge ${\cal R}=-\Delta$ \cite{Imamura:2011su} in representation $R$ of the gauge group.
This simplifies the notation, and we will indicate only at the end the results for the generalized index, which incorporates the background magnetic flux on $\mathbb S^2$ for flavor symmetries \cite{Kapustin:2011jm}.
As in the previous section where we studied the ellipsoid $\mathbb S^3_b$, we compute the one-loop determinant using the equivariant index.
We begin by explaining how to compute the superconformal index without a vortex loop in our approach.

\subsection{Partition function}

The geometry is defined by the metric
\begin{equation}
  ds^2=  d\tau^2+d\theta^2+\sin^2\theta d\varphi^2\,,
\end{equation}
with periodicity $\tau\sim\tau+\beta$.
The vielbein are
\begin{equation}
  e^1=d\tau\,,\quad e^2=d\theta\,,\quad e^3=\sin\theta d\varphi\,.
\end{equation}
Let us consider the supercharge generated by the following two conformal Killing spinors%
\footnote{%
There are four independent conformal Killing spinors on $\mathbb R\times \mathbb S^2$.
If $\epsilon_0$ is an arbitrary constant spinor, they are given by
$e^{\mp\frac 1 2\tau} e^{\pm \frac  i 2 \theta\gamma_{\hat\varphi}}
e^{\frac i 2 \varphi \gamma_{\hat\tau}} \epsilon_0\,.$
}
\begin{equation}
  \varepsilon=\frac{1}{\sqrt 2}
e^{-\tau/2}
  \begin{pmatrix}
    -e^{\frac{i}{2}(\theta-\varphi)}
\\
e^{\frac{i}{2}(-\theta-\varphi)}
  \end{pmatrix}\,,
\quad
\bar\varepsilon=\frac{1}{\sqrt 2}
e^{\tau/2}
  \begin{pmatrix}
e^{\frac{i}{2}(-\theta+\varphi)}
\\
e^{\frac{i}{2}(\theta+\varphi)}
  \end{pmatrix}\,.
\end{equation}
Note, however, that these are not periodic in $\tau$.
To understand the origin of non-periodicity, let us look at the definition of the index
\begin{equation}
\label{index-def}
Z_{\mathbb S^1\times\mathbb S^2}
=
{\rm Tr}(-1)^{\rm F}     
e^{-\beta_1(\mathcal H-{\cal R}-j_3)} e^{-\beta_2(\mathcal H+j_3)}
\,.
\end{equation}
Here $\mathcal H=-\partial_\tau$ is the Hamiltonian, ${\cal R}$ is the R-symmetry generator, and $j_3$ is a generator of the isometry group $SU(2)$ that acts as $-i\partial_\varphi$ on neutral scalars.
The index $\mathcal I$ should be independent of $\beta_1$ because $\mathcal H-{\cal R}-j_3=i\{\delta_{\varepsilon},\delta_{\bar\varepsilon}\}
$.
Formally, the operators in the trace require the fields to satisfy the quasi-periodic boundary conditions
\begin{equation}
\label{BC-S1S2}
  (\text{fields})_{\tau+\beta}=e^{\beta_1(-{\cal R}-j_3)+\beta_2j_3
}
  (\text{fields})_{\tau}\,,
\end{equation}
where $\beta=\beta_1+\beta_2$.
By assigning the R-charges $+1$ to $\epsilon$ and $-1$ to $\bar\epsilon$, we see that $\epsilon$ and $\bar\epsilon$ precisely satisfy the boundary conditions (\ref{BC-S1S2}).
Note, however, that the group action on the right hand side involves a rotation by an imaginary angle.
The way to make sense of this is to rewrite everything including the Lagrangian and the SUSY transformations in terms of the redefined periodic fields 
\begin{equation}
\label{redef}
 (\text{fields})_\text{new}:=e^{-(\tau/\beta)(\beta_1(-{\cal R}-j_3)+\beta_2j_3
)} (\text{fields})
\end{equation}
 that are periodic in $\tau$ \cite{Kim:2009wb}.
This is equivalent to replacing everywhere the time derivative $\partial_\tau=-\mathcal H$ by%
\footnote{%
Another way to understand the shift is that, after the redefinition that makes fields periodic, the shift in the derivatives cancels the twist in the trace and the index becomes ${\rm Tr}(-1)^F e^{\beta \partial_\tau}$.
}
\begin{equation}
  \partial_\tau+\beta^{-1}[(-{\cal R}-j_3)\beta_1+j_3\beta_2
]\,.
\end{equation}
In the new formulation, the spinors ($\epsilon,\bar\epsilon):=(\varepsilon_\text{new},\bar\varepsilon_\text{new})$ that generate supersymmetry become $\tau$-independent:
\begin{equation}
\label{S1S2-spinors}
  \epsilon=\frac{1}{\sqrt 2}
  \begin{pmatrix}
    -e^{\frac{i}{2}(\theta-\varphi)}
\\
e^{\frac{i}{2}(-\theta-\varphi)}
  \end{pmatrix}\,,
\quad
\bar\epsilon=\frac{1}{\sqrt 2}
  \begin{pmatrix}
e^{\frac{i}{2}(-\theta+\varphi)}
\\
e^{\frac{i}{2}(\theta+\varphi)}
  \end{pmatrix}\,.
\end{equation}

In the localization approach \cite{Kim:2009wb} to the computation of the index (\ref{index-def}), we deform the action by $t Q\cdot V$ for some fermionic functional $V$.   Our choice is again (\ref{Vchoice}).
In the limit $t\rightarrow +\infty$, the path integral localizes to configurations
\begin{equation}
  \begin{aligned}
\label{S1S2saddle}
 & A_\tau=-\frac{a}{\beta}\,,\quad
A^{\pm}_{\varphi}=\frac m 2 (\pm 1-\cos\theta)\,,
\quad
\sigma=
-\frac{m}{2}\,,\quad
\text{where }
a=\text{const.},\,
m=\text{const.},
\\
&\hspace{6cm}\phi=0\,,\quad     F=0\,.
  \end{aligned}
\end{equation}
The flux $m$ takes values in the Cartan subalgebra of the Lie group, and is further required to satisfy
\begin{equation}
\alpha(m)\,,  \rho(m)\in\mathbb Z
\end{equation}
for any root $\alpha$ and any weight $\rho$ in representation $R$.
The expressions $A^{\pm}_\varphi$ are valid in the standard two patches $U^+=\{\theta\neq\pi\}$ and $U^-=\{\theta\neq 0\}$ of $\mathbb S^2$.

The supercharge $Q=\delta_\epsilon+\delta_{\bar\epsilon}$ squares to
\begin{equation}
\label{Q2-S1S2}
  \begin{aligned}
  Q^2&=i {\cal L}_v+i(iv^\mu A_\mu+\sigma\bar\epsilon\epsilon)
+i{\cal R}
+i\beta^{-1}[(-{\cal R}-j_3)\beta_1+j_3\beta_2
]
\,,
\end{aligned}
\end{equation}
where
\begin{equation}
v\equiv(\bar\epsilon\gamma^\mu \epsilon)\partial_\mu=\partial_\tau-i\partial_\varphi
\end{equation}
and $\bar\epsilon\epsilon=-\cos\theta$.
In order to simplify the expression (\ref{Q2-S1S2}) further, we need to take into account the saddle point configurations (\ref{S1S2saddle}) and the representation of the $SU(2)$ in the monopole background.
On a scalar field with electric charge $+1$ in the background of monopole charge $\rho(m)$, the angular momentum operator $j_3$ acts as
  \cite{Wu:1976ge}:%
 \footnote{%
The other generators act as
$j_+= e^{i \varphi}( \partial_\theta+i \cot\theta (\partial_\varphi+i\rho( A^\pm_\varphi)))
+\frac {\rho(m)}{ 2} e^{i\varphi}\sin\theta$ 
and  $j_-=e^{- i \varphi}(- \partial_\theta+i \cot\theta (\partial_\varphi+i\rho( A^\pm_\varphi)))
+\frac{\rho(m)}{2} e^{-i\varphi}\sin\theta$.
}
\begin{equation}
\label{j3-monopole}
j_3=-i (\partial_\varphi+i\rho( A_{\varphi}^\pm))+\frac{\rho(m)}{2}\cos\theta
=-i\partial_\varphi \pm \frac {\rho(m)} 2\,.
 \end{equation}
The expression (\ref{Q2-S1S2}) can be rewritten as
\begin{equation}
\label{Q2-S1S2-simple}
\begin{aligned}
Q^2
&=
i{\cal L}_{\partial_\tau }
+\frac{a}{\beta}
+i\frac{\beta_2}\beta
(2 j_3+{\cal R}_0-\Delta \mathcal F)
  \end{aligned}
\end{equation}
at the saddle point (\ref{S1S2saddle}),%
\footnote{%
The right hand side is the precise expression of $\hat Q^2$ for any field configuration.
See Appendix~\ref{sec:index-details-S1S2}.
}
which is a linear combination of the generators of $G$ defined again by (\ref{G-def-S3b}), 
which acts on the coordinates $(h,t)\in H\times K$ by $(e^{2\pi i\tau/\beta},e^{i\varphi}) \mapsto (h\cdot e^{2\pi i\tau/\beta}, t\cdot e^{i\varphi})$.

As in the case of $\mathbb S^3_b$ in the previous section, we would like to compute the equivariant index for the relevant differential operator $D_{10}$ that appears in the fermionic functional $V$.
Some details are given in Appendix~\ref{sec:index-details-S1S2}.
Note that if we choose $c=i \beta$, the index is independent of $\beta_1$ as required by the definition (\ref{index-def}).
The operator $D_{10}$ is transversally elliptic with respect to the vector field $\partial_\tau$ that generates the free $U(1)$ action on $\mathbb S^1\times\mathbb S^2$ (this time in a trivial way as a translation along the circle), and thus reduces to a transversally elliptic operator on $\mathbb S^2$.

Let $\rho\in R$ denote the weights in representation $R$ of the gauge group.
We show in Appendix~\ref{sec:index-details-S1S2} that the equivariant index for the chiral multiplet in representation $R$ is
\begin{equation} \label{S1S2-ind-chi-text}
  \text{ind}_g(D_{10}^\text{chi})|_{g=e^{i Q^2}}=
\sum_{n\in \mathbb Z}\sum_{r=0}^\infty \sum_{\rho\in R}
\left( e^{i w(n,r,\rho)} +e^{-i w(n,r,\rho)}  - e^{i \tilde w(n,r,\rho)} -e^{-i \tilde w(n,r,\rho)} \right)
\end{equation}
with
 \begin{equation}
   \begin{aligned}
   i  w(n,r,\rho) &\propto (2r-\rho(m)+\Delta)\beta_2-2 \pi i n + i \rho(a)
 \,,
 \\
   i \tilde  w(n,r,\rho) &\propto -(2r-\rho(m)+ 2 -\Delta)\beta_2+2 \pi i n +i \rho(a)\,.
   \end{aligned}
 \end{equation}
The one-loop determinant that follows from the rule (\ref{rule}) is
\begin{equation}
\label{S1S2-1loop-chiral}
  \begin{aligned}
  Z_\text{1-loop}^\text{chi}&=
\prod_{\rho \in R}
\prod_{r=0}^\infty
 \prod_{n\in\mathbb Z}
\frac{-(r-\frac{\rho(m)}{2}+1-\Delta/2)\beta_2+\frac{i}{2}\rho(a)+\pi i n}
{-(r-\frac{\rho(m)}{2}+\Delta/2)\beta_2-\frac{i}{2}\rho(a)+\pi i n}
\\
&=
\prod_{\rho \in R}
\prod_{r=0}^\infty 
\frac{\sinh\left[-(r-\frac{\rho(m)}{2}+1-\Delta/2)\beta_2+\frac{i}{2}\rho(a)\right]}
{\sinh\left[-(r-\frac{\rho(m)}{2}+\Delta/2)\beta_2-\frac{i}{2}\rho(a)\right]}
\,.
  \end{aligned}
\end{equation}
This can be rewritten as follows:
 \begin{equation}
   \begin{aligned}
   Z_\text{1-loop}^\text{chi}&=
\prod_{\rho \in R}
\left(\prod_{r=0}^\infty 
 \frac{e^{ +(r-\frac{\rho(m)}{2}+1-\Delta/2)\beta_2-\frac{i}{2}\rho(a)}}
 {e^{+(r-\frac{\rho(m)}{2}+\Delta/2)\beta_2+\frac{i}{2}\rho(a)}}
 \times
 \prod_{r=0}^\infty 
 \frac{1-q^{r-\frac{\rho(m)}{2}+1-\Delta/2}e^{i\rho(a)}}
 {1-q^{r-\frac{\rho(m)}{2}+\Delta/2}e^{-i \rho(a)}}
\right)\,,
   \end{aligned}
 \end{equation}
 where 
 $q=e^{-2\beta_2}$.
After regularizing the infinite product,%
 \footnote{%
Following \cite{Kim:2009wb} (cf.  \cite{Imamura:2011su}),
we regularize the logarithm of the first factor as
\begin{equation}
  \begin{aligned}
&\quad\sum_{r=0}^\infty  [+(r-\frac{\rho(m)}{2}+1-\Delta/2)\beta_2-\frac{i}{2}\rho(a)]
-\sum_{r=0}^\infty [+(r-\frac{\rho(m)}{2}+\Delta/2)\beta_2+\frac{i}{2}\rho(a)]
\\
&=
 \lim_{x,y\rightarrow 1}
 (\beta_2\frac{\partial}{\partial x}+\frac{i}{2}\rho(a) \frac{\partial}{\partial y})
\left(
+\frac{x^{-\frac{\rho(m)}{2}+1-\Delta/2}y^{-1}}{1-x}
- \frac{x^{-\frac{\rho(m)}{2}+\Delta/2}y}{1-x}
 \right)
\\
 &= +\frac{\rho(m)}{2}((1-\Delta)\beta_2 -i\rho(a))-i\rho(a)
 \lim_{x\rightarrow 1}\left(\frac{1}{1-x}-1+{\cal O}(1-x)\right)\,.
  \end{aligned}
\end{equation}
 By dropping the $m$-independent terms, we renormalize this to $+\frac{\rho(m)}{2}((1-\Delta)\beta_2 -  i\rho(a))$ by taking the BF coupling as counter terms.
 }
the one-loop determinant is given by
 \begin{equation}
\label{S1S2-chi-loop}
   \begin{aligned}
   Z_\text{1-loop}^\text{chi}&=
\prod_{\rho\in R}
\left(
q^{ -\frac{\rho(m)}{4}(1-\Delta)} e^{-\frac i2 \rho(m)\rho(a)}
\prod_{r=0}^\infty 
 \frac{1-q^{r-\frac{\rho(m)}{2}+1-\Delta/2}e^{i\rho(a)}}
 {1-q^{r-\frac{\rho(m)}{2}+\Delta/2}e^{-i\rho(a)}}
\right)
\,.
   \end{aligned}
 \end{equation}

For the vector multiplet, the north and the south poles of $\mathbb S^2$ contribute identical amounts to the equivariant index, and 
in Appendix~\ref{sec:index-details-S1S2} we compute the equivariant index for the vector multiplet.
By applying the rule (\ref{rule}), the corresponding one-loop determinant is
\begin{equation}
\label{S1S2-gauge-loop}
  \begin{aligned}
Z_\text{1-loop}^\text{vec}
&=
\prod_{\alpha\in \text{adj}}  
\prod_{n\in\mathbb Z}
\left(
\frac{i}2 \alpha(a) +\frac{1}2 \alpha(m) \beta_2
+\pi i n
\right)^{1/2}
\left(
\frac{i}2 \alpha(a) - \frac{1}2 \alpha(m) \beta_2
+\pi i n
\right)^{1/2}
\\
&\sim
\prod_{\alpha>0}
\left[2\sinh\left(
\frac{i}2 \alpha(a) +\frac{1}2 \alpha(m) \beta_2
\right)\right]
\left[2\sinh\left(
\frac{i}2 \alpha(a) -\frac{1}2 \alpha(m) \beta_2
\right)\right]
\\
&=
\prod_{\alpha\in \text{adj}}  
q^{-|\alpha(m)|/4}
\left(
1-e^{-i\alpha(a)} q^{|\alpha(m)|/2}
\right)
\,.
  \end{aligned}
\end{equation}
Both (\ref{S1S2-chi-loop})%
\footnote{%
The corresponding formulas in \cite{Imamura:2011su} and \cite{Kapustin:2011jm} involve the absolute values $|\rho(m)|$.
As explained in \cite{Dimofte:2011py}, one can rewrite such an expression and eliminate $|\rho(m)|$ in favor of $-\rho(m)$.
} and (\ref{S1S2-gauge-loop})
agree with the results in the literature.

\subsection{Vortex loop expectation values}

Let us now insert vortex loop operators with vorticities $H^+$ and $H^-$ at the north and south poles of $\mathbb S^2$, respectively.
In the presence of background magnetic flux, the two loop operators are defined by the $Q$-invariant configurations
\begin{equation}
F_{\mu\nu}= \left[H^+ \delta(1-\cos\theta)+ H^- \delta(-1-\cos\theta) \right] \epsilon_{\mu\nu\rho}v^\rho 
+
\frac{m}{2}\epsilon_{\tau\mu\nu}
\,,
\end{equation}
\begin{equation}\label{S1S2-vortex-D-sigma}
  D=i H^+ \delta(1-\cos\theta) -i H^- \delta(-1-\cos\theta)\,,
\qquad
\sigma=-\frac{m}{2}\,.
\end{equation}
The gauge field is given as $A_\tau=-a/\beta$, $A_\theta=0$,
\begin{equation}
\label{S1S2-vortex-gauge}
A_{\varphi}^{\pm}=
\frac m 2 (\pm 1-\cos\theta) \pm H^\pm\,.
\end{equation}
In order for the gauge fields on the two patches to be glued by a well-defined transition function, we need that
$m+H^++H^-$ is a GNO charge that satisfies the Dirac quantization conditions $\alpha(m+H^++H^-), \rho(m+H^++H^-)\in\mathbb Z$.

It is clearest to restrict to the case $\beta_1=\beta_2=\beta/2$.
In this case $j_3$ does not enter the field redefinition (\ref{redef}), after which
$$
\begin{aligned}
i Q^2
& = 
 -\partial_\tau  +i \mathcal L_{\partial_\varphi } 
\mp\frac{m+H^++H^-}{2}
+  \frac{i}{\beta}
\left(
a
+\frac{i\beta }{2} (H^+-H^-)
 - \frac{i  \Delta}{2}\beta
\right)
-\frac{1}{2}\mathcal R_0
\,,
\end{aligned}
$$
where the combination $-i\mathcal L_\varphi \pm \frac{m+H^++H^-}{2}$ is precisely $j_3$.
The effect on $Z_\text{1-loop}^\text{chi}$ of the vortex loop is the shifts
\begin{equation}
\label{S1S2shifts}
  m\rightarrow m+H^++H^-\,,
\quad\quad
a \rightarrow a - \frac{i}{2} (\log q) (H^+-H^-)\,.
\end{equation}

We now specialize to the $U(1)$ gauge group and set $H^\pm=\eta^\pm$.
We need the on-shell value of a supersymmetric Chern-Simons action in the vortex loop background.   
It enters as $e^{i\mathcal S_{SCS}}$ in the path integral.
To evaluate it, we introduce a connection $ A'=r A^\text{new}_\tau d\tau + A^\text{new}_\theta d\theta+ A^\text{new}_\varphi d\varphi$ extended to the disk $\mathbb D^2=\{r e^{2\pi i\tau/\beta}|r\leq 1\}$ times $\mathbb S^2$ and put \cite{witten-cs,Dijkgraaf:1989pz,Kim:2009wb}
\begin{equation}
\mathcal S_{SCS}^\text{boson}=\frac{k}{4\pi}  \left(
\int_{\mathbb D^2\times \mathbb S^2}d A'\wedge d A'
+
\int_{\mathbb S^1\times \mathbb S^2}
2 D \sigma \cdot vol\right)\,.
\end{equation}
After some calculations we find%
\footnote{%
For the flavor Chern-Simons action, we get $\mathcal    S_{SCS}^\text{boson}
=
k \beta
\left(-\frac{\boldsymbol  a}{\beta} +\frac{i}{2}\Delta
-\frac{i}{2} (\boldsymbol \eta^+-\boldsymbol \eta^-)
\right)
(\boldsymbol m+ \boldsymbol \eta^+ + \boldsymbol \eta^-)
$, where bold fonts are used for quantities related to the flavor symmetry.
Dependence on $\Delta$ arises because after the field redefinition (\ref{redef}) ${\rm Im}\, \boldsymbol A_\tau$ contains $\frac{i}{2}\Delta$ as in the 4d case \cite{Festuccia:2011ws}.
}
\begin{equation}\label{S1S2-SCS}
\mathcal    S_{SCS}^\text{boson}
=
k \beta
\left(-\frac{ a}{\beta} 
-\frac{i}{2} (\eta^+-\eta^-)
\right)
(m+ \eta^+ + \eta^-)\,.
\end{equation}

Let us consider the BF coupling (\ref{SUSY-BF}) that appears as $e^{-\mathcal S_{BF}}$.
It may be evaluated via the relation (\ref{BF-CS}) between the BF and Chern-Simons terms.
If we use the full gauge multiplet configuration in (\ref{S1S2-vortex-D-sigma}) and  (\ref{S1S2-vortex-gauge}), we find that $\mathcal S_{BF}$ equals
\begin{equation}\label{SBF-naive}
\left(i\boldsymbol a + \frac{\Delta }{2} \beta 
\right)
\left(m+\eta^++\eta^-\right)
+
\left(i a
-\frac{\beta}{2} (\eta^+-\eta^-)
\right)
\boldsymbol m\,.
\end{equation}
If this were included in the path integral,
since the gauge vortex loops shift the parameters $m\in\mathbb Z$ and $a$ that are summed or integrated over, it would not affect the partition function at all.
From the point of view of the $SL(2,\mathbb Z)$ action \cite{Witten:2003ya} on superconformal theories, however, it is more natural to not include terms proportional to $\eta^\pm$ in the BF coupling, as follows from a discussion in \cite{Kapustin:2012iw}.
See also (\ref{BF-gauge-bosonic}).
Thus we should drop $\eta^\pm$ from (\ref{SBF-naive}).
Since all other contributions uniformly receive the shifts (\ref{S1S2shifts}), the only effect of the vortex loop is
the multiplication by an overall factor:
\begin{equation}\label{S1S2GaugeVortex}
  \langle 
V_{\eta^+}^\text{gauge}(\text{north})
V_{\eta^-}^\text{gauge}(\text{south})
\rangle_{\mathbb S^1\times \mathbb S^2}
=
  e^{+i(\eta^++\eta^-)\left( \boldsymbol a-\frac{i}{2}\Delta\right)}q^{+(\eta^+-\eta^-)\boldsymbol m/2}
Z_{\mathbb S^1\times\mathbb S^2}(\boldsymbol m, e^{i\boldsymbol a})\,.
\end{equation}
Here $\boldsymbol a=-\int_{\mathbb S^1} \boldsymbol A$ is the background holonomy along $\mathbb S^1$,
and ${\boldsymbol m}=(2\pi)^{-1}\int_{\mathbb S^2} \boldsymbol{F}$ is the background flux through $\mathbb S^2$.
For a non-Abelian gauge group, it is a possibility that there are non-perturbative contributions.

Next we consider an $\mathcal N=2$ theory with a flavor symmetry.
The generalized index is again a function of $\boldsymbol a=-\int_{\mathbb S^1} \boldsymbol A$ and ${\boldsymbol m}=(2\pi)^{-1}\int_{\mathbb S^2} \boldsymbol{F}$, though this time $\boldsymbol A$ is coupled to the flavor symmetry.
From the discussion above, we see that for two flavor vortex loops at the north and south poles, the correlator is given by the shifts in the partition function, {\it i.e.}, the generalized index:
\begin{equation}\label{S1S2FlavorVortex}
  \langle 
V_{H^+}^\text{flavor}(\text{north})
V_{H^-}^\text{flavor}(\text{south})
\rangle_{\mathbb S^1\times \mathbb S^2}
=Z_{\mathbb S^1\times\mathbb S^2}(\boldsymbol m+H^++H^-,e^{i \boldsymbol a} q^{\frac 1 2(H^+-H^-)})\,.
\end{equation}

\subsection{Wilson loop expectation values}

We can also compute the expectation values of Wilson loops.%
\footnote{%
We thank J.  Gomis for discussions on this calculation.
}
Consider an $\mathcal N=2$ theory with at least one $U(1)$ gauge group.
Such a theory possesses a global symmetry $U(1)_J$ generated by the conserved current $J^\mu= \epsilon^{\mu\nu\rho} \partial_\nu A_\rho$.
The generalized index of the theory involves a sum over the gauge fluxes $m$, and depends on the flux $\boldsymbol m$ of the background gauge field $\boldsymbol A_\mu$ for $U(1)_J$, whose coupling is given by the BF term (\ref{SUSY-BF}).
The partition function on $\mathbb S^1\times\mathbb S^2$ then takes the form 
\begin{equation}
Z_{\mathbb S^1\times\mathbb S^2}
=
\sum_{m\in\mathbb Z} (q^c e^{-i \boldsymbol a})^m \oint \frac{da}{2\pi} e^{-i \boldsymbol m a} f(m,a,q,\ldots)\,.
\end{equation}
Here $c$ parameterizes the contribution of the gauge flux to the R-charge \cite{Dimofte:2011py}, $\boldsymbol a$ is the chemical potential for $U(1)_J$, and $f$ is some function.
We now insert a Wilson loop of charge $\eta^+$ at the north pole, and another of charge $\eta^-$ at the south pole.
For $\beta_1=\beta_2$, we find that the value of the product of the Wilson loops in the saddle point configuration is
\begin{equation}
  e^{-i(\eta^++\eta^-) a}q^{-(\eta^+-\eta^-)m/2}\,,
\end{equation}
which is to be inserted inside the sum and the integral.
Thus the effect of the Wilson loops is the shifts
\begin{equation}
\label{S1S2shifts-Wilson}
 \boldsymbol m\rightarrow \boldsymbol m+\eta^++\eta^-\,,
\quad\quad
\boldsymbol a \rightarrow \boldsymbol a - \frac{i}{2} (\log q) (\eta^+-\eta^-)\,.
\end{equation}
Thus the correlation function of the two ordinary Wilson loops is given by the partition function (generalized index) whose arguments are shifted: 
\begin{equation}\label{S1S2GaugeWilson}
  \langle 
W_{\eta^+}^\text{gauge}(\text{north})
W_{\eta^-}^\text{gauge}(\text{south})
\rangle_{\mathbb S^1\times \mathbb S^2}
=Z_{\mathbb S^1\times\mathbb S^2}(\boldsymbol m+\eta^++\eta^-,e^{i \boldsymbol a} q^{\frac 1 2(\eta^+-\eta^-)})\,.
\end{equation}

In  an $\mathcal N=2$ theory with flavor symmetry, the correlator of two flavor Wilson loops with charges $\eta^+$ and $\eta^-$, inserted at the north and south poles respectively, is given by
\begin{equation}\label{S1S2FlavorWilson}
  \langle 
W_{\eta^+}^\text{flavor}(\text{north})
W_{\eta^-}^\text{flavor}(\text{south})
\rangle_{\mathbb S^1\times \mathbb S^2}
=
  e^{-i(\eta^++\eta^-) (\boldsymbol a - \frac{i}{2} \Delta \beta)}q^{-(\eta^+-\eta^-)\boldsymbol m/2}
Z_{\mathbb S^1\times\mathbb S^2}(\boldsymbol m, e^{i\boldsymbol a})\,.
\end{equation} 

\section{Abelian mirror symmetry}
\label{sec:discuss}

We have employed the supersymmetric localization method to obtain exact quantitative results for the expectation values and correlators of vortex loop operators.
Let us now discuss more qualitative and conceptual points regarding loop operators in three dimensional supersymmetric theories.

Any duality maps global symmetries of one theory to those of the other.
In particular abelian mirror symmetry 
\cite{Intriligator:1996ex, Hanany:1996ie, deBoer:1996mp, deBoer:1996ck, Kapustin:1999ha} 
by definition maps a topological symmetry $U(1)_J$ in one theory to a flavor symmetry in the dual theory.
It was explained in Section \ref{sec:flavor} that the vortex loop for $U(1)_J$ is the gauge Wilson loop, and that the Wilson loop for $U(1)_J$ is the gauge vortex loop.
Thus the transformations of loop operators under abelian mirror symmetry follow from those of global symmetries.
We can summarize the abelian mirror symmetry action on loop operators in $\mathcal N=2$ theories.
\begin{table}[h]
\centering
 \begin{tabular}{c|cc}
  & Theory A& Theory B
\\
 \hline
 Global symmetry 
 &
$U(1)_J$
  &
  Flavor symmetry
\\ \\
 \begin{tabular}{c}  Vortex loop for \\  global symmetry \end{tabular}
 & Gauge Wilson loop &
  Flavor vortex loop
\\ \\
 \begin{tabular}{c}  Wilson loop for \\  global symmetry  \end{tabular}
& Gauge vortex loop &
 \begin{tabular}{c}  Flavor Wilson loop \\  \end{tabular}
 \end{tabular}
\caption{\label{TableFirst}Abelian mirror symmetry action on global symmetries and loop operators.}
\end{table}

Let us illustrate the mapping of global symmetries and loop operators in a well-known $\mathcal N=2$ mirror pair \cite{Aharony:1997bx}.
As Theory A (SQED), we consider the $U(1)$ gauge theory with two chirals $(\Phi,\tilde\Phi)$ of charges $(1,-1)$.
This theory has a flavor symmetry $U(1)_\text{axial}$ for which the fields have charges $(1,1)$, as well as a topological symmetry $U(1)_J$.
As Theory B (XYZ model), we consider a theory of three chiral superfields $(X,Y,Z)$, interacting through the superpotential $W=XYZ$.
The superpotential is invariant under two symmetries $U(1)_1$ and $U(1)_2$, whose charges are given by $(2,-1,-1)$ and $(0,1,-1)$ respectively.
It is known that $U(1)_\text{axial}$ is identified with $U(1)_1$, and $U(1)_J$ with $U(1)_2$.
We summarize the symmetries and the loop operator spectra in Table \ref{TableSecond}.  
\begin{table}[h]
\centering
\begin{tabular}{c}
SQED
\\ \\
 \begin{tabular}{c|cc|l}
 & $\Phi$ & $\tilde \Phi$& \hspace{1.8mm} Loop operator
\\
\hline
$U(1)_\text{gauge}$ & $1$ & $-1$& 
\\
$U(1)_\text{axial}$ & $1$ & $1$&
$\left\{\hspace{-2mm}\text{\begin{tabular}{l}Flavor vortex\\Flavor Wilson\end{tabular}}\right.  \hspace{-10mm}$
\\
$U(1)_J$ & & &
$\left\{\hspace{-2mm}\text{\begin{tabular}{l}Gauge Wilson\\Gauge vortex\end{tabular}}\right.   \hspace{-10mm}$
 \begin{tabular}{c}\\\end{tabular}
\end{tabular}
\end{tabular}
\hspace{5mm}
\begin{tabular}{c}
XYZ model
\\ \\
 \begin{tabular}{c|ccc|l}
 & $X$ & $Y$& $Z$& \hspace{1.8mm} Loop operator
\\ 
\hline
\\
$U(1)_1$ & $2$ & $-1$&$-1$&
$\left\{\hspace{-2mm}\text{\begin{tabular}{l}Flavor vortex\\Flavor Wilson\end{tabular}}\right.  \hspace{-10mm}$
\\
$U(1)_2$ & $0$& $1$& $-1$ &
$\left\{\hspace{-2mm}\text{\begin{tabular}{l}Flavor vortex\\Flavor Wilson\end{tabular}}\right.  \hspace{-10mm}$
\end{tabular}
\end{tabular}
\caption{\label{TableSecond}%
Mirror symmetry for the two-flavor SQED and the XYZ model.}
\end{table}

Our localization results for loop operators provide a quantitative test of the mirror symmetry predictions.
On $\mathbb S^1\times\mathbb S^2$, the correlation function (\ref{S1S2GaugeWilson}) of two gauge Wilson loops is identical to the correlation function (\ref{S1S2FlavorVortex}) of two flavor vortex loops, confirming the correspondence on the middle row in Table \ref{TableFirst}.
Similarly, the equality between (\ref{S1S2GaugeVortex}) and (\ref{S1S2FlavorWilson}) verifies the mirror symmetry action on the bottom row of Table \ref{TableFirst}.  
The same checks can also be made using the results for Wilson and vortex loop operators on $\mathbb S^3_b$.

\section*{Acknowledgements}

We would like to thank Francesco Benini, Sergio Benvenuti, Jaume Gomis, Kazuo Hosomichi, Sungjay Lee, 
Marcos Mari\~no, Sara Pasquetti, Vasily Pestun and J\"org Teschner for useful discussions.  
We also thank Francesco Benini for pointing out an error in the first version.
N.D.  would like to thank the hospitality of the CERN theory devision, GGI Florence, 
Perimeter Institute, the Newton Institute, Nordita, the Israeli Institute for Advanced 
Studies and The Collaborative Research Center SFB 676 at The University of Hamburg 
and DESY 
for their hospitality and financial support during the course of this work.
T.O.  would like to acknowledge the hospitality of  the KIAS, the Yukawa Institute, the Perimeter Institute and the Simons Center.  
F.P.  would like to thank the hospitality of the KITP, the Newton Institute, Nordita, 
the Simons Center and The Perimeter Institute.  
The work of N.D.  is underwritten by an advanced fellowship of the 
Science \& Technology Facilities Council.
 The research of T.O.  is supported in part by Grant-in-Aid for Young Scientists (B) No.  23740168 and by Grant-in-Aid for Scientific Research (B) No.  20340048.  
Since October 2013, the work of F.P.  is supported by a Marie Curie International Outgoing Fellowship  FP7-PEOPLE-2011-IOF, Project n\textsuperscript{o} 298073 (ERGTB).

\appendix
\section{Metric and vielbein on $\bS^3$}
\label{sec:metrics3}

The three dimensional sphere $\bS^3$ with radius $R$ can be represented by a pair 
of complex coordinates $(u,v)\in\bC^2$ by the equation 
\bal
\label{uv}
u\bar u+v\bar v =R^2\,.
\eal
The manifold is invariant under $SO(4)\cong SU(2)_L\times SU(2)_R$
symmetry.  The generators of the two $SU(2)$ factors are denoted as $L_1^L$, $L_2^L$, 
$L_3^L$ and $L_1^R$, $L_2^R$, $L_3^R$ and they satisfy the following commutation relations 
\beq
{}[ L^L_a,L^L_b ]=i\varepsilon_{abc}L^L_c\,,
\qquad
[ L^R_a,L^R_b ]=i\varepsilon_{abc}L^R_c\,,
\qquad
[ L^L_a,L^R_b ]=0\,.
\eeq
We define raising operators $L_+^L=L_1^L+iL_2^L$, $L_+^R=L_1^R+iL_2^R$ 
and lowering operators $L_-^L=L_1^L-iL_2^L$, $L_-^R=L_1^R-iL_2^R$.  
The representation of the generators in the $(u,v)$ coordinates is given by
\beq
\label{left}
L^L_-=\bar u\, \partial_{v}-\bar v\, \partial_{u}\,,
\qquad
L^L_+=- u\, \partial_{\bar v}+ v\, \partial_{\bar u}\,,
\qquad
L^L_3=\frac{1}{2}(u\,\partial_u +v\,\partial_v-\bar u\, \partial_{\bar u} -\bar v\, \partial_{\bar v})\,,
\eeq
and 
\beq
\label{right}
L^R_-=\bar u\, \partial_{\bar v}- v\, \partial_{u}\,,
\qquad
L^R_+=- u\, \partial_{ v}+ \bar v\, \partial_{\bar u}\,,
\qquad
L^R_3=\frac{1}{2}(u\,\partial_u -v\,\partial_v-\bar u\, \partial_{\bar u} +\bar v\, \partial_{\bar v})\,.
\eeq
In the main text we use two different parameterization of the $\bS^3$, the Hopf fibration and the torus fibration.

\subsection{Hopf fibration }
\label{sec:hopf}

The Hopf fibration of $\bS^3$ is given by the parameterization $u=R\sin{\textstyle\frac{\theta}{2}}\,e^{i(\psi-\phi)/2}$ 
and $v=R \cos{\textstyle\frac{\theta}{2}}\,e^{i(\psi+\phi)/2}$ where $0\leq\theta\leq\pi$, $0\leq\phi\leq2\pi$ 
and $0\leq\psi\leq4\pi$.  The metric in the Hopf fibration is given by
\bal
\label{hopf}
ds^2=g_{\mu\nu} dx^\mu dx^\nu
&=\frac{R^2}{4}(d\theta^2+ \sin\theta^2d\phi^2 +(d\psi+\cos\theta\, d\phi)^2)
\\&=\frac{R^2}{4}(d\theta^2+ d\phi^2 +d\psi^2+2\cos \theta\, d\phi\, d \psi)\,.
\eal
This metric can be derived considering that $\bS^3=SU(2)$, as shown also in appendix~A of 
\cite{Marino:2011nm}.  The left invariant vielbein basis is
\bal
\label{viel}
e^{1}&=\frac{R}{2}(\cos \psi\,d\theta+\sin\psi\sin\theta\, d\phi )\,,\\
e^{2}&=\frac{R}{2}(\sin \psi\,d\theta-\cos\psi\sin\theta\, d\phi )\,,\\
e^{3}&=\frac{R}{2}(\cos\theta\,d\phi+d\psi )\,,
\eal 
and the inverse vielbein defined as $e_{a}{}^{\mu}=e^{b}{}_{\nu}g^{\nu\mu}\delta_{ab}$ is given by 
\beq
\label{inviel}
e_{a}{}^{\mu}= \frac{2}{R}\left(
\begin{array}{ccc}
\cos \psi & \frac{ \sin \psi }{ \sin \theta } & -\cot \theta \sin\psi
\\
\sin\psi & -\frac{ \cos\psi }{\sin \theta} & \cot \theta \cos \psi
\\
0 & 0 &1
\end{array}
\right)\,.
\eeq

\subsection{Torus fibration }
\label{sec:torus}

The torus fibration is obtained parameterizing $u$ and $v$ as $u=R \sin\vartheta e^{i\varphi_1}$ and 
$v=R \cos\vartheta e^{i\varphi_2}$, where $0\leq\vartheta\leq\pi/2$ and $0\leq\varphi_1,\varphi_2\leq2\pi$.  
Torus fibration and Hopf fibration parameters are related by $\theta=2\vartheta$, 
$\phi=\varphi_2-\varphi_1$ and $\psi=\varphi_2+\varphi_1$.
The metric is given by 
\beq
\label{torusf}
ds^2=R^2(d\vartheta^2+\sin^2\vartheta\,d\varphi_1^2+\cos^2\vartheta\,d\varphi_2^2)\,,
\eeq
and a natural frame is
\beq
\label{torusviel}
e^1=R\,d\vartheta\,,
\qquad
e^2=R\sin\vartheta\,d\varphi_1\,,
\qquad
e^3=R\cos\vartheta\,d\varphi_2\,.
\eeq
The vortex loop operator is located at $\vartheta=0$ and extended along 
$\varphi_2$.  For this field configuration, the holonomy is constant when computed along a 
curve linked to the vortex loop, therefore the monodromy along the $\varphi_1$ circle will 
be independent from $\vartheta$.

In the frame described above, the solution of the Killing spinor equation on $\bS^3$ is given by 
\beq\label{kiltorus}
\epsilon=e^{\frac{i}{2}\vartheta \gamma_1}e^{\frac{i}{2}(\varphi_1+\varphi_2) \gamma_3}\epsilon_0\,.
\eeq

\section{SUSY on 3D Euclidean manifolds}
\label{sec:susys3}
\subsection{Conventions}
\label{sec:conv}
We follow the conventions as in \cite{Marino:2011nm}.  
The curved space gamma matrices $\gamma_{\mu}$ are 
defined as $\gamma_{\mu}=\gamma_{a}\,e^{a}{}_{\mu}$ 
where $\gamma_a$ are Pauli matrices and $e^{a}{}_{\mu}$ is a vielbein.  It follows
\beq
\{ \gamma_\mu, \gamma_\nu \} =2g_{\mu\nu},
\eeq
where $g_{\mu\nu}$ is the spacetime metric.
Some useful relations for Pauli matrices are 
\bal
&\gamma_{ab}=\frac{1}{2}[\gamma_{a}, \gamma_{b}]=i\varepsilon_{abc}\gamma_{c}
\qquad\text{with}\qquad
\varepsilon_{123}=\varepsilon^{123}=1,
\\
&\gamma_{1}\gamma_{2}\gamma_{3}=i.
\eal
The spinors $\psi$ and $\bar\psi$ are independent and have the same index structure, 
{\em i.e.}, $\psi^\alpha$ and $\bar\psi^\alpha$.  Spinor indices are omitted in the main text and contracted as
\bal
\label{spinco}
\bar\psi \psi=\bar\psi^\alpha C_{\alpha\beta} \psi^\beta\,,
\qquad
\bar\psi \gamma^\mu \psi
=\bar\psi^\alpha C_{\alpha\beta} (\gamma^\mu)^\beta{}_\gamma \psi^\gamma\,.
\eal
We take $C=\left(\begin{array}{cc}
0 & -1 \\
1 & 0\end{array}\right)$.  Given that $C_{\alpha\beta}$ is antisymmetric and 
$(C\gamma^\mu)_{\alpha\beta}$ is symmetric, considering Grassmann-odd spinors it follows 
\bal
\bar\psi \psi=\psi \bar\psi\,,
\qquad
\bar\psi \gamma^\mu \psi=-\psi \gamma^\mu \bar\psi\,,
\qquad
(\gamma^\mu\bar\psi)\psi=-\bar\psi\gamma^\mu\psi\,.
\eal

\subsection{Vector multiplet}
\label{sec:vector-susy}

The field content of Euclidean ${\cal N}=2$ vector multiplet is given by the gauge field $A_\mu$, 
two complex Dirac spinors $\lambda$ and $\bar{\lambda}$   and two 
auxiliary real scalar fields $D$ and $\sigma$.  The supersymmetry variations are parameterized  by two independent complex spinors $\epsilon$ and $\bar{\epsilon}$ and they  are given by \cite{Marino:2011nm, Hama:2010av, Hama:2011ea}
\bal
\label{susyV}
\delta A_\mu &=
\frac{i}{2} (\bar\epsilon\gamma_\mu\lambda-\bar\lambda\gamma_\mu\epsilon), 
\\
\delta\sigma &=
\frac{1}{2} (\bar\epsilon\lambda-\bar\lambda\epsilon),
\\
\delta\lambda &=
-\frac{1}{2} \gamma^{\mu\nu}\epsilon F_{\mu\nu}-D\epsilon
+i\gamma^\mu\epsilon D_\mu\sigma +\frac{2 i}{3}\sigma\gamma^\mu D_\mu\epsilon,
\\
\delta\bar\lambda &=
-\frac{1}{2}\gamma^{\mu\nu}\bar\epsilon F_{\mu\nu}+D\bar\epsilon
-i\gamma^\mu\bar\epsilon D_\mu\sigma -\frac{2 i}{3}\sigma\gamma^\mu D_\mu\bar\epsilon,
\\
\delta D &=
-\frac{i}{2} \bar\epsilon\gamma^\mu D_\mu\lambda
-\frac{i}{2} D_\mu\bar\lambda\gamma^\mu\epsilon
+\frac{i}{2}[\bar\epsilon\lambda,\sigma]
+\frac{i}{2}[\bar\lambda\epsilon,\sigma] -\frac{i}{6}(D_\mu\bar\epsilon\gamma^\mu\lambda
+\bar\lambda\gamma^\mu D_\mu\epsilon)\,.
\eal
$D_\mu$ is the covariant derivative with respect spacetime and    gauge connection.  For the $\bS^3_b$ metric, $D_\mu$ is covariant also with respect to an R-symmetry gauge field%
\footnote{%
For a chiral scalar of R-charge $-\Delta$, we have $D_\mu \phi=(\nabla_\mu +i A_\mu -i \Delta V_\mu)\phi$.
}   
$
V=-\frac{1}{2}\left(1-\frac{b}{f}\right)d\varphi_1 -\frac{1}{2}\left(1-\frac{1}{bf}\right)d\varphi_2
$ \cite{Hama:2011ea}.
Denoting as  $\delta_\epsilon$ and  $\delta_{\bar\epsilon}$ the supersymmetry generated  by $\epsilon$ and $\bar{\epsilon}$, it results  $[\delta_\epsilon,\delta_{\epsilon'}]=[\delta_{\bar\epsilon},\delta_{\bar\epsilon'}]=0$ and \cite{Marino:2011nm, Hama:2010av, Hama:2011ea} 

\bal
\label{commuvec}
[\delta_\epsilon,\delta_{\bar\epsilon}] A_\mu &=
i  v^\nu\partial_\nu A_\mu + i \partial_\mu v^\nu A_\nu
 -D_\mu\Lambda,
  \\
 [\delta_\epsilon,\delta_{\bar\epsilon}]\sigma &=
 i v^\mu\partial_\mu\sigma+i[\Lambda,\sigma]+\rho\sigma,
  \\
 [\delta_\epsilon,\delta_{\bar\epsilon}]\lambda &=
 i v^\mu \partial_\mu\lambda+\frac{i}{4}\Theta_{\mu\nu}\gamma^{\mu\nu}\lambda
 +i[\Lambda,\lambda]+\frac{3}{2} \rho\lambda
 +\alpha\lambda,
  \\
 [\delta_\epsilon,\delta_{\bar\epsilon}]\bar\lambda &=
i  v^\mu \partial_\mu\bar\lambda
 +\frac{i}{4}\Theta_{\mu\nu}\gamma^{\mu\nu}\bar\lambda
 +i[\Lambda,\bar\lambda]+\frac{3}{2}\rho\bar\lambda
 -\alpha\bar\lambda,
  \\
 [\delta_\epsilon,\delta_{\bar\epsilon}]D &=
i v^\mu\partial_\mu D+i[\Lambda,D]+2\rho D  +\cW,
\eal               
where 
\bal\label{dw}
\cW=\frac{1}{3}\sigma(\bar\epsilon\gamma^\mu\gamma^\nu D_\mu D_\nu\epsilon
-\epsilon\gamma^\mu\gamma^\nu D_\mu D_\nu\bar\epsilon).
\eal
Therefore, for all the fields except the scalar $D$, the commutator is a sum of a translation by $v^\mu$, a rotation by $ \Theta^{\mu\nu}$, a $R$-symmetry rotation by $\alpha$, a gauge transformation by $\Lambda$ and a dilation by $\rho$.  The explicit expression of the symmetry generators is 
\bal
 v^\mu &= \bar\epsilon\gamma^\mu\epsilon,
  \\
 \Theta^{\mu\nu} &= D^{[\mu}v^{\nu]}+v^\lambda\omega_\lambda^{\mu\nu},
  \\
 \Lambda &= v^\mu i A_\mu +\sigma\bar\epsilon\epsilon,
  \\
 \rho &= \frac{i}{3}
 (\bar\epsilon\gamma^\mu D_\mu\epsilon
 +D_\mu\bar\epsilon\gamma^\mu\epsilon),
  \\
 \alpha &= \frac{i}{3}
(D_\mu\bar\epsilon\gamma^\mu\epsilon-\bar\epsilon\gamma^\mu D_\mu\epsilon) +v^\mu V_\mu,
\eal
where $\omega_\lambda^{\mu\nu}$ is the spin connection.   The supersymmetry parameters satisfy the Killing spinor equations
\bal\label{genk}
D_\mu \epsilon =\gamma_\mu \tilde \epsilon\,,\qquad\qquad D_\mu \bar\epsilon = \gamma_\mu \tilde{\bar{\epsilon}}\,.
\eal 
The explicit expression for the spinors  $\tilde\epsilon$ and $\tilde{\bar{\epsilon}}$ for   $\bS^3$ is \cite{kwy1}
\bal
\tilde \epsilon=\frac{i}{2R} \epsilon,    \qquad \tilde{\bar{\epsilon}} =\frac{i}{2R}  \bar \epsilon.
\eal
For  $\bS^3_b$ is \cite{ Hama:2011ea}
\bal
\tilde \epsilon=\frac{i}{2R f(\vartheta)}  \epsilon,    \qquad \tilde{\bar{\epsilon}} =\frac{i}{2R f(\vartheta)} \bar \epsilon
\eal
where $f(\vartheta)$ is defined in the main text and for $\bS^1\times \bS^2$ \cite{Imamura:2011su}
\bal
\tilde \epsilon=-\frac{1}{2} \gamma_{\hat\tau} \epsilon,    \qquad \tilde{\bar{\epsilon}} =\frac{1}{2} \gamma_{\hat\tau} \bar \epsilon.
\eal
With these supersymmetry generators, it follows that for all the spaces that we consider, it results $\cW=0$, where $\cW$ is defined in (\ref{dw}).  This implies that the supersymmetry closes off-shell on all the fields.   It also results $\rho=0$ for all the spaces,    that implies that the commutator  $[\delta_\epsilon,\delta_{\bar\epsilon}]$ does not include a dilation.

\subsection{Chiral multiplet}
\label{sec:chiral-susy}

The field content of the chiral multiplet is given by two complex scalars $\phi$ and $F$ and spinors $\psi$ and $\bar\psi$ with two complex components.  These fields are in a generic representation of the gauge group.  The supersymmetry variations are given by 
 \cite{Marino:2011nm, Hama:2010av, Hama:2011ea}
\bal
\label{chiralsusy}
\delta\phi &= \bar\epsilon\psi, 
\\
\delta\bar\phi &= \epsilon\bar\psi, 
\\
\delta\psi &= i\gamma^\mu\epsilon D_\mu\phi +i\epsilon\sigma\phi
+\frac{2\Delta i}{ 3}\gamma^\mu D_\mu\epsilon\phi+\bar\epsilon F,
\\
\delta\bar\psi &= i\gamma^\mu\bar\epsilon D_\mu\bar\phi
+i\bar\phi\sigma\bar\epsilon+\frac{2\Delta i}{ 3}\bar\phi\gamma^\mu D_\mu\bar\epsilon
+\bar F\epsilon,
\\
\delta F &=
\epsilon(i\gamma^\mu D_\mu\psi-i\sigma\psi-i\lambda\phi)
+\frac{i}{ 3}(2\Delta-1)D_\mu\epsilon\gamma^\mu\psi,
\\
\delta\bar F &=
\bar\epsilon(i\gamma^\mu D_\mu\bar\psi-i\bar\psi\sigma+i\bar\phi\bar\lambda)
+\frac{i}{3}(2\Delta-1)D_\mu\bar\epsilon\gamma^\mu\bar\psi
\eal
and the commutators give the following off-shell result \cite{Marino:2011nm, Hama:2010av, Hama:2011ea}  
\bal\,
[\delta_\epsilon,\delta_{\bar\epsilon}]\phi &=
 i v^\mu\partial_\mu\phi+i\Lambda\phi+\Delta\rho\phi-\Delta\alpha\phi,
 \\
[\delta_\epsilon,\delta_{\bar\epsilon}]\bar\phi &=
i v^\mu\partial_\mu\bar\phi-i\bar\phi\Lambda+\Delta\rho\bar\phi
 +\Delta\alpha\bar\phi,
\\
[\delta_\epsilon,\delta_{\bar\epsilon}]\psi &=
i v^\mu \partial_\mu\psi
 +\frac{i}{4}\Theta_{\mu\nu}\gamma^{\mu\nu}\psi
 +i\Lambda\psi+\left(\Delta+\frac{1}{2}\right)\rho\psi+(1-\Delta)\alpha\psi,
\\
[\delta_\epsilon,\delta_{\bar\epsilon}]\bar\psi &=
 i v^\mu \partial_\mu\bar\psi
 +\frac{i}{4}\Theta_{\mu\nu}\gamma^{\mu\nu}\bar\psi
 -i\bar\psi\Lambda+\left(\Delta+\frac{1}{2} \right)\rho\bar\psi
 +(\Delta-1)\alpha\bar\psi,
 \\
[\delta_\epsilon,\delta_{\bar\epsilon}]F &=
 i v^\mu\partial_\mu F+i\Lambda F+(\Delta+1)\rho F+(2-\Delta)\alpha F,
  \\
[\delta_\epsilon,\delta_{\bar\epsilon}]\bar F &=
 iv^\mu\partial_\mu\bar F-i\bar F\Lambda+(\Delta+1)\rho\bar F
 +(\Delta-2)\alpha\bar F,
\eal
that is for generic supersymmetry parameters, as for the vector multiplet, the commutator  $[\delta_\epsilon,\delta_{\bar\epsilon}]$ is a sum of a translation, a rotation, a $R$-symmetry rotation, a gauge transformation and a dilation.  The commutator of two generic unbarred supersymmetries is different from zero for the scalar $F$ \cite{Marino:2011nm, Hama:2010av, Hama:2011ea} 
\bal
 \,[\delta_\epsilon,\delta_{\epsilon'}]F &=
 \epsilon\gamma^{\mu\nu}\epsilon'(2D_\mu D_\nu\phi+iF_{\mu\nu}\phi)
+\frac{2\Delta}{3}\phi
 (\epsilon \gamma^\mu\gamma^\nu D_\mu D_\nu\epsilon'
 -\epsilon'\gamma^\mu\gamma^\nu D_\mu D_\nu\epsilon).
\eal
However, considering the supersymmetry generators that we described in the previous section, for all  the spaces it results 
 $[\delta_\epsilon,\delta_{\epsilon'}]F =0$.  A similar results holds also for the commutators of two barred supersymmetries on the field $\bar F$.  We can therefore conclude that,  for the spaces considered in the main text, also for the chiral multiplet two unbarred supersymmetries  and two barred supersymmetries commute.   The commutator  $[\delta_\epsilon,\delta_{\bar\epsilon}]$ is a sum of a translation, a rotation, a $R$-symmetry rotation and  a gauge transformation.

\subsection{Background gauge multiplet}
In three-dimensional theories, one often considers a background non-dynamical gauge field $\boldsymbol A_\mu$ that couples to a global symmetry.
In $\mathcal N=2$ theories, one introduces the SUSY partners ($\boldsymbol\sigma, \boldsymbol D,\ldots$), on which the path integral depends.
The gauginos are set to zero, and in order to preserve supersymmetry, we require that their variation vanish.  
Suppose that on $\mathbb S^3_b$ we have vortex loop of charge $\eta$ at $\vartheta=0$.
Supersymmetry requires that 
\begin{equation}
\boldsymbol A=\eta d\varphi_1\,,\quad \boldsymbol D=-\frac{\boldsymbol \sigma}{Rf}\,,
\quad
\boldsymbol\sigma=\text{constant}\,.
\end{equation}
On $\mathbb S^1\times \mathbb S^2$, when we have a (anti-)vortex loop of charge $\eta^+$ ($\eta^-$) at the north (south) pole, the configuration  preserving SUSY is given by
\begin{equation}
\boldsymbol A^\pm =-\frac{ a}{\beta}d\tau +\frac{m}{2} (\pm 1 -\cos\theta)d\varphi
\pm \eta^\pm d\varphi
\,,\quad
\boldsymbol\sigma=-\frac{ m}{2}\,,\quad
 a=\text{constant},\,\quad
 m\in\mathbb Z
\end{equation}
on the two patches $U^+=\{\theta\neq\pi\}$ and $U^-=\{\theta\neq 0\}$.

The supersymmetric BF coupling between the background and dynamical gauge multiplets is given by the insertion of $e^{-\mathcal S_{BF}}$ in the path integral where
\begin{equation}
\label{SUSY-BF}
 \mathcal S_{BF}=-\frac i {2\pi} \int \boldsymbol A\wedge dA
-\frac i {2\pi} \int  d^3x \sqrt g 
(\boldsymbol D\sigma+\boldsymbol \sigma  D)\,.
\end{equation}
The invariance of the BF term under $Q$ follows from that of the CS term because
\begin{equation}\label{BF-CS}
\mathcal S_{BF}(\boldsymbol A,\ldots;A,\ldots)=-i \left[
\mathcal S_{SCS}(\boldsymbol A+A,\ldots)- \mathcal S_{SCS}(\boldsymbol A,\ldots)- \mathcal S_{SCS}(A,\ldots)\right]_{k=1}\,.
\end{equation}

On $\mathbb S^3_b$, the scalar $\boldsymbol \sigma=\zeta/R$ is nothing but the FI parameter, and the second term in (\ref{SUSY-BF}) is the standard FI term $\mathcal S_\text{FI}$ that enters the path integral as
\begin{equation}\label{SUSY-FI}
e^{-\mathcal S_\text{FI}}\,,
\quad
\mathcal S_\text{FI}=
-\frac{i\zeta}{ 2\pi R} \int d^3x \sqrt g \left(D-\frac{\sigma}{Rf}\right)
\,.
\end{equation}

\section{Boundary terms on the round sphere  $\bS^3$}
\label{sec:boundaries}

In the presence of the vortex loop operator one needs to keep track of delta function 
contributions at the singularity, or alternatively of boundary terms arising from an 
excised tubular region of the loop operator.  There are three main reasons why these 
terms are important:
\begin{enumerate}
\item
Without these terms the localizing actions are not $Q$-exact.
\item
Without these terms the vortex loop operators would seem not to break any 
supersymmetry at all, while the boundary terms ensure they preserve only 
one half.
\item
The boundary terms may contribute to the value of the action evaluated at the 
saddle points of the localizing action.
\end{enumerate}
In this appendix we study the boundary terms for the different pieces of the 
$\cN=2$ actions in three dimensions, focusing for simplicity on the case 
of the round $\bS^3$.

For the round sphere $\bS^3$, the supersymmetry variations for the vector multiplet spinors 
\eqn{susyV} simplify to  
\bal
\label{susyspinor}
\delta\lambda &=
-\frac{1}{2} \gamma^{\mu\nu}\epsilon F_{\mu\nu}
+i\gamma^\mu\epsilon D_\mu\sigma -\left(D+\frac{\sigma}{R}\right) \epsilon,
\\
\delta\bar\lambda &=
-\frac{1}{2}\gamma^{\mu\nu}\bar\epsilon F_{\mu\nu}
-i\gamma^\mu\bar\epsilon D_\mu\sigma +\left(D+\frac{\sigma}{R}\right)\bar\epsilon,
\eal
and the variations for the spinors and the auxiliary scalars in the chiral multiplet 
\eqn{chiralsusy} are 
\bal
\delta\psi &= i\gamma^\mu\epsilon D_\mu\phi +i\epsilon\sigma\phi
-\frac{\Delta }{ R}\epsilon\phi+\bar\epsilon F,
\\
\delta\bar\psi &= i\gamma^\mu\bar\epsilon D_\mu\bar\phi
+i\bar\phi\sigma\bar\epsilon-\frac{\Delta }{ R}\bar\epsilon\bar\phi
+\bar F\epsilon,
\\
\delta F &=
\epsilon(i\gamma^\mu D_\mu\psi-i\sigma\psi-i\lambda\phi)
+\frac{1}{ 2R}(2\Delta-1)\epsilon\psi,
\\
\delta\bar F &=
\bar\epsilon(i\gamma^\mu D_\mu\bar\psi-i\bar\psi\sigma+i\bar\phi\bar\lambda)
+\frac{1}{2R}(2\Delta-1)\bar\epsilon\bar\psi .
\eal

\subsection{Boundary terms for supersymmetric Yang-Mills action}
\label{QSYM}

When using the supersymmetric Yang-Mills action \eqn{SYMact} as a localizing term 
it should be written as a total superderivative \cite{Hama:2010av}.  Keeping track of total derivative terms it is 
possible to show that for generic Killing spinors $\epsilon$ and $\bar\epsilon$
\bal
\delta_{\bar\epsilon}\delta_{\epsilon}\Tr\left(\frac{1}{2}\bar\lambda\lambda-2D\sigma\right)
=\bar\epsilon\epsilon\, \cL_{SYM} 
+D_\mu \Tr\bigg(& i\bar\epsilon\gamma_\nu\epsilon\, F^{\nu\mu}\sigma 
- \bar\epsilon\epsilon\,\sigma D^\mu \sigma 
-\bar\epsilon \gamma^{\nu\mu}\epsilon\, \sigma D_\nu\sigma
\\ &
+i\bar\epsilon \gamma^\mu\epsilon \left(D+\frac{\sigma}{R}\right)\sigma
+\frac{i}{2}\bar\lambda \gamma^\mu\epsilon (\bar\epsilon \lambda)\bigg).
\eal
Considering a boundary at small $\vartheta=\vartheta_0$ in the coordinate 
system \eqn{torusf}, we obtain 
\bal
\int d^3x\sqrt{g}\,
\delta_{\bar\epsilon}\delta_{\epsilon}&\Tr\left(\frac{1}{2}\bar\lambda\lambda-2D\sigma\right)
\\ 
=\bar\epsilon\epsilon\, &g_\text{YM}^2\cS_{SYM}
-R^3\int d\varphi_1 d\varphi_2 \cos\vartheta_0\sin\vartheta_0
\Tr\bigg( i\bar\epsilon\gamma_\nu\epsilon\, F^{\nu\vartheta}\sigma
-\bar\epsilon\epsilon\,\sigma D^\vartheta \sigma
\\&\hskip2cm
-\bar\epsilon \gamma^{\nu\vartheta}\epsilon\, \sigma D_\nu\sigma
+i\bar\epsilon \gamma^\vartheta\epsilon \left(D+\frac{\sigma}{R}\right)\sigma
+\frac{i}{2}\bar\lambda \gamma^\vartheta\epsilon (\bar\epsilon \lambda)\bigg)\,.
\eal
With $\epsilon$ and $\bar\epsilon$ satisfying (\ref{susy-S3}), the previous expression simplifies to 
\bal
&\label{S-SYM-B}
\int d^3x\sqrt{g}\,\delta_{\bar\epsilon}\delta_{\epsilon}\Tr\left(\frac{1}{2}\bar\lambda\lambda-2D\sigma\right)
\\ &
=g_\text{YM}^2\cS_{SYM}
+\int d\varphi_1 d\varphi_2 \cos\vartheta_0\sin\vartheta_0
\Tr \left(\frac{i}{\cos\theta_0}\sigma F_{\vartheta\varphi_2}
+\frac{R}{2}D_{\vartheta}\sigma^2
+\frac{iR^2}{4} \bar\lambda (\gamma_1-i\gamma_2)\lambda\right)\,.
\eal

Being a supersymmetry variation automatically implies that the sum of the bulk 
and boundary actions are invariant under supersymmetry.  This can also be verified 
directly, as the supersymmetry variations of the SYM Lagrangian (\ref{SYMact}) are 
total derivatives
\beq
\delta_{\epsilon}\cS_\text{SYM}
=\frac{1}{g_\text{YM}^2}\int d^3 x \sqrt{g}\,D_\mu\Tr 
\left[-\frac{i}{2}\bar\lambda\gamma_\nu\epsilon F^{\mu\nu}
-\frac{1}{2}\bar\lambda\epsilon D^{\mu}\sigma
-\frac{i}{2}\bar\lambda\gamma^\mu\epsilon \left(D+\frac{\sigma}{R}\right)\right]\,.
\eeq
Putting a cutoff at small $\vartheta_0$ gives the boundary term
\bal
\delta_{\epsilon}\cS_\text{SYM}
=\frac{1}{g_\text{YM}^2}\int d\varphi_1\,d\varphi_2
\Tr \bigg[&\frac{i}{2} \cos\vartheta_0\bar\lambda\gamma_2\epsilon F_{\vartheta\varphi_1}
+\frac{i}{2}\sin\vartheta_0\bar\lambda\gamma_3\epsilon F_{\vartheta\varphi_2}
\\&
+\frac{1}{2} \cos\vartheta_0\sin\vartheta_0\left(R\bar\lambda\epsilon\, D_{\vartheta}\sigma
+iR^2\bar\lambda\gamma_1\epsilon \left(D+\frac{\sigma}{R}\right)\right)\bigg]\,,
\eal
likewise
\bal
\delta_{\bar\epsilon}\cS_\text{SYM}
&=\frac{1}{g_\text{YM}^2}\int d^3 x \sqrt{g}\,D_\mu\Tr 
\left[-\frac{1}{4}\varepsilon^{\mu\nu\rho}\bar\epsilon\lambda F_{\nu\rho}
+\frac{R}{2}\bar\epsilon\gamma^{\mu\nu}\lambda D_{\nu}\sigma\right]
\\
&=\frac{1}{g_\text{YM}^2}\int d\varphi_1\,d\varphi_2
\left[\frac{1}{2}\bar\epsilon\lambda F_{\varphi_1\varphi_2}
-\frac{1}{2}\cos\vartheta_0
\bar\epsilon\gamma_{12}\lambda D_{\varphi_1}\sigma
-\frac{1}{2}\sin\vartheta_0
\bar\epsilon\gamma_{13}\lambda D_{\varphi_3}\sigma\right]\,.
\eal
By adding the boundary term in \eqn{S-SYM-B} 
and using the projection equations \eqn{susy-S3} it is possible to check that 
$\delta_\epsilon(S_\text{SYM}+S_\text{SYM}^B)
=\delta_{\bar\epsilon}(S_\text{SYM}+S_\text{SYM}^B)=0$.  Since this statement 
is true only assuming the projection equations, this also confirms 
that the vortex loop operators break half of the supersymmetry at the singularity.

Note also that the boundary action vanishes on the BPS solutions \eqn{vor} with 
arbitrary~$\sigma_0$.

\subsection{Boundary terms for Chern-Simons action}
\label{sec:boundarycs}

The supersymmetric CS action \eqn{CS} is not a total superderivative, still 
it is possible to add boundary terms such that its variation vanishes.  
The supersymmetry variation of this term is of course also a total derivative, 
which reads
\bal
\delta_\epsilon \cS_\text{SCS}&=\frac{ik}{4\pi}\int d^3x\sqrt{g}\,\Tr
\left[\frac{1}{2}\epsilon^{\mu\nu\rho}\partial_\mu(A_\nu\bar\lambda\gamma_\rho\epsilon)
-\partial_\mu(\sigma\bar\lambda\gamma^\mu\epsilon)\right]
\\&
=\frac{ik}{4\pi}\int d^2x\sqrt{g}\,\Tr
\left[-\frac{1}{2}\epsilon^{n\mu\nu}A_\mu\bar\lambda\gamma_ \nu\epsilon
+\sigma\bar\lambda\gamma^n\epsilon\right],
\eal
where the integral on the second line is on the boundary and 
$\gamma^n$ is in the normal direction.

Let us use the metric \eqn{torusf} and frame \eqn{torusviel} with a boundary 
at small $\vartheta_0$, so we get explicitly
\beq
\delta_\epsilon \cS_\text{SCS}
=\frac{ikR}{4\pi}\int d\varphi_1\,d\varphi_2\,\cos\vartheta_0\sin\vartheta_0
\Tr\left[-\frac{1}{2\sin\vartheta_0}A_{\varphi_1}\bar\lambda\gamma_3\epsilon
+\frac{1}{2\cos\vartheta_0}A_{\varphi_2}\bar\lambda\gamma_2\epsilon
+R \sigma\bar\lambda\gamma_1\epsilon\right].
\eeq

Let us consider the boundary term
\beq
\label{boundary}
\cS_\text{SCS}^{B}
=\frac{k}{4\pi}\int d\varphi_1\,d\varphi_2\,
\Tr\left[A_{\varphi_1}(A_{\varphi_2}-2iR\sigma)\right].
\eeq
Given that near the singularity $\gamma_3\epsilon=\epsilon$ \eqn{susy-S3} it follows
\beq
\delta_\epsilon (\cS_\text{SCS}+\cS_\text{SCS}^B)
=0.
\eeq

Likewise for the $\bar\epsilon$ variation
\bal
\delta_{\bar\epsilon} \cS_\text{SCS}&=\frac{ik}{4\pi}\int d^3x\sqrt{g}\,\Tr
\left[-\frac{1}{2}\epsilon^{\mu\nu\rho}\partial_\mu(A_\nu\bar\epsilon\gamma_\rho\lambda)
-\partial_\mu(\bar\epsilon\gamma^\mu\lambda \sigma)\right]
\\&
=\frac{ik}{4\pi}\int d^2x\sqrt{g}\,\Tr
\left[\frac{1}{2}\epsilon^{n\mu\nu}A_\mu\bar\epsilon\gamma_ \nu\lambda
+\bar\epsilon\gamma^n\lambda \sigma\right]\,,
\eal
and given that near the singularity $\gamma_3\bar\epsilon=-\bar\epsilon$, it results 
\beq
\delta_{\bar\epsilon} (\cS_\text{SCS}+\cS_\text{SCS}^B)
=0\,.
\eeq

Let's now consider gauge transformations.  Given an element $g$ of the gauge group, the gauge vector $A_\mu$ transforms as
\beq
A_\mu\rightarrow A_\mu^g=gA_\mu g^{-1} -i g \partial_\mu g^{-1}
\eeq  
and the  remaining fields in the vector multiplet transform in the adjoint representation, for instance
\beq
\sigma\rightarrow \sigma^g=g\sigma g^{-1} \,.
\eeq 
The super Chern-Simons Lagrangian 
\beq
\cL_\text{SCS}=\Tr \left[
\varepsilon^{\mu\nu\rho}\left(A_\mu\partial_\nu A_\rho+\frac{2i}{3}A_\mu A_\nu A_\rho\right)
-\bar\lambda\lambda+2D\sigma\right]
\eeq
transforms as
\beq
\cL_\text{SCS}\rightarrow\cL_\text{SCS}^g
=\cL_\text{SCS}+i\varepsilon^{\mu\nu\rho} \partial_\mu\Tr \left[\partial_\nu g^{-1} g A_\rho\right]
-\frac{1}{3}\varepsilon^{\mu\nu\rho}\Tr \left[g^{-1}\partial_\mu g\,  g^{-1}\partial_\nu g\, g^{-1}\partial_\rho g\right]\,.
\eeq
Using this result, it follows that the sum of the bulk and boundary actions transforms to
\beq
(\cS_{SCS}+\cS_{SCS}^B)^g=(\cS_{SCS}+\cS_{SCS}^B)-\frac{k}{2\pi}\int d\varphi_1 d \varphi_2\,
\Tr\left[\partial_{\varphi_1}g^{-1}g\left(iA_{\varphi_2}+\cos\theta_0 R\sigma \right)\right]  +T\,,
\eeq
where $T$  is given by 
\beq
\begin{aligned}
T&=-\frac{k}{12\pi}\int_{\vartheta\geq\vartheta_0} d^3 x \sqrt{g}\,\varepsilon^{\mu\nu\rho}
\Tr \left[g^{-1}\partial_\mu g\,  g^{-1}\partial_\nu g\, g^{-1}\partial_\rho g\right]
\\
&\qquad \qquad \qquad
-\frac{k}{4\pi}\int_{\vartheta=\vartheta_0} d\varphi_1 d\varphi_2 \Tr \left[g^{-1}\partial_{\varphi_1} g\,  g^{-1}\partial_{\varphi_2}  g\right]\,.
\end{aligned}
\eeq
In the limit $\vartheta_0\rightarrow 0$, $g(\vartheta=\vartheta_0)$ is independent of $\varphi_1$, and we recover the usual gauge transformation of the Chern-Simons action.

\subsection{Boundary terms for Fayet-Iliopoulos term}
\label{QFI}

Under a generic supersymmetry variation the Fayet-Iliopoulos action \eqn{FI} transforms as
\bal
\delta \cS_\text{FI}=-\frac{\zeta}{4\pi R}\int d^3x\sqrt{g}
D_\mu(\bar\lambda\gamma^\mu\epsilon+\bar\epsilon\gamma^\mu\lambda)\,,
\eal
and placing a boundary at small $\theta=\theta_0$, it results
\bal
\delta\cS_\text{FI}=\frac{\zeta R^2}{4\pi }\int d\varphi_1 d\varphi_2 \cos\vartheta_0\sin\vartheta_0
(\bar\lambda\gamma^\theta\epsilon+\bar\epsilon\gamma^\theta\lambda)\,.
\eal

If we add the boundary term
\bal
\label{FI-B}
\cS_\text{FI}^B=-\frac{\zeta }{2\pi }\int d\varphi_1 d\varphi_2 \cos\vartheta_0 A_{\varphi_1}\,,
\eal
and consider the supersymmetry generators that satisfy the condition (\ref{susy-S3}) we find
\bal
\delta(\cS_\text{FI}+\cS_\text{FI}^B)=0\,.
\eal

\subsection{Boundary terms for chiral action}
\label{Qchiral}

Like the SYM action, the chiral action \eqn{chi} is a total superderivative \cite{Hama:2010av} and can be 
used as a localizing term.  In the presence of a boundary or a singularity we need to consider 
boundary terms.

For generic Killing spinors $\epsilon$ and $\bar\epsilon$ the double variation is
\bal
&\delta_{\bar\epsilon}\delta_{\epsilon}
\left(\bar\psi \psi - 2i \bar\phi \sigma\phi+\frac{2(\Delta-1)}{R}\bar\phi\phi\right)
\\&\qquad
=\bar\epsilon\epsilon\, \cL_\text{chiral} 
+D_\mu \bigg(\bar\epsilon\gamma^{\mu\nu}\epsilon\, \bar\phi D_\nu \phi 
- \bar\epsilon\gamma^{\mu}\epsilon\, \bar\phi \sigma \phi
+\frac{i(2-\Delta)}{R} \bar\epsilon\gamma^{\mu}\epsilon\, \bar\phi \phi
+i\,\bar\epsilon \gamma^\mu\bar\psi\,(\epsilon \psi)\bigg)\,.
\eal
Placing a boundary at small $\vartheta=\vartheta_0$ in the coordinate system \eqn{torusf}, we obtain 
\bal
\int & d^3x\sqrt{g}\,\delta_{\bar\epsilon}\delta_{\epsilon}\left(\bar\psi \psi 
- 2i \bar\phi \sigma\phi+\frac{2(\Delta-1)}{R}\bar\phi\phi\right)
=\bar\epsilon\epsilon\, \cS_\text{chiral}
\\ &
- R^3 \int d\varphi_1 d\varphi_2 \cos\vartheta_0\sin\vartheta_0
\bigg( \bar\epsilon\gamma^{\vartheta\nu}\epsilon\, \bar\phi D_\nu \phi 
- \bar\epsilon\gamma^{\vartheta}\epsilon\, \bar\phi \sigma \phi 
+\frac{i(2-\Delta)}{R} \bar\epsilon\gamma^{\vartheta}\epsilon\, \bar\phi \phi
+i\,\bar\epsilon \gamma^\vartheta\bar\psi\,(\epsilon \psi)\bigg)\,.
\eal
With $\epsilon$ and $\bar\epsilon$ satisfying (\ref{susy-S3}), the previous expression simplifies to 
\bal
\label{mat-b}
\int  d^3x\sqrt{g}\,\delta_{\bar\epsilon}\delta_{\epsilon}
&\left(\bar\psi \psi - 2i \bar\phi \sigma\phi+\frac{2(\Delta-1)}{R}\bar\phi\phi\right)
=\cS_\text{chiral}
\\ &
- R \int d\varphi_1 d\varphi_2 
\cos\vartheta_0\bigg(i \bar\phi D_{\varphi_1} \phi
+\frac{i\,R}{2}\sin\vartheta_0\bar\psi(\gamma_1+i\gamma_2)\psi\bigg)\,.
\eal
Clearly the variation of the sum of bulk and boundary actions will vanish for all the 
supercharges parameterized by $\epsilon$ and $\bar\epsilon$ satisfying \eqn{susy-S3}.  
This can be verified by an explicit calculation as done for the SYM action above.

\section{Kinetic operators on $\bS^3$}
\label{sec:1-loop}

In this appendix we study the kinetic operators arising from expanding the localizing 
actions on $\bS^3$ and calculate their spectra.

\subsection{Vector multiplet}

In order to localize the vector multiplet we add to the Lagrangian a total superderivative 
$t\,\cS_\text{loc}=t\delta_\epsilon\delta_{\bar\epsilon}\int(\bar\lambda\lambda-4D\sigma)$, which 
is proportional to the Yang-Mills action \eqn{SYMact} \cite{Hama:2010av,Marino:2011nm} 
and the boundary term \eqn{S-SYM-B}.

To compute the one-loop contribution we consider fluctuations around the BPS configuration, i.e.
\beq
\sigma=\sigma_0+\frac{\sigma'}{\sqrt{t}}\,,
\qquad
D=-\frac{\sigma_0}{R}+\frac{D'}{\sqrt{t}}\,,
\qquad
A_\mu=A^{(0)}_\mu+\frac{A'_\mu}{\sqrt{t}}\,,
\qquad
\lambda=\frac{\lambda'}{\sqrt{t}}\,,
\eeq
and expand the Yang-Mills action up to quadratic order in the fluctuations.  Considering the 
background gauge $D^{(0)\mu}A'_\mu=0$, where $D^{(0)}$ is defined using the connection $A_\mu^{(0)}$
\bal
D^{(0)}_\mu=\nabla_\mu+i[A_\mu^{(0)},\,\,],
\eal
we obtain
\bal
\label{SYM-quad}
t\,\cL_\text{SYM} = \frac{1}{2}
\Tr \bigg[&
{-}A^{\prime\mu}D^{(0)\nu}D^{(0)}_\nu A'_\mu
-[\sigma_0,A_\mu']^2
-\sigma' D^{(0)\mu}D^{(0)}_\mu\sigma'
+\left(D'+\frac{\sigma'}{R}\right)^2
\\&
+i\bar\lambda'\gamma^\mu D^{(0)}_\mu\lambda'
+i\bar\lambda'[\sigma_0,\lambda']
-\frac{1}{2R} \bar\lambda'\lambda'
\bigg]
\eal
and this quadratic action is invariant under the supersymmetry variations
\bal
\label{susyVp}
\delta A_\mu' &=
\frac{i}{2} (\bar\epsilon\gamma_\mu\lambda'-\bar\lambda'\gamma_\mu\epsilon),
\\
\delta\sigma' &=
\frac{1}{2} (\bar\epsilon\lambda'-\bar\lambda'\epsilon),
\\
\delta\lambda' &=
-\gamma^{\mu\nu}\epsilon D^{(0)}_\mu A_{\nu}'
+i\gamma^\mu\epsilon (D_\mu^{(0)}\sigma'+i[A_\mu',\sigma_0]) -\left(D'+\frac{\sigma'}{R}\right)\epsilon,
\\
\delta\bar \lambda' &=
-\gamma^{\mu\nu}\bar\epsilon D^{(0)}_\mu A_{\nu}'
-i\gamma^\mu\bar\epsilon (D_\mu^{(0)}\sigma'+i[A_\mu',\sigma_0]) +\left(D'+\frac{\sigma'}{R}\right)\bar\epsilon,
\\
\delta D' &=
-\frac{i}{2} \bar\epsilon\gamma^\mu D_\mu^{(0)}\lambda'
-\frac{i}{2} D_\mu^{(0)}\bar\lambda'\gamma^\mu\epsilon
+\frac{i}{2}[\bar\epsilon\lambda',\sigma_0]
+\frac{i}{2}[\bar\lambda'\epsilon,\sigma_0] -\frac{1}{4R}\bar\epsilon\lambda'
+\frac{1}{4R}\bar\lambda'\epsilon,
\eal
where $\epsilon$ and $\bar\epsilon$ are the supersymmetry preserved by the vortex.  
Using the Cartan decomposition of the gauge group, a generic fluctuation field $\Phi'$ is written as 
\bal
\Phi'=\Phi^{\alpha}X_\alpha+\Phi^iK_i\,.
\eal
In the following we ignore the contribution of the Cartan components $\Phi^i$, since their actions 
do not depend on $\sigma_0$.  The action $t\cL_\text{SYM}$ is therefore written as 
\bal
\label{qacp}
\frac{1}{2}\int d^3 x \sqrt{g}
\sum_\alpha\Big[ & g^{\mu\nu}A^{-\alpha}_\mu 
\left(-\nabla_\mu^{(0)}\nabla^{\mu(0)}+\alpha(\sigma_0)^2\right) A^{\alpha}_\nu
+\bar\lambda^{-\alpha} 
\left(i\gamma^\mu \nabla_\mu^{(0)} +i \alpha(\sigma_0)
-\textstyle \frac{1}{2R} \right) \lambda^{\alpha}\\
&-\sigma^{-\alpha} \nabla^{(0)\mu}\nabla^{(0)}_\mu\sigma^{\alpha}
+\left(D^{-\alpha}+\textstyle\frac{\sigma^{-\alpha}}{R}\right)\left( D^{\alpha}
+\textstyle\frac{\sigma^{\alpha}}{R}\right)\Big]
\eal
where we defined the operator 
\beq
\nabla_\mu^{(0)}=\nabla_\mu+i\alpha(A^{(0)}_\mu)
\eeq
and used the fact that $A^{(0)}_\mu$ is in the Cartan of the gauge group.  The supersymmetry 
transformation for any $\Phi^\alpha$ can be easily obtained projecting on the $X_\alpha$ 
generator the expressions in ($\ref{susyVp}$).  It results that the effect of the vortex on 
the localizing action corresponds to replacing $\nabla_\mu$ with $\nabla_\mu^{(0)}$.  It is 
therefore convenient to redefine the generic field $\Phi^\alpha$ as 
\beq
\label{redef0}
\Phi^\alpha=e^{-i\alpha(A^{(0)}_\mu)x^\mu}\tilde\Phi^\alpha
\eeq
so that $\nabla_\mu^{(0)}\Phi^\alpha=e^{-i\alpha(A^{(0)}_\mu)x^\mu}\nabla_\mu\tilde\Phi^\alpha$.  
Since $\Phi^\alpha$ should be a periodic function, it follows that $\tilde\Phi^\alpha$ satisfies
\bal
\tilde\Phi^\alpha(\vartheta,\varphi_1+2\pi,\varphi_2)
&=e^{2\pi i \alpha(A^{(0)}_{\varphi_1})}\tilde\Phi^\alpha(\vartheta,\varphi_1,\varphi_2)
=e^{2\pi i \alpha (H)}\tilde\Phi^\alpha(\vartheta,\varphi_1,\varphi_2),\\
\tilde\Phi^\alpha(\vartheta,\varphi_1,\varphi_2+2\pi)&=\tilde\Phi^\alpha(\vartheta,\varphi_1,\varphi_2),
\eal
or in terms of $u$ and $v$ \eqn{uv}
\bal
\label{per}
\tilde\Phi^\alpha(e^{2\pi i}u,v)&=e^{2\pi i \alpha (H)}\tilde\Phi^\alpha(u,v),\\
\tilde\Phi^\alpha(u,e^{2\pi i}v)&=\tilde\Phi^\alpha(u,v).
\eal 
Considering the redefinition (\ref{redef0}), the action (\ref{qacp}) is written in terms of 
$\tilde\Phi^\alpha$ fields as
\bal
\label{qact}
\frac{1}{2}\int d^3 x \sqrt{g}
\sum_\alpha\Big[ & g^{\mu\nu}\tilde A^{-\alpha}_\mu 
\left(-\nabla_\mu\nabla^{\mu}+\alpha(\sigma_0)^2\right)\tilde A^{\alpha}_\nu
+\tilde{\bar{\lambda}}^{-\alpha} 
\left(i\gamma^\mu \nabla_\mu+i \alpha(\sigma_0)
-\textstyle \frac{1}{2R} \right)\tilde \lambda^{\alpha}\\
&-\tilde\sigma^{-\alpha} \nabla^{\mu}\nabla_\mu\tilde\sigma^{\alpha}
+\left(\tilde D^{-\alpha}+\textstyle\frac{\tilde\sigma^{-\alpha}}{R}\right)
\left( \tilde D^{\alpha}+\textstyle\frac{\tilde \sigma^{\alpha}}{R}\right)\Big]
\eal
and the supersymmetry transformations are given by 
\bal
\label{susyVt}
\delta \tilde A_\mu^\alpha &=
\frac{i}{2} (\bar\epsilon\gamma_\mu\tilde\lambda^{\alpha}
-\tilde{\bar\lambda}^\alpha\gamma_\mu\epsilon), 
\\
\delta\tilde\sigma^\alpha &=
\frac{1}{2} (\bar\epsilon\tilde\lambda^\alpha-\tilde{\bar\lambda}^\alpha\epsilon),
\\
\delta\tilde\lambda^\alpha &=
-\gamma^{\mu\nu}\epsilon \nabla_\mu \tilde A_{\nu}^{\alpha}
+i\gamma^\mu\epsilon (\nabla_\mu\tilde\sigma^\alpha-i\alpha(\sigma_0)\tilde A_\mu^{\alpha}) 
-\left(\tilde D^{\alpha}+\frac{\textstyle \tilde\sigma^{\alpha}}{R}\right)\epsilon,
\\
\delta\tilde{\bar\lambda}^\alpha &=
-\gamma^{\mu\nu}\bar\epsilon \nabla_\mu \tilde A_{\nu}^{\alpha}
-i\gamma^\mu\bar\epsilon (\nabla_\mu\tilde\sigma^\alpha-i\alpha(\sigma_0)\tilde A_\mu^{\alpha})
+\left(\tilde D^{\alpha}+\frac{\textstyle \tilde\sigma^{\alpha}}{R}\right)\bar\epsilon,
\\
\delta \tilde D^\alpha &=
-\frac{i}{2} \bar\epsilon\gamma^\mu \nabla_\mu\tilde \lambda^\alpha+
\frac{i}{2} \epsilon\gamma^\mu \nabla_\mu\tilde{\bar\lambda}^\alpha
-\left(\frac{i}{2}\alpha(\sigma_0)+\frac{1}{4R}\right)\bar\epsilon\tilde\lambda^\alpha
+\left(-\frac{i}{2}\alpha(\sigma_0)+\frac{1}{4R}\right)\tilde{\bar\lambda}^\alpha\epsilon\, .
\eal

The one-loop contribution of the vector multiplet that depends on $\sigma_0$ therefore is given by 
\beq
Z^{\text{vector}}_{\text{1-loop}}(\sigma_0)
=\prod_\alpha\frac{{\det}_\alpha\left(i \gamma^\mu \nabla_\mu
+i \alpha(\sigma_0) - \frac{1}{2R}\right)}
{\det_\alpha\left(-\nabla_\mu\nabla^{\mu}+\alpha(\sigma_0)^2\right)^{1/2}},
\eeq
where ${\det}_\alpha(\cal{O})$ is the determinant of the operator $\cal{O}$, evaluated in 
a space of fields $\tilde{\Phi}^\alpha$ that satisfy the boundary condition (\ref{per}).

In appendix~\ref{sec:harmonics} we have constructed a basis of harmonics satisfying 
these boundary conditions.  There are many such harmonics, most of which are non-normalizable 
and should not be included in the spectrum.  We discuss the states in  detail in 
Appendix~\ref{sec:harmonics} and try to make an educated guess which modes should 
be included in the spectrum.  The result is
\beq
\label{ts}
Z^{\text{vector}}_{\text{1-loop}}(\sigma_0)
=\prod_{\alpha>0}\prod_{n}^\infty\left(n^2+\alpha(R\sigma_0+iH)^2\right)\,.
\eeq

\subsection{Chiral multiplet}
\label{sec:chiral}

We now focus on the matter sector of the theory.  

Since $\cL_\text{chiral}$ \eqn{chi} is a total superderivative 
\cite{Hama:2010av,Marino:2011nm}, multiplying the matter action 
by an arbitrary parameter $t$, the result of the path integral remain unchanged.  
Therefore, the contribution of the matter sector is given by the quadratic fluctuation 
with respect to the classical configuration that minimize (\ref{chi}), {\em i.e.}, 
$\phi=\bar\phi=F=\bar F=\psi=\bar\psi=0$.

As for the vector multiplet, we consider fluctuation with respect to the localizing configuration
\bal
\phi=\frac{1}{\sqrt{t}}\phi',
\qquad
F=\frac{1}{\sqrt{t}}F',
\qquad
\psi=\frac{1}{\sqrt{t}}\psi',
\qquad
\bar\psi=\frac{1}{\sqrt{t}}\bar\psi',
\eal
and in the large $t$ limit,
\bal
\label{cq}
t\cL_\text{chiral} =&\,
D_\mu^{(0)}\bar\phi' D^{\mu(0)}\phi'
+\bar\phi'\sigma_0^2\phi'
+\frac{i2(\Delta-1)}{R} \bar\phi'\sigma_0\phi' +\frac{\Delta (2-\Delta)}{R^2} \bar\phi'\phi'
+\bar F'F' 
\\ &\,
-i\bar\psi'\gamma^\mu D_\mu^{(0)}\psi'
+i\bar\psi'\sigma_0\psi'
- \frac{2\Delta-1}{2R}\bar\psi'\psi'
\eal
where we defined 
\bal
D_\mu^{(0)}=\nabla_\mu+iA_\mu^{(0)i}K_{i}^{R},
\eal
where $K_{i}^{R}$ are the elements of the Cartan subalgebra written in the $R$ 
representation.  Also the constant field $\sigma_0$ is expressed in the $R$ 
representation of the Cartan subalgebra, {\em i.e.}, $\sigma_{0}=\sigma_{0}^iK_{i}^{R}$.  
The Lagrangian (\ref{cq}) is invariant under the following supersymmetry transformations
\bal
\label{chiralsusyq1}
\delta\phi' &= \bar\epsilon\psi',
\\
\delta\bar\phi' &= \epsilon\bar\psi',
\\
\delta\psi' &= i\gamma^\mu\epsilon D_\mu^{(0)}\phi' +i\epsilon\sigma_0\phi'
-\frac{\Delta }{ R}\epsilon\phi'+\bar\epsilon F',
\\
\delta\bar\psi '&= i\gamma^\mu\bar\epsilon D_\mu^{(0)}\bar\phi'
+i\bar\phi'\sigma_0\bar\epsilon-\frac{\Delta }{ R}\bar\epsilon\bar\phi'
+\bar F'\epsilon,
\\
\delta F '&=
\epsilon(i\gamma^\mu D_\mu^{(0)}\psi'-i\sigma_0\psi')
+\frac{1}{ 2R}(2\Delta-1)\epsilon\psi',
\\
\delta\bar F '&=
\bar\epsilon(i\gamma^\mu D_\mu^{(0)}\bar\psi'-i\bar\psi'\sigma_0)
+\frac{1}{2R}(2\Delta-1)\bar\epsilon\bar\psi' .
\eal
Expanding all the fields in the weights $\rho$ of the representation $R$ 
(see Section~\ref{sec:nonabel}) the Lagrangian becomes
\bal
\label{cq1}
t\cL_\text{chiral} =\sum_{\rho}\Bigg[
&\bar\phi^\rho\left(\nabla_\mu^{(0)} \nabla^{\mu(0)}
+\left(\rho(\sigma_0)+i\frac{\Delta-1}{R}\right)^2+\frac{1}{R^2}\right)\phi^\rho
+\bar F^\rho F^\rho 
\\ &
+\bar\psi^\rho\left({-i}\gamma^\mu \nabla_\mu^{(0)}+i\rho(\sigma_0)
-\frac{2\Delta-1}{2R}\right)\psi^\rho\Bigg]
\eal
where we defined 
\bal
\nabla_\mu^{(0)}=\nabla_\mu+i\rho(A_\mu^{(0)}).
\eal
Like for the vector multiplet, we redefine the fields as 
\beq
\Phi^\rho=e^{-i\rho(A^{(0)}_\mu)x^\mu}\tilde\Phi^\rho,
\eeq
so that 
$\nabla_\mu^{(0)}\Phi^\rho=e^{-i\rho(A^{(0)}_\mu)x^\mu}\nabla_\mu\tilde\Phi^\rho$ 
and $\tilde\Phi^\rho$ satisfy
\beq
\label{boundc}
\tilde\Phi^\rho(e^{2\pi i}u,v)=e^{2\pi i \rho(H)}\tilde\Phi^\rho(u,v),
\qquad
\tilde\Phi^\rho(u,e^{2\pi i}v)=\tilde\Phi^\rho(u,v).
\eeq
The fields $\tilde{\bar\Phi}^\rho$ can be thought as the complex conjugate of $\Phi^\rho$ fields, although in the Euclidean formulation of the theory $\tilde{\bar\Phi}^\rho$ and $\Phi^\rho$ are independent.  It is however natural to redefine $\tilde{\bar\Phi}^\rho$ as
\beq
\bar\Phi^\rho=e^{i\rho(A^{(0)}_\mu)x^\mu}\tilde{\bar\Phi}^\rho
\eeq
so that 
\beq
\tilde{\bar\Phi}^\rho(e^{2\pi i}u,v)
=e^{-2\pi i \rho(H)}\tilde{\bar\Phi}^\rho(u,v),
\qquad
\tilde{\bar\Phi}^\rho(u,e^{2\pi i}v)=\tilde{\bar\Phi}^\rho(u,v).
\eeq
In terms of $\tilde\Phi^\rho$ and $\tilde{\bar\Phi}^\rho$ fields, the Lagrangian reads
\bal
\label{cqr}
t\cL_\text{chiral} 
=\sum_{\rho}\Bigg[&
\tilde{\bar\phi}^\rho\left(-\nabla_\mu \nabla^{\mu}
+\left(\rho(\sigma_0)
+i\frac{\Delta-1}{R}\right)^2+\frac{1}{R^2} \right)\tilde{\phi}^\rho
+\tilde{\bar F}^\rho\tilde{F}^\rho
\\ &\hskip1in
+\tilde{\bar\psi}^\rho\left(-i\gamma^\mu \nabla_\mu
+i\rho(\sigma_0)
- \frac{2\Delta-1}{2R}\right)\tilde{\psi}^\rho\Bigg]
\eal
and it is invariant under the following supersymmetry transformations
\bal
\label{chiralsusyq2}
\delta\tilde\phi^\rho &= \bar\epsilon\tilde\psi^\rho,
\\
\delta\tilde{\bar\phi}^\rho &= \epsilon\tilde{\bar\psi}^\rho,
\\
\delta\tilde{\psi}^\rho &= i\gamma^\mu\epsilon \nabla_\mu\tilde{\phi}^\rho +i\epsilon\sigma_0\tilde{\phi}^\rho
-\frac{\Delta }{ R}\epsilon\tilde{\phi}^\rho+\bar\epsilon \tilde{F}^\rho,
\\
\delta\tilde{\bar\psi}^\rho&= i\gamma^\mu\bar\epsilon \nabla_\mu\tilde{\bar\phi}^\rho
+i\tilde{\bar\phi}^\rho\sigma_0\bar\epsilon-\frac{\Delta }{ R}\bar\epsilon\tilde{\bar\phi}^\rho
+\tilde{\bar F}^\rho\epsilon,
\\
\delta \tilde F^\rho&=
\epsilon(i\gamma^\mu \nabla_\mu\tilde\psi^\rho-i\sigma_0\tilde\psi^\rho)
+\frac{1}{ 2R}(2\Delta-1)\epsilon\tilde\psi^\rho,
\\
\delta\tilde{\bar F}^\rho&=
\bar\epsilon(i\gamma^\mu \nabla_\mu\tilde{\bar\psi}^\rho-i\tilde{\bar\psi}^\rho\sigma_0)
+\frac{1}{2R}(2\Delta-1)\bar\epsilon\tilde{\bar\psi}^\rho.
\eal
The one-loop contribution of the chiral multiplet that depend on $\sigma_0$ is given by 
\bal
Z^{\text{chiral}}_{\text{1-loop}}(\sigma_0)
=\prod_\rho\frac{\det_\rho\left( -i\gamma^\mu \nabla_\mu
+i\rho(\sigma_0)- \frac{2\Delta-1}{2R}\right)}
{\det_\rho\left(-\nabla_\mu \nabla^{\mu}
+\left(\rho(\sigma_0)+i\frac{\Delta-1}{R}\right)^2+\frac{1}{R^2} 
\right)},
\eal
where $\det_\rho({\cal O})$ is computed on the space of fields that satisfy the boundary 
conditions (\ref{boundc}).

As in the case of the vector multiplet we have to consider spherical harmonics 
with modified periodicity, discussed in Appendix~\ref{sec:harmonics}.  Our result is
\beq
\label{chiraldet}
Z_{\text{1-loop}}^{\text{chiral}}(\sigma_0)
=\prod_{n=1}^\infty\prod_\rho\left(\frac{n+1-\Delta+i\rho(R\sigma_0+iH)}
{n-1+\Delta-i\rho(R\sigma_0+iH)}\right)^{n}
=\prod_\rho s_{b=1}(i-i\Delta -\rho(R\sigma_0 +iH)),
\eeq
where $s_b(x)$ is the double sine function.  There are many subtleties in this 
expression which are discussed in Section \ref{sec:spectral}.

\section{Spherical harmonics with non-standard periodicity}
\label{sec:harmonics}

As explained in Appendix~\ref{sec:1-loop} we are interested in computing the spectrum of operators using a basis that does not satisfy the standard periodicity 
condition.  In particular, we consider eigenfunctions $\Phi(\theta,\phi,\psi)$ that satisfy
\beq
\label{per1}
\Phi(e^{2\pi i} u,v)=e^{2\pi i \alpha (H)}\Phi(u,v)=e^{2\pi i \eta}\Phi(u,v).
\eeq
Here $\alpha(H)=\eta$ is the value of $H$ for one of the weights, appropriate for a 
field in the adjoint representation.  For other representations it is replaced by 
$\rho(H)$, which to avoid clutter we will also denote by $\eta$.

\subsection{Scalar harmonics}
\label{sec:scalar-har}

Let us recall the construction of the usual scalar harmonics on $\bS^3$.  
The scalar Laplacian $-\nabla^2$ can be expressed in terms of the $SU(2)_L$ or 
$SU(2)_R$ angular momentum operators (\ref{left}), (\ref{right}) as
\bal
\label{deltaz}
-\nabla^2=\frac{4}{R^2}({L}^L)^2=\frac{4}{R^2}({ L}^R)^2.
\eal
The spherical harmonics are classified by representations of $SU(2)_L\times SU(2)_R$ 
which obviously should have the same quadratic Casimir.  The states $S(n,m,m')$ 
are labeled by three integers $n,m,m'$, such that $j=n/2\geq|m|,|m'|$.  $n$ is the principal quantum number and $m$ and $m'$ are eigenvalues of the operators $L^L_3$ and $L^R_3$.

These spherical harmonics can be written in terms of homogenous polynomials of degree $n$ 
in the four coordinates $u$, $\bar u$, $v$ and $\bar v$.  We can construct a highest weight 
state%
\footnote{The normalization is not important for our purposes, so we will ignore it.}
\beq
S(n,n/2,n/2)\propto u^{n}\,,
\eeq
which is annihilated by $L^L_+$ and $L^R_+$.  The full 
multiplet with $(n+1)^2$ states can be constructed by acting with $L^L_-$ and $L^R_-$
\beq
S(n,m,m')\propto (L^L_-)^{n/2-m}(L^R_-)^{n/2-m'} u^{n} \,.
\eeq
The lowest weight state is reached by acting $n$ times with both $L^L_-$ and $L^R_-$ and it has the 
form
\beq
S(n,-n/2,-n/2)\propto \bar u^{n}\,.
\eeq
We would like now to generalize this to functions which satisfy the periodicity condition (\ref{per1}).  
We require that the functions vanish at $u=0$ and are regular at $v=0$.  
$m$ and $m'$ are shifted by $\eta/2$ and a natural highest weight state is
\beq
S_H({\textstyle n+\eta,\frac{n+\eta}{2},\frac{n+\eta}{2}})
\propto u^{n+\eta}\,.
\eeq
This function has the desired periodicity conditions, for $\eta>-n$ it vanishes at $u=0$ and is regular 
at $v=0$.  Acting with the Laplacian on it gives
\beq
\label{scalar-casimir}
-\nabla^2 S_H\left({\textstyle n+\eta,\frac{n+\eta}{2},\frac{n+\eta}{2}}\right)
=\frac{(n+\eta)(n+\eta+2)}{R^2}\,S_H\left({\textstyle n+\eta,\frac{n+\eta}{2},\frac{n+\eta}{2}}\right).
\eeq
We can create other states solving the same equation by acting on this state with any 
number of $L^L_-$ and $L^R_-$.  For non-integer $\eta$ they form a non-unitary representation of 
$SO(4)$ which is infinite dimensional, so it is not clear how many of the states in this 
representation we should include.  We are not required to include the full representation, 
since the loop operator breaks the $SO(4)$ symmetry.  Note that both $L_-^L$ and 
$L_-^R$ include a $\partial_u$ derivative, so acting with a total of $k$ lowering operators 
will give a term proportional to $u^{n-k+\eta}$.  For $k>n+\eta$ this mode is singular at $u=0$.  
We rely on the analysis of supersymmetry multiplets in Appendix~\ref{sec:susyrepre} 
to determine which states should be included.  

In addition to the states which are descendants of the highest weight 
state there are more regular states that we can construct by starting with the modified 
lowest weight state
\beq
S_L\left({\textstyle n-\eta,\frac{-n+\eta}{2},\frac{-n+\eta}{2}}\right) 
\propto \bar u^{n-\eta}\,.
\eeq
Note that it has the same periodicity $e^{2\pi i\eta}$ under $u\to e^{2\pi i}u$, but a different Casimir
\beq
-\nabla^2 S_L\left({\textstyle n-\eta,\frac{-n+\eta}{2},\frac{-n+\eta}{2}}\right) 
=\frac{(n-\eta)(n-\eta+2)}{R^2}\,S_L\left({\textstyle n-\eta,\frac{-n+\eta}{2},\frac{-n+\eta}{2}}\right) .
\eeq
Acting on this state up to $k$ times with either $L^L_+$ or $L^R_+$ 
will generate $(k+1)(k+2)/2$ states.  If $k<n-\eta$ then these states are regular at $u=0$.

The space of scalar harmonics with non-trivial periodicity is equipped with a scalar product defined as for standard scalar harmonics 
\bal
\langle S(n_1,m_1,m_1')\,,S(n_2,m_2,m_2')\rangle=\int d\Omega\, \bar S(n_1,m_1,m_1')\,S(n_2,m_2,m_2')
\eal
where $d\Omega$ is the volume element on the $\bS^3$\footnote{For instance, in the torus fibration coordinates it is given by $d\Omega=dx^3\sqrt{g}=R^3\sin\vartheta\cos\vartheta d\vartheta d\varphi_1 d\varphi_2$.} and $\bar S$ is the complex conjugate of $S$.  It results 
\bal
\langle S_H(n_1,m_1,m_1')\,,S_H(n_2,m_2,m_2')\rangle&\propto \delta_{n_1,n_2}\delta_{m_1,m_2}\delta_{m_1',m_2'}\\
\langle S_L(n_1,m_1,m_1')\,,S_L(n_2,m_2,m_2')\rangle&\propto \delta_{n_1,n_2}\delta_{m_1,m_2}\delta_{m_1',m_2'}\\
\langle S_H(n_1,m_1,m_1')\,,S_L(n_2,m_2,m_2')\rangle&=0
\eal
where all the functions have the same deformation parameter $\eta$.

\subsection{Vector harmonics}
\label{sec:vec-har}

There are two sets of divergenceless vector harmonics on $\bS^3$ \cite{Sen:1985dc,Cutkosky:1983jd}: 
${\bf V}^{+}(n, m, m')$ that form a representation $(\frac{n+1}{2},\frac{n-1}{2})$ of the symmetry group 
$SU(2)_L\times SU(2)_R$, and ${\bf V}^{-}(n, m, m')$ that form a representation 
$(\frac{n-1}{2},\frac{n+1}{2})$.\footnote{We use bold characters for vectors in the four dimensional embedding space.} The two sets satisfy
\beq
\label{vectorlaplace}
-\nabla_\mu\nabla^\mu\,{\bf V}^\pm(n,m,m')=(n+1)^2\,{\bf V}^\pm(n,m,m')\,,
\eeq
and are related to each other by the parity operator $P$ as 
\beq
P\, {\bf V}^{+}(n, m, m')=(-1)^{n+1} {\bf V}^{-}(n, m', m),
\eeq
where the action of the parity on the complex variables $u$, $v$ is given by 
\beq
P \, u=-u,
\qquad
P\, v=-\bar v.
\eeq
Complex conjugation acts as 
\beq
\bar {\bf V}^{\pm}(n, m, m')=(-1)^{m+m'+1} {\bf V}^{\pm}(n, -m, -m').
\eeq
In each of the ${\bf V}^{+}(n, m, m')$ and ${\bf V}^{-}(n, m, m')$ multiplet there are $n(n+2)$ states.  
To explicitly write the states it is convenient to consider the scalar product of the vector harmonics 
with an auxiliary vector ${\bf r}'=(u',v')$ defined in the embedding space $\bC^2$.\footnote{ The scalar product for four dimensional vectors is defined as $A\cdot B=A_1B_1+A_2B_2+A_3B_3+A_4B_4=A_uB^u+A_{\bar{u}}B^{\bar{u}}+A_{v}B^{v}+A_{\bar{v}}B^{\bar{v}}=2A_uB_{\bar{u}}+2A_{\bar{u}}B_{u}+2A_{v}B_{\bar{v}}+2A_{\bar{v}}B_{v}$.} The highest 
weight state ${\bf V}^{+}(n,\frac{n+1}{2},\frac{n-1}{2})$ is given (ignoring normalizations) by 
\beq
\label{hig}
\textstyle{\bf r}'\cdot{\bf V}^{+}(n,\frac{n+1}{2},\frac{n-1}{2})\propto(-u)^{n-1}(v u' -u v')
\eeq
and the other ${\bf r}'\cdot{\bf V}^{+}(n, m, m')$ in the multiplet are obtained applying the 
annihilation operators $L^L_-+L^{\prime\, L}_-$ and $L^R_-+L^{\prime\, R}_-$, where 
$L^{\prime\, L}_-$ and $L^{\prime\, R}_-$ are generators acting on the auxiliary 
variables $u'$ and $v'$.  ${\bf V}^-$ are gotten by acting with the parity operator $P$.

We would like to construct vector harmonics that satisfy the periodicity condition (\ref{per1}).  
As with the scalars, we take the (unnormalized) modified highest weight states
\bal
\textstyle{\bf r}'\cdot{\bf V}^{+}_{H}(n+\eta,\frac{n+1+\eta}{2},\frac{n-1+\eta}{2})\propto u^{n-1+\eta}(v u' -u v'),
\\
\textstyle{\bf r}'\cdot{\bf V}^{-}_{H}(n+\eta,\frac{n-1+\eta}{2},\frac{n+1+\eta}{2})\propto u^{n-1+\eta}(\bar{v} u' -u \bar{v}').
\eal
We can act on these states with lowering operators to create other states
all of which satisfy
\beq
\label{veclaplace1}
-\nabla_\mu\nabla^\mu\,{\bf V}^\pm(n+\eta,m,m')=(n+\eta+1)^2\,{\bf V}^\pm(n+\eta,m,m')\,.
\eeq
Let us examine their behavior at $u\to0$.  The lowest power of $u$ in the descendants are
\bal
\label{hplusminus}
\textstyle{\bf r}'\cdot{\bf V}^{+}_{H}(n+\eta,\frac{n+1+\eta}{2}-l,\frac{n-1+\eta}{2}-r)
&\sim u'\,\bar v^l\, v^{r+1} (-\partial_u)^{l+r} u^{n-1+\eta}+\cdots
\\
\textstyle{\bf r}'\cdot{\bf V}^{-}_{H}(n+\eta,\frac{n-1+\eta}{2}-\tilde l,\frac{n+1+\eta}{2}-\tilde r)
&\sim u'\,\bar v^{\tilde{l}+1}\, v^{\tilde r} (-\partial_u)^{\tilde l+\tilde r} u^{n-1+\eta}+\cdots
\eal
We find that for $0<\eta<1$ there are 
$\frac{n(n+1)}{2}$ modes of each of ${\bf V}_H^\pm$ which are regular as $u\to0$.  The 
same statement holds true when considering singularities of the field strength rather 
than the gauge field.

In a similar fashion to before we can also start with the lowest weight states
\bal
\textstyle{\bf r}'\cdot{\bf V}^{+}_{L}(n-\eta,\frac{-n-1+\eta}{2},\frac{-n+1+\eta}{2})
\propto\bar u^{n-1-\eta}(\bar v \bar u' -\bar u \bar v'),
\\
\textstyle{\bf r}'\cdot{\bf V}^{-}_{L}(n-\eta,\frac{-n+1+\eta}{2},\frac{-n-1+\eta}{2})
\propto\bar u^{n-1-\eta}( v \bar u' -\bar u v'),
\eal
and act on them with raising operators, giving eigenstate of the vector Laplacian
\beq
\label{veclaplace2}
-\nabla_\mu\nabla^\mu\,{\bf V}^\pm(n-\eta,m,m')=(n-\eta+1)^2\,{\bf V}^\pm(n-\eta,m,m')\,.
\eeq
Their leading behavior at $u\to0$ is
\bal
\textstyle{\bf r}'\cdot{\bf V}^{+}_{L}(n-\eta,\frac{-n-1+\eta}{2}+l,\frac{-n+1+\eta}{2}+r)
&\sim \bar u'\, v^l\, \bar v^{r+1}\, (\partial_{\bar{u}})^{l+r} (\bar u)^{n-1-\eta}+\cdots
\\
\textstyle{\bf r}'\cdot{\bf V}^{-}_{L}(n-\eta,\frac{-n+1+\eta}{2}+\tilde l,\frac{-n-1+\eta}{2}+\tilde r)
&\sim \bar u'\, v^{\tilde{l}+1}\, \bar v^{\tilde r} (\partial_{\bar{u}})^{\tilde l+\tilde r} (\bar u)^{n-1-\eta}+\cdots
\eal

\subsection{Spinor Harmonics}
\label{sec:spinor-har}

We now study the spectrum of the Dirac operator 
\beq
-i \slashed{\nabla}=-i\gamma^\mu\nabla_\mu
\eeq
where the covariant derivative for spinors in the left-invariant frame is given by 
\beq
\nabla_\mu=\partial_\mu+\frac{i}{2R}\gamma_\mu.
\eeq
In terms of ${\bf L}^L$, the left-invariant angular momentum generators \eqn{left} 
and the spin operator $S^a=\frac{1}{2}\gamma^a$ this is
\beq
-i \slashed{\nabla}=\frac{1}{R}\left(4{\bf L}^L\cdot {\bf S}+\frac{3}{2}\right).
\eeq
Considering ${\bf J}={\bf L}^L+{\bf S}$, we have 
\beq
\label{dirac}
-i \slashed{\nabla}=\frac{1}{R}\left(2({\bf J}^2-({\bf L}^L)^2-{\bf S}^2)+\frac{3}{2}\right).
\eeq
Given that ${\bf S}$ has spin $s=1/2$, then for ${\bf L}$ with spin $l=n/2$, the spin of 
${\bf J}$ is $j=(n\pm1)/2$ and we label the eigenstate as $\chi^\pm(n,m,m')$.  It follows that for $\chi^+(n,m,m')$ states $n\ge0$, and for $\chi^-(n,m,m')$ states $n\ge1$.  The eigenvalues 
are $\pm(j+1)/R$, or explicitly
\bal
\label{diracvalues}
-i\slashed{\nabla}\chi^+(n,m,m')&=\frac{1}{R}(n+3/2)\,\chi^+(n,m,m')\,,
\\
-i\slashed{\nabla}\chi^-(n,m,m')&=-\frac{1}{R}(n+1/2)\,\chi^-(n,m,m')\,.
\eal
The multiplicity for $j=(n+1)/2$ is $(n+2)(n+1)$ and for $j=(n-1)/2$ is $n(n+1)$.

The highest and lowest states are given in \cite{Sen:1985dc} (in slightly 
different notations)
\bal
\label{chiplusminus}
j&={\textstyle\frac{n+1}{2}}:\qquad&
\textstyle\chi^+(n,\frac{n+1}{2},\frac{n}{2})&=\begin{pmatrix}u^n\\0\end{pmatrix}
\qquad&
\textstyle\chi^+(n,-\frac{n+1}{2},-\frac{n}{2})&=\begin{pmatrix}0\\\bar u ^n\end{pmatrix}
\\
j&={\textstyle\frac{n-1}{2}}:\qquad&
\textstyle\chi^-(n,\frac{n-1}{2},\frac{n}{2})&=\begin{pmatrix}\bar v u^{n-1}\\u^n\end{pmatrix}
\qquad&
\textstyle\chi^-(n,-\frac{n-1}{2},-\frac{n}{2})&=\begin{pmatrix}\bar u ^n\\v\bar u^{n-1}\end{pmatrix}
\eal

As with the other harmonics, we can deform them for $\eta\neq0$ as
\begin{align}
j={\textstyle\frac{n+1}{2}}:\qquad\qquad\qquad&
\nonumber\\
\textstyle\chi^+_H(n+\eta,\frac{n+1+\eta}{2},\frac{n+\eta}{2})
&=\begin{pmatrix}u^{n+\eta}\\0\end{pmatrix}\,
\quad&
\textstyle\chi^+_L(n-\eta,\frac{-n-1+\eta}{2},\frac{-n+\eta}{2})
&=\begin{pmatrix}0\\\bar u ^{n-\eta}\end{pmatrix}
\nonumber\\
j={\textstyle\frac{n-1}{2}}:\qquad\qquad\qquad&
\\\nonumber
\textstyle\chi^-_H(n+\eta,\frac{n-1+\eta}{2},\frac{n+\eta}{2})
&=\begin{pmatrix}\bar v u^{n-1+\eta}\\u^{n+\eta}\end{pmatrix}\,
\quad&
\textstyle\chi^-_L(n-\eta,\frac{-n+1+\eta}{2},\frac{-n+\eta}{2})
&=\begin{pmatrix}\bar u ^{n-\eta}\\v\bar u^{n-1-\eta}\end{pmatrix}\,
\end{align}
More modes are obtained by applying the annihilators $J_-=L^L_-+S_-$ and $L^R_-$ 
to highest weight states or the creators $J_+=L^L_++S_+$ and $L^R_+$ to the lowest 
weight states.  All these modes are eigenstate of the Dirac operator \eqn{dirac} with eigenvalues 
\bal
\label{diraceta}
-i\slashed{\nabla}\chi_H^\pm(n+\eta,m,m')&=\frac{1}{R}(\pm(n+\eta+1)+1/2)\,\chi_H^\pm(n+\eta,m,m')\,,
\\
-i\slashed{\nabla}\chi_L^\pm(n-\eta,m,m')&=\frac{1}{R}(\pm(n-\eta+1)+1/2)\,\chi_L^\pm(n-\eta,m,m')\,.
\eal
As before, these states will become singular when acting with too many creation/annihilation 
operators.  Some of these states have to be included to complete the supersymmetry multiplets 
analyzed now.

\subsection{Supersymmetry multiplets}
\label{sec:susyrepre}

In order to determine which fluctuation modes one should include in the calculation 
of the determinant it is helpful to consider the multiplets they form under 
the supercharges preserved by the vortex loop operator.  This is analyzed here 
and the final expressions for the determinants determined.

\subsubsection{Chiral multiplet}
\label{sec:chiral-har-susy}

The fluctuations of bosons and fermions can be expressed in terms of fields that satisfy non trivial 
boundary conditions, as discussed in previous sections.  These fields are related by supersymmetry 
transformations \eqn{susyVt}, \eqn{chiralsusyq2}, that when written in terms of symmetry generators 
$D_\mu=\frac{2i}{R}L_\mu^L$ \eqn{left} and $\gamma_\mu=2S_\mu$ are
\bal
\label{ctapp}
\delta\tilde\phi^\rho &= \bar\epsilon\tilde\psi^\rho,
\\
\delta\tilde{\psi}^\rho &= -\frac{4}{R}L_3^L\tilde{\phi}^\rho S_3\epsilon
-\frac{2}{R}L_+^L\tilde{\phi}^\rho S_-\epsilon -\frac{2}{R}L_-^L\tilde{\phi}^\rho S_+\epsilon 
+\left(i\sigma_0\tilde{\phi}^\rho
-\frac{\Delta }{ R}\tilde{\phi}^\rho\right)\epsilon+ \tilde{F}^\rho\bar\epsilon,
\\
\delta \tilde F^\rho&=
-\frac{4}{R}\,\epsilon({\bf L}^L\cdot{\bf S})\tilde\psi^\rho
-i\sigma_0\epsilon\tilde\psi^\rho
+\frac{1}{ R}(\Delta-2)\epsilon\tilde\psi^\rho,
\\
\delta\tilde{\bar\phi}^\rho &= \epsilon\tilde{\bar\psi}^\rho,
\\
\delta\tilde{\bar\psi}^\rho&= -\frac{4}{R}L_3^L\tilde{\bar\phi}^\rho S_3\bar\epsilon 
-\frac{2}{R}L_+^L\tilde{\bar\phi}^\rho S_-\bar\epsilon 
-\frac{2}{R}L_-^L\tilde{\bar\phi}^\rho S_+\bar\epsilon 
+\left(i\tilde{\bar\phi}^\rho\sigma_0
-\frac{\Delta }{ R}\tilde{\bar\phi}^\rho\right)\bar\epsilon
+\tilde{\bar F}^\rho\epsilon,
\\
\delta\tilde{\bar F}^\rho&=
-\frac{4}{R}\,\bar\epsilon({\bf L}^L\cdot{\bf S})\tilde{\bar\psi}^\rho
-i\bar\epsilon\tilde{\bar\psi}^\rho\sigma_0
+\frac{1}{R}(\Delta-2)\bar\epsilon\tilde{\bar\psi}^\rho.
\eal
For the supersymmetry preserved by the loop operators (\ref{susy-S3}), the parameter 
$\epsilon$ has spin $+1/2$ and $\bar\epsilon$ spin $-1/2$, therefore 
$\gamma_+\epsilon=\gamma_-\bar\epsilon=0$.  They can be written as 
$\epsilon=\left(\begin{smallmatrix}\epsilon_1\\0\end{smallmatrix}\right)$ and 
$\bar\epsilon=\left(\begin{smallmatrix}0\\\bar\epsilon_2\end{smallmatrix}\right)$ 
and a few terms of (\ref{ctapp}) drops out.

Focusing on specific components $\rho$ of the fluctuation fields with $\rho(H)=\eta$, 
they can be expanded in the harmonic bases as 
\bal
\label{chexp}
\tilde\phi=&\,\sum_{n,m,m'} \phi^H_{n+\eta,m,m'}\,S_H(n+\eta,m,m')
+\sum_{n,m,m'} \phi^L_{n-\eta,m,m'}\,S_L(n-\eta,m,m'),
\\ 
\tilde\psi=&\,\sum_{n,m,m'} \psi^{H+}_{n+\eta,m,m'}\,\chi_H^+(n+\eta,m,m')
+ \sum_{n,m, m'} \psi^{H-}_{n+\eta,m,m'}\,\chi_H^-(n+\eta,m,m'),
\\
&\,+\sum_{n,m,m'} \psi^{L+}_{n-\eta,m,m'}\,\chi_L^+(n-\eta,m,m')
+ \sum_{n,m, m'} \psi^{L-}_{n-\eta,m,m'}\,\chi_L^-(n-\eta,m,m'),
\\
\tilde F=&\,\sum_{n,m,m'} F^H_{n+\eta,m,m'}\,S_H(n+\eta,m,m')
+\sum_{n,m,m'} F^L_{n-\eta,m,m'}\,S_L(n-\eta,m,m').
\eal 
The expansions of $\tilde{\bar\phi}$, $\tilde{\bar\psi}$ and $\tilde{\bar F}$ are similar 
and from orthogonality of the states and the eigenvalues calculated in the previous 
sections \eqn{scalar-casimir}, \eqn{diraceta} one sees that the action \eqn{cqr} is\footnote{ This result follows from the relation $\bar S_{H,L}(n+\eta,m,m')=S_{L,H}(n+\eta,-m, -m')$ and similar relations for the other type of harmonics, and the scalar products discussed in the previous sections.  }
\bal
\label{modeaction}
\cS_\text{chiral}
&=\sum_{n,m,m'}
\bigg(\frac{(n+\eta)(n+\eta+2)+(R\rho(\sigma_0)+i(\Delta-1))^2+1}{R^2}\,
\bar \phi^L_{n+\eta,-m,-m'}\phi^H_{n+\eta,m,m'}
\\&\hskip15mm
+\frac{(n-\eta)(n-\eta+2)+(R\rho(\sigma_0)+i(\Delta-1))^2+1}{R^2}\,
\bar \phi^H_{n-\eta,-m,-m'}\phi^L_{n-\eta,m,m'}
\bigg)\\&\quad
+\sum_{n,m,m',\pm}
\bigg(\frac{(\pm(n+\eta+1)+iR\rho(\sigma_0)-\Delta+1)}{R}\,
\bar \psi^{L\pm}_{n+\eta,-m,-m'}\psi^{H\pm}_{n+\eta,m,m'}
\\&\hskip15mm
+\frac{(\pm(n-\eta+1)+iR\rho(\sigma_0)-\Delta+1)}{R}\,
\bar \psi^{H\pm}_{n-\eta,-m,-m'}\psi^{L\pm}_{n-\eta,m,m'}
\bigg)
\\&\quad
+\sum_{n,m,m'}
\left(\bar F^L_{n+\eta,-m,-m'}F^H_{n+\eta,m,m'}
+\bar F^H_{n-\eta,-m,-m'}F^L_{n-\eta,m,m'}\right)\,.
\eal
Note that $\bar{\tilde\phi}$ is in the conjugate representation to $\tilde\phi$, so the allowed 
values of $\eta$, which are the eigenvalues of the weights $\rho$ have the opposite signs.  
This matches with the fact that the shift of $n$ in the states arising from the highest and 
lowest weight states have the opposite signs.

To see the supermultiplet structure we can plug the expansion \eqn{chexp} into 
\eqn{ctapp}.  If we project the variation $\delta\tilde\phi^\rho$ into eigenstates of the total 
angular momentum ${\bf L}^2$, $L^L_3$ and $L^R_3$ and find that 
\bal
\delta \phi^H_{n+\eta,m,m'}
&\sim \psi^{H+}_{n+\eta,m+1/2,m'}+\psi^{H-}_{n+\eta,m+1/2,m'}\, ,
\\
\delta \phi^L_{n-\eta,m,m'}
&\sim \psi^{L+}_{n-\eta,m+1/2,m'}+\psi^{L-}_{n-\eta,m+1/2,m'}\, .
\eal
In the last expression we ignored numerical factors and assumed the 
states on the right hand side exist.

Likewise, when projecting $\delta\tilde\psi$ on eigenstates of 
${\bf L}\cdot{\bf S}$, $J^L_3$ and $L^R_3$ we find
\beq
\delta \psi^{H\pm}_{n+\eta,m+1/2,m'}
\sim \phi^H_{n+\eta,m,m'}+F^H_{n+\eta,m+1,m'}
\eeq
(and likewise for $\psi^{L\pm}$).  
The variation of the modes of $F$ give back the same modes $\psi$ as above.

We therefore conclude that the states
\beq
\label{long-chiral}
\big\{\phi^H_{n+\eta,m,m'}\,,\ \psi^{H+}_{n+\eta,m+1/2,m'}\,,\ 
\psi^{H-}_{n+\eta,m+1/2,m'}\,,\ F^H_{n+\eta,m+1,m'}\big\}
\eeq
are multiplets of the unbroken supersymmetry and likewise
\beq
\big\{\phi^L_{n-\eta,m,m'}\,,\ \psi^{L+}_{n-\eta,m+1/2,m'}\,,\ 
\psi^{L-}_{n-\eta,m+1/2,m'}\,,\ F^L_{n-\eta,m+1,m'}\big\}\,.
\eeq
For each such multiplet there is another multiplet of the barred fields, which couple to them in 
the action \eqn{modeaction}.  The contribution of each multiplet in \eqn{long-chiral} to the 
determinant is
\beq
\frac{(n+\eta+2+iR\rho(\sigma_0)-\Delta)(-n-\eta+iR\rho(\sigma_0)-\Delta)}
{(n+\eta)(n+\eta+2)+(R\rho(\sigma_0)+i(\Delta-1))^2+1}=-1\,.
\eeq
So up to minus signs, which we will not try to keep track of, the determinant is trivial.

The only exception to this statement is when the full multiplet does not exist, rather it 
gets shortened, in which case the determinant is nontrivial.

The largest value of $m$ for which the state $\phi^H_{n+\eta,m,m'}$ exists is 
$m=\frac{n+\eta}{2}$.  In these cases the multiplets get shortened, as 
the states $\psi^{H-}_{n+\eta,\frac{n+1+\eta}{2},m'}$ and $F^H_{n+\eta,1+\frac{n+\eta}{2},m'}$ 
do not exist.  Likewise there is a state $\psi^{L+}_{n-\eta,-\frac{n+1-\eta}{2}}$ 
but no modes $\phi^L$ and $\psi^{L-}$ with the relevant quantum numbers, only $F^L$.  
The shortened multiplets are therefore associated to $m=\frac{n+\eta}{2}$ and 
$m=-\frac{n-\eta}{2}-1$ and are respectively given by 
\beq
\label{short-chiral}
\begin{gathered}
\big\{
\phi^H_{n+\eta,\frac{n+\eta}{2},m'}\,,\ \psi^{H+}_{n+\eta,\frac{n+1+\eta}{2},m'}\big\}\,,
\\
\big\{\psi^{L+}_{n-\eta,-\frac{n+1-\eta}{2},m'}\,,\ F^L_{n-\eta,-\frac{n-\eta}{2},m'}\big\}\,.
\end{gathered}
\eeq
Of course a similar statement applies to $\bar \phi$, $\bar \psi^\pm$ and $\bar F$.

Each of the multiplets on the first line of \eqn{short-chiral} contributes to the 
determinant a factor of
\beq
\frac{(n+\eta+2+iR\rho(\sigma_0)-\Delta)}
{(n+\eta)(n+\eta+2)+(R\rho(\sigma_0)+i(\Delta-1))^2+1}
=\frac{1}{n+\eta-iR\rho(\sigma_0)+\Delta}\,,
\eeq
and each multiplet on the second line
\beq
n-\eta+2+iR\rho(\sigma_0)-\Delta\,.
\eeq
For $\eta=0$ there are $n+1$ copies of each of these multiplets, which we expect to 
not change when turning on $\eta\neq0$.  The only question is how many states get a shift 
$n\to n+\eta$ and how many $n\to n-\eta$, which is answered by the supersymmetry 
analysis above and the assumption of minimal singularities.  
We finally find that the determinant for the full chiral multiplet including the full representation $R$ is
\bal
\label{zc}
Z_{\text{1-loop}}^{\text{chiral}}(\sigma_0)
&=\prod_\rho\prod_{n=0}^\infty\left(\frac{n+2-\Delta+i\rho(R\sigma_0+iH)}{n+\Delta-i\rho(R\sigma_0+iH)}\right)^{n+1}
\\&
=\prod_\rho\prod_{n=1}^\infty\left(\frac{n+1-\Delta+i\rho(R\sigma_0+iH)}{n-1+\Delta-i\rho(R\sigma_0+iH)}\right)^{n}
\\
&=\prod_\rho s_{b=1}(i-i\Delta -\rho(R\sigma_0 +iH))
\eal
where $s_b(x)$ is the double sine function.

\subsubsection{Vector multiplet}
\label{sec:vec-har-susy}

We can repeat the same analysis for the vector multiplet.  We expand the fluctuation fields as%
\footnote{For brevity we omit the indices of the harmonic functions which match the modes they multiply.}
\bal
\label{vecexp}
A_\mu'=&\,\sum_{n,m,m'} \left(
A^{H+}_{n+\eta,m,m'}\,V_{H\mu}^+
+ A^{H-}_{n+\eta,m,m'}\,V_{H\mu}^-
+A^{L+}_{n-\eta,m,m'}\,V_{L\mu}^+
+A^{L-}_{n-\eta,m,m'}\,V_{L\mu}^-\right),
\\
\lambda'=&\, \sum_{n,m,m'} \left(\lambda^{H+}_{n+\eta,m,m'}\,\chi_H^+
+\lambda^{H-}_{n+\eta,m,m'}\,\chi_H^-
+\lambda^{L+}_{n-\eta,m,m'}\,\chi_L^+
+\lambda^{L-}_{n-\eta,m,m'}\,\chi_L^-\right),
\\
\bar\lambda'=&\, \sum_{n,m,m'}\left(\bar \lambda^{H+}_{n+\eta,m,m'}\,\chi_H^+
+\bar \lambda^{H-}_{n+\eta,m,m'}\,\chi_H^-
+\bar \lambda^{L+}_{n-\eta,m,m'}\,\chi_L^+
+\bar \lambda^{L-}_{n-\eta,m,m'}\,\chi_L^-\right),
\\ 
\sigma'=&\,\sum_{n,m,m'} \left(\sigma^H_{n+\eta,m,m'}\,S_H+\sigma^L_{n-\eta,m,m'}\,S_L\right),
\\
D'=&\,\sum_{n,m,m'} \left(D^H_{n+\eta,m,m'}\,S_H+D^L_{n-\eta,m,m'}\,S_L\right),
\\
c=&\,\sum_{n,m,m'} \left(c^H_{n+\eta,m,m'}\,S_H+c^L_{n-\eta,m,m'}\,S_L\right),
\eal 
and we included the ghost field $c$, which thus far has been ignored, but should be included in a 
full analysis of the theory.

Expanding the supersymmetry transformations \eqn{susyVt} as
\bal
\label{susyVm2}
\delta \tilde A_\mu^\alpha &=
i e_\mu{}^{a}(\bar\epsilon S_a\tilde\lambda^{\alpha}-\tilde{\bar\lambda}^\alpha S_a\epsilon),
\\
\delta\tilde\sigma^\alpha &=
\frac{1}{2} (\bar\epsilon\tilde\lambda^\alpha-\tilde{\bar\lambda}^\alpha\epsilon),
\\
\delta\tilde\lambda^\alpha &=
\frac{4}{R}\varepsilon^{abc}e_a{}^\nu S_b\epsilon L^L_c \tilde A_{\nu}^{\alpha}
-\frac{4}{R}S^a\epsilon L^L_a \tilde\sigma^\alpha
+2\alpha(\sigma_0) S^a \epsilon e_a{}^{\mu}\tilde A_\mu^{\alpha}
-\left(\tilde D^{\alpha}+\frac{\textstyle \tilde\sigma^{\alpha}}{R}\right)\epsilon,
\\
\delta\tilde{\bar\lambda}^\alpha 
&=\frac{4}{R}\varepsilon^{abc}e_a{}^\nu S_b\bar\epsilon L^L_c \tilde A_{\nu}^{\alpha}
+\frac{4}{R}S^a\bar\epsilon L^L_a \tilde\sigma^\alpha
-2\alpha(\sigma_0) S^a \bar\epsilon e_a{}^{\mu}\tilde A_\mu^{\alpha} 
+\left(\tilde D^{\alpha}+\frac{\textstyle \tilde\sigma^{\alpha}}{R}\right)\bar\epsilon,
\\
\delta \tilde D^\alpha &=
\frac{1}{2R} \bar\epsilon(4 {\bf L}^L\cdot{ \bf S}+\frac{3}{2})\tilde \lambda^\alpha
-\frac{1}{2R} \epsilon(4 {\bf L}^L\cdot{ \bf S}+\frac{3}{2})\tilde{\bar\lambda}^\alpha
-\left(\frac{i}{2}\alpha(\sigma_0)+\frac{1}{4R}\right)\bar\epsilon\tilde\lambda^\alpha ,
\\&\quad
+\left(-\frac{i}{2}\alpha(\sigma_0)+\frac{1}{4R}\right)\tilde{\bar\lambda}^\alpha\epsilon,
\eal
one finds that 
the modes that belong to the same multiplets have quantum numbers
\bal
\big\{
&A^{H+}_{n+1+\eta,m,m'}\,,\ A^{H-}_{n-1+\eta,m,m'}\,,\ 
\sigma^H_{n+\eta,m,m'}\,,\ D^H_{n+\eta,m,m'}\,,\ c^H_{n+\eta,m,m'}\,,\ 
\\&
\lambda^{H+}_{n+\eta,m+1/2,m'}\,,\ \lambda^{H-}_{n+\eta,m+1/2,m'}\,,\ 
\bar \lambda^{H+}_{n+\eta,m-1/2,m'}\,,\ \bar \lambda^{H-}_{n+\eta,m-1/2,m'}
\big\}
\eal
and likewise for those arising from the lowest weight states.  Though we did not 
include the ghosts in the explicit SUSY transformations, it is clear that they 
should appear in the off-shell multiplets as above.  Note that 
for the vectors the principle quantum number $n$ is shifted by $\pm1$.  

As with the chiral multiplet, a full multiplet contributes nothing to the 
determinant of the associated kinetic operators, and the only multiplets which 
contribute are shortened ones.%
\footnote{Note that the ghost $c$ cancels the contribution of the scalar $\sigma$.}

The shortest multiplets for the $H$ modes are obtained with $m=\frac{n+\eta}{2}+1$ 
and for the $L$ modes with $m=-\frac{n-\eta}{2}-1$.  They are of the form
\bal
&\big\{A^{H+}_{n+1+\eta,1+\frac{n+\eta}{2},m'}\,,\ 
\bar \lambda^{H+}_{n+\eta,\frac{n+1+\eta}{2},m'}\big\},
\\
&\big\{A^{L+}_{n+1-\eta,-1-\frac{n-\eta}{2},m'}\,,\ 
\lambda^{L+}_{n-\eta,-\frac{n+1-\eta}{2},m'}\big\}.
\eal
The action \eqn{qact} couples a mode arising from the highest weight state and a mode from the lowest 
weight state with opposite weights $\pm\alpha$, hence with opposite signs of 
$\eta$.  Thus a pair of short multiplets as above is coupled by 
({\em cf.}, \eqn{veclaplace1}, \eqn{diraceta})
\bal
&((n+2+\alpha(H))^2+R^2\alpha(\sigma_0)^2)\,
A^{H+}_{n+1+\eta,1+\frac{n+\eta}{2},m'}
A^{L+}_{n+1+\eta,-1-\frac{n-\eta}{2},m'}
\\&
-(n+2+i\alpha(R\sigma_0-iH))\,
\bar \lambda^{H+}_{n+\eta,\frac{n+1+\eta}{2},m'}
\lambda^{L+}_{n+\eta,-\frac{n+1-\eta}{2},m'}\,.
\eal
The contribution of such pair of multiplets to the determinant is therefore
\beq
\label{vshortdet}
-\frac{1}{n+2-i \alpha(R\sigma_0+iH)}\,,
\eeq

Other longer, but still short multiplets are
\bal
\label{vecshort}
&\big\{A^{H+}_{n+1+\eta,\frac{n+\eta}{2},m'}\,,\ 
\sigma^H_{n+\eta,\frac{n+\eta}{2},m'}\,,\ D^H_{n+\eta,\frac{n+\eta}{2},m'}\,,\ c^H_{n+\eta,\frac{n+\eta}{2},m'}\,,\ 
\\&\qquad\qquad\qquad
\lambda^{H+}_{n+\eta,\frac{n+1+\eta}{2},m'}\,,\ 
\bar \lambda^{H+}_{n+\eta,\frac{n-1+\eta}{2},m'}\,,\ \bar \lambda^{H-}_{n+\eta,\frac{n-1+\eta}{2},m'}
\big\}
\\
&\big\{A^{L+}_{n+1-\eta,-\frac{n-\eta}{2},m'}\,,\ 
\sigma^L_{n-\eta,-\frac{n-\eta}{2},m'}\,,\ D^L_{n-\eta,-\frac{n-\eta}{2},m'}\,,\ c^L_{n-\eta,-\frac{n-\eta}{2},m'}\,,\ 
\\&\qquad\qquad\qquad
\lambda^{L+}_{n-\eta,-\frac{n-1-\eta}{2},m'}\,,\ \lambda^{L-}_{n-\eta,-\frac{n-1-\eta}{2},m'}\,,\ 
\bar \lambda^{L+}_{n-\eta,-\frac{n+1-\eta}{2},m'}\big\}\,.
\eal
These together give (ignoring overall signs)
\beq
\label{shortdet}
(n-i \alpha(R\sigma_0+iH)).
\eeq
As before we assume that the number of short multiplets is the same as for $\eta=0$, 
which gives $n+1$.  Multiplying \eqn{vshortdet} and \eqn{shortdet} each with 
multiplicity $n+1$ gives
\bal
\label{app-infprod}
Z^{\text{vector}}_{\text{1-loop}}(\sigma_0)
&=\prod_\alpha\prod_{n=0}^\infty
\left(\frac{n-i \alpha(R\sigma_0+iH)}{n+2-i \alpha(R\sigma_0+iH)}\right)^{n+1}\
\\&=
\prod_{\alpha>0}\prod_{n=0}^\infty
\left(\frac{n^2+ \alpha(R\sigma_0+iH)^2}
{(n+2)^2+\alpha(R\sigma_0+iH)^2}\right)^{n+1}
\\&=
\prod_{\alpha>0}\left(\alpha(R\sigma_0+iH)^2
\prod_{n=1}^\infty
(n^2+\alpha(R\sigma_0+iH)^2)^2\right).
\eal
The first term on the last line arises from the numerator of the $n=0$ case in the line 
above, all the rest comes from the numerator of the $n=1$ case and the difference 
arising from shifting the index $n+2\to n$ in the denominator.

It should be pointed out that for $\eta=0$ the product in the numerator starts at 
$n=1$, not $n=0$.  The relevant multiplet is \eqn{vecshort} with $n=0$, where 
the state $\bar\lambda^{H-}_{0,-\frac{1}{2},0}$ does not exist, without which the 
determinant of this multiplet is one.  For $0<\eta<1$ this mode is singular, but 
we think it should still be included to complete the multiplet, and because other 
modes that are equally singular are also included.  Physically this mode should be 
thought of as an almost goldstino mode due to the broken supersymmetry induced 
by the vortex, which exists only in the vortex background.

Regularizing the infinite product in \eqn{app-infprod} we obtain 
\bal
\label{app-vec-result}
Z^{\text{vector}}_{\text{1-loop}}(\sigma_0)
&=\prod_{\alpha>0}\alpha(R\sigma_0+iH)^2
\frac{\sinh^2(\pi\alpha(R\sigma_0+iH))}{(\pi\alpha(R\sigma_0+iH))^2}
\\
&=\prod_{\alpha>0}\frac{1}{\pi^2}\sinh^2(\pi\alpha(R\sigma_0+iH)).
\eal
For $\alpha(H)=0$ the product should start at $n=1$ and the denominator is instead 
$\pi^2\to\pi^2\alpha(\sigma_0)^2$.

\section{Index    theory  calculations for $\mathbb S^3_b$}
\label{sec:index-details}

In this Appendix, we provide some details on the localization computation by the index  theory  for $\mathbb S^3_b$.
To avoid cluttering equations we will suppress the mass parameters that are associated with flavor symmetries.
They can be easily restored by the replacement $\sigma\rightarrow \sigma+\text{mass}$.

\subsection{Vanishing theorem}

Let us show that in the limit $t\rightarrow +\infty$, the path integral weighted by $e^{-t Q\cdot V}$ with $V$ given in (\ref{Vchoice}) and (\ref{Vvecchi}) localizes to the configurations (\ref{S3b-config}).
We will also show that in the presence of a vortex loop, the fields acquire, on top of the smooth configurations (\ref{S3b-config}),
the appropriate singular parts that characterize the operator.
Given our choice of $V$, this will be done by solving the equations $Q\cdot\text{fermion}=0$.

For a vector multiplet, noting that our SUSY parameters (\ref{S3b-spinors}) satisfy $\epsilon=C^{-1}\bar\epsilon^*$, let us compute%
\footnote{%
  The symbol $*$ acts as complex conjugation on Grassmann-even and -odd numbers, and as hermitian conjugation on fields: $\phi^*=\bar\phi, \bar\phi^*=\phi, \sigma^*=\sigma$, $D^*=D^\dagger$.
}
\begin{equation}
  \begin{aligned}
0=C^{-1}(Q\bar\lambda)^* &=
C^{-1}(-\frac{1}{2}\gamma^{\mu\nu}\bar\epsilon F_{\mu\nu}+D\bar\epsilon
-i\gamma^\mu\bar\epsilon D_\mu\sigma  +\frac 1 {Rf} \sigma{\bar\epsilon})^*
\\
&=-\frac 12 \gamma^{\mu\nu}\epsilon F_{\mu\nu} + D^\dagger \epsilon -i\gamma^\mu\epsilon D_\mu\sigma +\frac 1 {Rf} \sigma \epsilon\,.
  \end{aligned}
\end{equation}
and compare it with
\begin{equation}
0=     \delta\lambda =
-\frac{1}{2} \gamma^{\mu\nu}\epsilon F_{\mu\nu}-D\epsilon
+i\gamma^\mu\epsilon D_\mu\sigma - \frac{1}{Rf} \sigma\epsilon\,.
\end{equation}
We find that
\begin{equation}
\label{delta-lambda}
  -\frac  12 \gamma^{\mu\nu}\epsilon F_{\mu\nu} -i ({\rm Im}\,D) \epsilon=0\,,
\quad
({\rm Re}\,D)\epsilon -i\gamma^\mu\epsilon D_\mu\sigma +\frac 1 {Rf} \sigma \epsilon=0\,.
\end{equation}
In the absence of a vortex loop, we take $D$ to be real (hermitian), and hence obtain $F_{\mu\nu}=0$ from the first equation.
If we have vortex loops at $\vartheta=0$ or $\pi/2$,%
\footnote{%
In the following, we will set $A_\mu=0$ without a vortex loop, and $A_{\vartheta}=0, A_{\varphi_1}=H_1$, $A_{\varphi_2}=H_2$ with vortex loops, where $H_1$ and $H_2$ are the vorticities of the operators at $\vartheta=0$ and $\vartheta=\pi/2$.
}
then $\text{Im}\,D$ develops delta function singularities there so that the first equation
in (\ref{delta-lambda}) is satisfied.

Contracting the second equation with $\epsilon^\dagger$, we deduce from the real part that ${\rm Re}\,D+ \frac 1 {Rf} \sigma=0$.
Then we have $0=||\gamma^\mu\epsilon D_\mu\sigma||^2=||D_\mu\sigma||^2$, thus $D_\mu\sigma=0$.

For a chiral multiplet we compute
\begin{equation}
  \begin{aligned}
C^{-1}(Q\bar\psi)^* &=
i\gamma^\mu\epsilon D_\mu\phi
-i\sigma\phi\epsilon
-\frac \Delta {Rf} \phi \epsilon 
-F\bar\epsilon\,.
  \end{aligned}
\end{equation}
Comparing this with
\begin{equation}
  \begin{aligned}
 0= Q\psi &= i\gamma^\mu\epsilon D_\mu\phi +i\epsilon\sigma\phi
-\frac{\Delta }{Rf}\epsilon\phi+\bar\epsilon F\,,
  \end{aligned}
\end{equation}
we find that $\sigma\phi=F=0$, and we are left with
\begin{equation}
\label{eq-chiral}
  \begin{aligned}
 0= i\gamma^\mu\epsilon D_\mu\phi 
-\frac{\Delta }{Rf}\epsilon\phi\,.
  \end{aligned}
\end{equation}
Let us substitute the explicit expression for $\epsilon$ given in (\ref{S3b-spinors}), and take linear combinations of the two components in (\ref{eq-chiral}).
From one combination we get $\frac{ib}{R} D_{\varphi_2}\phi +\frac{i}{Rb} D_{\varphi_1}\phi-\frac{\Delta}{Rf}\phi=0$, or, by taking gauge and  R-symmetry backgrounds into account, obtain
\begin{equation}
[ib\partial_{\varphi_2} +ib^{-1} \partial_{\varphi_1}-(\Delta+b H_2+ b^{-1}H_1)]\phi=0\,.
\end{equation}
A non-zero solution to this is the matter vortex configuration discussed in Section~\ref{sec:m-vortex}.
It is singular for generic values of $\Delta$ and does not contribute to the path integral unless we choose to insert the corresponding disorder operators.
We will not include the contributions from the configuration in this paper.

\subsection{Gauge fixing}
\label{sec:gauge-fix}
In order to perform  the one-loop calculation, we need to fix the gauge for the field configurations around the chosen saddle point.
As usual, we introduce ghosts $(c,\bar c)$ and a bosonic auxiliary field $B$ and require that by the BRST charge $Q_\text{B}$ acts as%
\footnote{%
Also note that $ [c,c]\equiv c_\alpha c_\beta [T^\alpha,T^\beta]$ if  $c=c_\alpha T^\alpha$.
}
\begin{equation}
\label{QB-ghosts}
  \begin{aligned}
Q_\text{B}\cdot c=&-\frac i 2 [c,c]\,,\\
Q_\text{B} \cdot\bar c=&B\,,\\
Q_\text{B}\cdot B=&0\,,
  \end{aligned}
\end{equation}
on $(c,\bar c, B)$, and as%
\footnote{%
Here $\mathcal G(c)\,\cdot$ is the gauge transformation with parameter $c$.   For example $Q_\text{B} A_\mu=D_\mu c$, $Q_\text{B}\lambda= -i \{c,\lambda\}$.
}
\begin{equation}\label{QB-fields}
  Q_\text{B}\cdot (\text{field})=- \mathcal G(c)\cdot (\text{field})\,,
\end{equation}
on the original fields.
It is also standard to define the functional
\begin{equation}\label{Vgh}
  V_\text{gh}=\int d^3x \sqrt h \Tr\left(\bar c(G(\tilde A)+\frac{\xi}{2}B)\right)
\end{equation}
with a choice of gauge fixing term $G$.%
\footnote{%
As in the standard $R_\xi$ gauge, $\xi$ is an arbitrary parameter that does not affect the result of path integral.
}
Let us indicate by $(0)$ objects defined at the saddle point, and by tilde the difference of the dynamical field from its background value.
For example $\tilde A=A-H_1d\varphi_1-H_2d\varphi_2\,,\tilde\sigma=\sigma-\sigma^{(0)}$.%
\footnote{%
In the text the saddle point value $\sigma^{(0)}$ is simply denoted as $\sigma$, and is only distinguished by the context.
}
The standard choice of $G(\tilde A)$ is $D^\mu_{(0)} \tilde A_\mu$, but we will make a slightly different choice below.
In the familiar background field gauge, we would gauge-fix by adding a gauge-fixing Lagrangian $Q_\text{B}\cdot V_\text{gh}$.
For localization, we need to modify  $Q_\text{B}\cdot V_\text{gh}$ so that it is compatible with the supercharge $Q$.
We do this by defining the $Q$ transformations of $(c,\bar c, B)$ as
\begin{equation}\label{Q-ghosts}
  \begin{aligned}
&  Q\cdot c=
\tilde \sigma  \bar\epsilon\epsilon  +iv^\mu \tilde A_\mu\,,
\quad\quad Q\cdot \bar c=0\,,\\
&\hspace{10mm}
Q\cdot B= i v^\mu D_\mu^{(0)} \bar c   + i [\sigma^{(0)},\bar c] \,.
  \end{aligned}
\end{equation}
One can check that on all fields including $(c,\bar c, B)$,  $\hat Q\equiv Q+Q_\text{B}$ acts as the bosonic symmetry
\begin{equation}
\label{Qhat-squared}
\hat  Q^2=i {\cal L}_v +i \sigma^{(0)}\bar\epsilon \epsilon - v^\mu A^{(0)}_\mu+\frac{1}{2R}(b+b^{-1}){\cal R}\,,
\end{equation}
which is the same as (\ref{Q-squared}) except that the fields take values at the saddle point.
With gauge-fixing, the localization procedure involves adding to the action the term $t\hat Q\cdot \hat V$ instead of $tQ\cdot V$, 
where $\hat V=V+V_\text{gh}$.

As explained in Section \ref{sec:S3bPF}, $e^{c \hat Q^2}$ represents the action on the fields of  the group $G$, which is the product of $H$, $K$, the maximal torus of the gauge group, and the flavor $U(1)$.
For $c=-iR$, the corresponding group elements are parameterized as%
\footnote{\label{read-off-group}%
This is obtained by identifying $(g\cdot \Phi)(x)\equiv g\cdot \Phi(g^{-1}\cdot x)$ with $e^{c \hat Q^2}\Phi(x)$, where $\Phi$ is an arbitrary field.
The action of $H\times K$ on coordinates is defined as $(h,t)\cdot (e^{i\psi/2},e^{i\phi/2})=(h e^{i\psi/2},t e^{i\phi/2})$.
}
\begin{equation}\label{S3b-group-para}
g=(h, t, e^{ia},f) \in G\quad \longrightarrow \quad ( e^{-\frac{i}{2}(b+b^{-1})}, e^{-\frac{i}{2}(b-b^{-1})}, e^{ \hat\sigma},e^{\frac i2 \Delta (b+b^{-1})})\in G_{\mathbb C}\,.
\end{equation}

\subsection{Cohomological organization of fields}

From now on we will set $R$ to one.
As in \cite{pestun}, we want to organize fields in the cohomological form.
We begin with a vector multiplet on $\mathbb S^3_b$.
Let us define%
\footnote{%
Throughout this section, the symbol $\Lambda$ denotes a component of the gaugino, and should not be confused with a gauge parameter in Section~\ref{sec:vector-susy}.
}
\begin{equation}
  \Lambda_\mu\equiv
\bar\epsilon \gamma_\mu\lambda+\epsilon\gamma_\mu\bar\lambda
\,,
\quad
\Lambda\equiv \bar\epsilon \lambda+\epsilon\bar\lambda\,.
\end{equation}
On the space of fields, we define bosonic and fermionic coordinates $(X_0,X_1)$ as\footnote{%
For the spinors (\ref{S3b-spinors}) and vector (\ref{vS3b}), we have $\bar\epsilon\epsilon=1$, $v_\mu\bar\epsilon\gamma^\mu=\bar\epsilon$, $v_\mu\epsilon\gamma^\mu=-\epsilon$,  $v^\mu v_\mu=1$.
}
 \begin{equation}
\label{X0X1def}
X_0=(X_0^\text{vec};X_0^\text{chi})\equiv(
\tilde A_\mu
;\phi, \bar\phi
)\,,
\quad
 X_1=(X_1^\text{vec};X_1^\text{chi})\equiv (
\Lambda
,c,\bar c\,;
\epsilon\psi,\bar\epsilon\bar\psi
)\,.
 \end{equation}
The field $\sigma$ is a dynamical equivariant parameter.
The remaining fields are interpreted as the differentials $ \hat Q X_0$ and $\hat Q X_1$.
In the following, we pick a saddle point and expand the action up to the quadratic order.
We can write the quadratic part $\hat V^{(2)}$ of $\hat V$ in the form
\begin{equation}\label{hatV2}
\hat  V^{(2)}=
  \begin{pmatrix}
\hat Q X_0{}&X_1
  \end{pmatrix}
\mathcal D
\begin{pmatrix}
X_0\\
\hat QX_1 
\end{pmatrix}\,,
\qquad
\mathcal D=
  \begin{pmatrix}
D_{00}&D_{01}\\
D_{10}&D_{11}
  \end{pmatrix}\,.
\end{equation}
Then we have $\hat Q \cdot \hat V^{(2)}=X_\text{bos}K_\text{bos}X_\text{bos}+X_\text{ferm}K_\text{ferm}X_\text{ferm}$, where
\begin{equation}
K_\text{bos}=
\begin{pmatrix}
  -\hat Q^2 & \\
& 1
\end{pmatrix}
\mathcal D + \mathcal D^T \begin{pmatrix}
\hat Q^2 & \\
& 1
\end{pmatrix}
\quad \text{ and } \quad
K_\text{ferm}=
-
\mathcal D 
\begin{pmatrix}
1& \\
& \hat Q^2 
\end{pmatrix}
+ 
 \begin{pmatrix}
1& \\
& -\hat Q^2 
\end{pmatrix}
\mathcal D^T
\end{equation}
can be viewed as infinite dimensional real matrices that are symmetric and anti-symmetric, respectively.
The invariance of $\hat V$ under $\hat Q^2$ implies that $\mathcal D$ commutes with $\hat Q^2$.   Then
$$
\begin{pmatrix}
  1&\\ &-\hat Q^2
\end{pmatrix}
K_\text{bos}=K_\text{ferm}
\begin{pmatrix}
  \hat Q^2&\\ &1
\end{pmatrix}\,.
$$
The one-loop determinant is thus given, up to a sign,  by
\begin{equation}
Z_\text{1-loop}=\left(\frac{\det K_\text{ferm}}{\det K_\text{bos}}\right)^{1/2}
=\left(\frac{\det_{\text{Coker}D_{10}}\hat Q^2}{\det_{\text{Ker}D_{10}}\hat Q^2}\right)^{1/2}
\end{equation}
and is related to the equivariant index (\ref{eq-index-def}) by the rule (\ref{rule}).

 The fermionic functional $\hat V=\hat V_\text{vec}+\hat V_\text{chi}$ is given by%
\footnote{%
It is useful to define $
  \varepsilon^0:=
  \begin{pmatrix}
  \epsilon\\
\bar\epsilon
  \end{pmatrix}
$ and 
$
\varepsilon^\mu:=
  \begin{pmatrix}
\gamma^\mu    &\\
&\gamma^\mu
  \end{pmatrix}\varepsilon_0$,
which satisfy
$  \varepsilon_m^\dagger  \varepsilon^n=2 \delta_m^n$
and $ \varepsilon^m \varepsilon^\dagger_m = 
2  {\bf 1}_{4\times 4}$ for $m, n=0,\ldots, 3$.
We then have $\Lambda_\mu= \varepsilon^\dagger_\mu
  \begin{pmatrix}
    \lambda\\-\bar\lambda
  \end{pmatrix}
$
and
$
\Lambda= \varepsilon^\dagger_0  \begin{pmatrix}
    \lambda\\-\bar\lambda
  \end{pmatrix}
$.
We also use the identities
$  \bar\epsilon\gamma^\mu\lambda=
\frac{1}{2}
 \Lambda^\mu-\frac{i}{2}\epsilon^\mu_{~\nu\rho} v^\nu \Lambda^\rho
+\half v^\mu \Lambda$, 
$\epsilon\gamma^\mu\bar\lambda=
\frac{1}{2} \Lambda^\mu
 +
\frac{i}{2}\epsilon^\mu_{~\nu\rho} v^\nu \Lambda^\rho
 -\half v^\mu \Lambda
$,
$  \bar\epsilon\lambda=\half v^\mu \Lambda_\mu+\half \Lambda$,
and
$\epsilon\bar\lambda=
 -
\half v^\mu \Lambda_\mu
 + \half \Lambda
$.
}
\begin{equation}\label{loc-action-lambda}
  \begin{aligned}
\hat V_\text{vec}&=(Q\lambda)^\dagger\lambda 
+(Q\bar\lambda)^\dagger\bar\lambda
+V_\text{gh}
\\
&=
- F_{\nu\rho}v^\nu \Lambda^\rho- \frac i 2 v^\mu \epsilon_{\mu\nu\rho} F^{\nu\rho} \Lambda
-i D_\mu\sigma
\Lambda^\mu
-D \Lambda
- (\sigma/f) v^\mu \Lambda_\mu 
+\bar c (G(\tilde A)+\frac{\xi}{2}B)
  \end{aligned}
\end{equation}
and
 \begin{equation}\label{loc-action-psi}
   \begin{aligned}
\hat V_\text{chi}   &=
 (Q\psi)^\dagger\psi 
 +(Q\bar\psi)^\dagger\bar\psi
 \\
 &=-i D_\mu\bar\phi (\bar\epsilon\gamma^\mu\psi)_\perp -i D_\mu\bar\phi v^\mu \bar\epsilon\psi
 -i \bar\phi(\sigma+m)\bar\epsilon\psi-\frac{\Delta}{f}\bar\phi\bar\epsilon\psi-F\epsilon\psi
 \\
&\quad\quad
+i(\epsilon\gamma^\mu\bar\psi)_\perp D_\mu\phi
+i\epsilon\bar\psi v^\mu D_\mu\phi
 +i (\sigma+m)\phi \epsilon\bar\psi
+\frac{\Delta}{f}\phi \epsilon\bar\psi+ F\bar\epsilon\bar\psi
\,.
   \end{aligned}
 \end{equation}
The symbol $\perp$ indicates the projection orthogonal to $v^\mu$.   We have $\bar\epsilon\gamma^\mu\psi=(\bar\epsilon\psi) v^\mu+(\bar\epsilon\gamma^\mu\psi)_\perp$.

\subsection{Differential operator $D_{10}$}

Given a differential operator $\mathcal D$ on space $X$, its symbol $\sigma(\mathcal D;x,p)$ is a function of $x\in X$ and $p=p_\mu dx^\mu\in T^*_x X$, defined by replacing $\partial_\mu$ by $i p_\mu$ everywhere in $\mathcal D$, and collecting the terms that have the highest order in $p$.
The operator $\mathcal D$ is called elliptic if its symbol is invertible for any $x\in X$ and $p\neq 0$.
When a compact Lie group $G$ acts on $X$, there is a weaker notion of ellipticity.
We say that $\mathcal D$ is transversally elliptic if its symbol is invertible,  at each $x\in X$, for all non-zero cotangent vectors $p \neq 0$ that are orthogonal to the $G$ directions \cite{MR0482866}.%
\footnote{%
We say that $p$ is orthogonal to the $G$ directions if $p_\mu V^\mu=0$ for any $V^\mu \partial_\mu$ whose flow is an action of $G$.  
}
When $\mathcal D$ is transversally elliptic with respect to the $G$-action, the equivariant index is well-defined as a distribution on $G$.
In this section we will compute the differential operator $D_{10}$ that appears in the fermionic functional $\hat V^{(2)}$, and show that it is transversally elliptic with respect to group $G$ defined in (\ref{G-def-S3b}).

Let us now study the differential operator $D_{10}$ more closely.
We begin with the vector multiplet.
It is convenient to split the fluctuation $\tilde A_\mu\equiv A_\mu-A^{(0)}_\mu$ into the components parallel and orthogonal to the vector field $v_\mu$: $\tilde A_\mu=a_\mu+v_\mu b$, $v^\mu a_\mu=0$.
The bosonic and fermionic coordinates $(X_0, X_1)$ on the space of fields are defined in (\ref{X0X1def}).
A technical complication is that the ghost $c$ appears with a derivative in
\begin{equation}
  \Lambda_\mu=
 -2i  \hat Q\tilde A_\mu+2 iD_\mu c\,.
\end{equation}
Although $\hat V^{(2)}_\text{vec}$ in terms of the original fields involves only terms with a single  derivative, several terms end up with two derivatives when we express $\hat V^{(2)}_\text{vec}$ in terms of $X_0$, etc.
Indeed, showing only the integrand, we find up to total derivatives%
\footnote{%
The square bracket $[\,\,]$ indicates that the derivatives act only on the functions inside.
If not in a square bracket, derivatives are understood to act on all the factors on the right.
}
\begin{align}
X_1^\text{vec} D_{10}^\text{vec} X_0^\text{vec}
&=-2 i
 [D_\nu(a_\mu+v_\mu b)]  (v^{\nu} D^{\mu}-v^\mu D^\nu) c
- i\epsilon^{\mu\nu\rho} [D_\mu (a_\nu+v_\nu b)] v_\rho \Lambda 
+\bar c \, G(a+v b)
\nonumber\\
&=
  \begin{pmatrix}
2i
c & \bar c & \Lambda
  \end{pmatrix}
  \begin{pmatrix}
(D^\mu v^\nu-D^\nu v^\mu)
 D_\nu v_\mu 
&
 (D^\mu v^\nu - D^\nu v^\mu)
D_\nu  \\
G(x,\partial)^\mu v_\mu & G(x,\partial)^\mu \\
-i \epsilon^{\mu\nu\rho} (D_\mu v_\nu)v_\rho& i \epsilon^{\mu\nu\rho} v_\rho D_\nu
  \end{pmatrix}
  \begin{pmatrix}
b \\
a_\mu
\end{pmatrix}
\,.
\end{align}
The operator is effectively a  square matrix because $a_\mu$ has two independent components.
Since the symbol is defined using the terms with most derivatives, superficially $D_{10}^\text{vec}$ is neither elliptic nor transversally elliptic.
We can, however, make a field redefinition so that  $D_{10}^\text{vec}$ is block diagonal, with one block being second order and the rest first order.%
\footnote{%
The same issue arises in for localization in four dimensions.
The authors of \cite{pestun} and \cite{Hama:2012bg}, working with momenta rather than derivatives, showed that highest order terms can be block diagonalized.
Here we are pedantic and demonstrate that the whole differential operator can be block diagonalized.
}

Let us introduce
\begin{equation}
{\mathcal D}_\mu=\frac{Rf}{2} \epsilon_{\mu\nu\rho}v^\nu  D^\rho
\end{equation}
and take the gauge fixing term to be%
\footnote{%
See \cite{Hama:2012bg} for a similar choice of the gauge fixing term $G$ in the four-dimensional case.
} 
\begin{equation}
G(\tilde A)=  G^\mu\tilde A_\mu= (D^\mu+ (D^\nu {\mathcal D}_\nu-v^\nu D_\nu)  v^\mu)\tilde A_\mu\,.
\end{equation}
After some calculation, we find that the differential operator $D_{10}$ for the vector multiplet can be block diagonalized:
\begin{equation}
\label{block-vec}
  \begin{aligned}
  &\quad  X_1^\text{vec} D_{10}^\text{vec} X_0^\text{vec}
\\
&=
  \begin{pmatrix}
2i
    c\, & \bar c-
2i
  v^\nu D_\nu c\,   & \Lambda
  \end{pmatrix}
  \begin{pmatrix}
2 D^{[\mu} v^{\nu]}
D_\nu v_\mu - D_\nu v^\nu G^\mu v_\mu 
&
0
\\
0
& D^\mu \\
0
& i \epsilon^{\mu\nu\rho} v_\rho D_\nu
  \end{pmatrix}
  \begin{pmatrix}
b \\
a_\mu + \mathcal D_\mu b
\end{pmatrix}\,.
\end{aligned}
\end{equation}

Thus at the quadratic order in fluctuations, $b=v^\mu\tilde A_\mu$ and $c$ decouple from the other combinations of fields in (\ref{block-vec}).  
The corresponding differential operator appears on the upper left corner of the matrix.
Its symbol has determinant $p^2 -(p\cdot v)^2$, which vanishes for $p$ parallel to $v$, but is non-zero for any non-zero $p$ satisfying $p\cdot v=0$.
Since the vector field $v$ given in (\ref{v-decomposition-S3b}) is a linear combination of the vector fields generating $H$ and $K$, the operator for $(b,c)$ is transversally elliptic, and the equivariant index is well-defined as a distribution on $G$.
The equivariant index is however trivial because the differential operator maps the space of scalars to itself, and its kernel and cokernel are identical.

The differential operator in the lower right block of the matrix in (\ref{block-vec}) is first order, and its symbol has determinant $i(p^2-(v\cdot p)^2)$.
Thus the operator is again not elliptic, but is transversally elliptic.

For the chiral multiplet, we obtain from (\ref{loc-action-psi})
\begin{equation}
\label{D10chiral}
\begin{aligned}
  X_1^\text{chi} D_{10}^\text{chi} X_0^\text{chi}
&=
i(\epsilon\gamma^\mu\bar\psi)_\perp D_\mu\phi
+c.c.
\\
&=
i (\bar\epsilon\bar\psi) (\epsilon\gamma^\mu\epsilon) D_\mu\phi
+c.c.
  \end{aligned}
\end{equation}
where complex conjugation $*$ acts formally as $\psi^*=-C\bar\psi$.
The vector field that appears in (\ref{D10chiral}) can be decomposed as $\epsilon\gamma^\mu\epsilon=
 e^{i(\varphi_1+\varphi_2)} (w^\mu+i u^\mu) $, where $w^\mu$ and $u^\mu$ are both real.%
\footnote{%
Explicitly, $w=(Rf)^{-1}\partial_\vartheta$, $u=(Rb)^{-1} \cot\vartheta \partial_{\varphi_1} - R^{-1}b \tan\vartheta \partial_{\varphi_2}$.
}
It can be checked that $(u,v,w)$ form an orthonormal basis of the tangent space.  
The symbol $\sigma$ of the differential operator $i \epsilon\gamma^\mu\epsilon D_\mu$ then satisfies $|\sigma|^2=(w\cdot p)^2+(u\cdot p)^2=p^2-(v\cdot p)^2$, so the operator is not elliptic but is transversally elliptic.
At the north and the south poles ($\vartheta=0$ and $\pi/2$) of the base $\mathbb S^2$, the operator acts as the Dolbeault operator in the directions orthogonal to $v$.  

\subsection{Computation of $\text{ind}\,D_{10}$}
\label{sec:comp-ind}

Let $T^*_G X|_p$ be the space of cotangent vectors at $p\in X$ conormal to the $G$-orbit.
Let $T^*_G X$ be the collection of $T^*_G X|_p$ for all $p\in X$, and let $\pi: T^*_G X \rightarrow X$ be the projection.
The symbol of a $G$-equivariant (pseudo)differential operator that maps a section of $E_0$ to a section of $E_1$ then defines a map $\pi^* E_0\rightarrow \pi^* E_1$.
The operator and its symbol are by definition transversally elliptic if the map is invertible on $T^*_G X$ away from the zero section.
Such a symbol defines a class in $K_G(T^*_G X)$, and the index depends only on this class \cite{MR0482866}.

We wish to compute the index of $D_{10}$.
The index is determined by the homotopy class of the symbol of $D_{10}$.
The key tool, when a factor $H$ in $G=H\times K$ acts freely on $X$, is the  following theorem (Corollary 3.3 of \cite{MR0482866}).

Let us label the irreducible representations of $H$ by $R$, and let $E_R^*$  be the dual of the vector bundle $E_R$ over $X/H$ induced from $R$.%
\footnote{%
The induced bundle $E_R$ is the quotient of $X\times R$ by the $H$-action.
} 
For a symbol $\sigma_0: \pi^* E_0\rightarrow \pi^* E_1$ on $X/H$ transversally elliptic with respect to $K$, the pull-back $p^*\sigma_0$ is a transversally elliptic symbol with respect to $H\times K$, where $p:X\rightarrow X/H$ is the projection.
The index of $p^*\sigma_0$ is given by
\begin{equation}
\label{index-reduce}
  \text{ind}_{(h,k)}(p^*\sigma_0)=\sum_{R} \text{ind}_{k}(\sigma_0\otimes E_R^*)\cdot \chi_R(h)\,,
\quad (h,k)\in H\times K\,,
\end{equation}
where $R$ labels irreducible representations of $H$, $E_R^*$ is the dual of the vector bundle $E_R$ induced from $R$, and $\chi_R$ is the character of $R$.

Let us apply this theorem to $X=\mathbb S^3_b$.
Since we only need to know the K theory class of the symbol and $H$ acts freely, let us set $p_\psi$ to zero.
Then $p^2 -(p\cdot v)^2$ cannot vanish unless $p=0$.
Thus the symbol of $D_{10}$ reduces to an elliptic symbol $\sigma_0$ on $X/H=\mathbb S^2$, and we have $\sigma(D_{10})=p^*\sigma_0$.
For the vector multiplet, we obtain the de Rham complex on $\mathbb S^2$ with a degree shifted by one.
The equivariant index of $D_{10}^\text{vec}$, obtained from (\ref{index-reduce}) and the Atiyah-Bott formula, is
\begin{equation}
\text{ind}_g( D^\text{vec}_{10})=  - 
\sum_{n\in\mathbb Z} (t^n+t^{-n}) \chi_\text{adj}(e^{ia})  h^n  \,,
\end{equation}
where $h\in H=U(1)$, $t\in K=U(1)$, and $\chi_\text{adj}$ is the character in the adjoint representation of the gauge group.
The identification (\ref{S3b-group-para}) leads to (\ref{S3b-vec-north}) and (\ref{S3b-vec-south}).
For the chiral multiplet $  \text{ind}_g D_{10}^\text{chi}  =   \text{ind}_g D_{10,\mathbb C}^\text{chi} +   \text{ind}_{g^{-1}} D_{10,\mathbb C}^\text{chi}$, where $D_{10,\mathbb C}^\text{chi}$ reduces to the Dolbeault complex on $\mathbb S^2$, whose index is
\begin{equation}
\label{Dolb-index-S2}
\sum_{n\in\mathbb Z} \left(\frac{t^n}{1-t^{-2}} + \frac{t^{-n}}{1-t^{2}}  \right)  \chi_R(e^{ia}) h^{n}  f 
\end{equation}
with $\chi_R$ the character of the matter representation $R$ of the gauge group.
Substitution (\ref{S3b-group-para}) gives (\ref{theta-zero-contribution}) and (\ref{theta-halfpi-contribution}).
Since the reduced symbol is elliptic, its index, the bracket in (\ref{Dolb-index-S2}), is a polynomial.
Thus there is no ambiguity in the index as long as we expand (\ref{theta-zero-contribution}) and (\ref{theta-halfpi-contribution}), we need to expand both in $t$ or $t^{-1}$, so that the sum for fixed $n$ is a finite polynomial.

\section{Index   theory calculations for $\mathbb S^1\times\mathbb S^2$}
\label{sec:index-details-S1S2} 

In this Appendix we repeat the steps in Appendix~\ref{sec:index-details} for $\mathbb S^1\times\mathbb S^2$.

First let us determine the saddle point configurations that contribute in the localization calculation.
For the vector multiplet, we again compare $Q\lambda=0$ with a complex conjugate of $(Q\bar\lambda)^*=0$.   Defining $B_\mu\equiv \frac  1 2 \epsilon_{\mu}{}^{\nu\rho}F_{\nu\rho}$, we obtain $Q\lambda=\mathcal E_1+\mathcal E_2$ and $\gamma_{\hat\tau}C^{-1}(Q\bar\lambda)^*=\mathcal E_1-\mathcal E_2$, where
\begin{equation}\label{S1S2-gaugino-var}
  \begin{aligned}
\mathcal E_1&\equiv-i\gamma_\tau\epsilon ({\rm Re}B_\tau) +({\rm Im}B^j)\gamma_j \epsilon -i ({\rm Im}D)\epsilon
+i\gamma^j \epsilon D_j\sigma -i\sigma \gamma_{\hat\tau}\epsilon\,,
\\
\mathcal E_2&\equiv  ({\rm Im}B_\tau) \gamma^\tau\epsilon
-i\gamma^j \epsilon ({\rm Re} B_j) -({\rm Re}\,D)\epsilon +i \gamma^\tau\epsilon D_\tau\sigma\,.
 \end{aligned}
\end{equation} 
Here we used the properties $C^{-1}\epsilon^*=\gamma_{\hat\tau} \bar\epsilon\,,
C^{-1}\bar \epsilon^*=\gamma_{\hat\tau} \epsilon$, $D_\mu\epsilon=-\frac 12 \gamma_\mu\gamma_\tau \epsilon$, and $D_\mu\bar\epsilon=\frac 12 \gamma_\mu\gamma_\tau \bar\epsilon$.
Thus $\mathcal E_{1,2}$ must vanish separately.
Let us set ${\rm Im}D={\rm Im}B_\mu=0$ here; turning on the imaginary parts corresponds to inserting vortex loops.
Contract $\mathcal E_2=0$ with $\epsilon^\dagger$.
The real part of the equation implies that ${\rm Re}\,D=0$.
Let us also consider contracting $\mathcal E_2=0$ with $\bar \epsilon^T C$.
Since $\bar\epsilon\gamma^\mu \epsilon=(1,0,-i)$ with $\mu=\tau,\theta,\varphi$, the real part implies that $B_\varphi=0$, and the imaginary part implies that $D_\tau\sigma=0$.
It then follows that $B_\theta=0$.
Next contract $\mathcal E_1=0$ with $\bar \epsilon^T C$.
We find that $D_\varphi\sigma=0$, and that
\begin{equation}
  B_\tau+\sigma =0\,.
\end{equation}
We also obtain $D_\theta\sigma=0$.
Thus we have $B_\tau+\sigma=0$, where  $B_\tau$ is the smooth part of $B_\tau$.
Diagonalizing $\sigma$, we find that $B_\tau=\frac{m}{2}$, $\sigma=-\frac m 2$, where $m$ is a quantized GNO charge.
For the chiral multiplet, the same reasoning applied to $Q\psi=0$ and $(Q\psi)^*=0$ leads to
\begin{equation}
  i\gamma^\tau\epsilon D_\tau\phi  + F\bar\epsilon=0\,,
\quad\quad
i\sum_{j=\theta,\varphi}\gamma^j\epsilon D_j\phi
+i\sigma\phi\epsilon -i\Delta \phi \gamma_{\hat\tau}\epsilon=0\,.
\end{equation}
The first equation contracted with $\bar\epsilon^TC$ implies that $D_\tau\phi=F=0$.
Using the explicit expression (\ref{S1S2-spinors}),
the remaining equations can be solved:
\begin{equation}
  \phi =\beta\left(e^{\mp i\varphi }\tan\frac\theta 2 \right)^{m/2}
\left(e^{-i\varphi}\sin\theta\right)^{-\Delta}
e^{\mp\frac i 2 H^\pm \varphi}\,,
\end{equation}
where the upper and lower signs are respectively for the two patches $U^\pm$ on $\mathbb S^2$, and $\beta$ is a constant.
Again the configuration represents matter vortex loops, and we do not include the contributions from such a configuration in this paper.
Thus the path integral localizes to the field configurations (\ref{S1S2saddle}) in the absence of a vortex loop, and to the same configurations on top of singular backgrounds in the presence of vortex loops.

The BRST transformations and the localization action are the same as in the $\mathbb S^3_b$ case.
In particular, the expressions (\ref{QB-ghosts})-(\ref{Q-ghosts}) are valid.
The square of $\hat Q^2$ is given precisely by the right hand side of (\ref{Q2-S1S2-simple}).
Thus for $c=i\beta$, the group element $g=e^{c\hat Q^2}$ is parameterized as%
\footnote{%
The parameterization can be found as in footnote \ref{read-off-group}.
The group action on coordinates is given by $(h,t)\cdot (e^{2\pi i \tau/\beta},e^{i\varphi})=(h e^{2\pi i \tau/\beta},t e^{i\varphi})$.
}
\begin{equation}\label{S1S2-group-para}
g=(h, t, e^{ia},f) \in G\quad \longrightarrow \quad ( e^{2\pi i}, e^{2\beta_2}, e^{ia}, e^{\beta_2 \Delta} )\in \hat G_{\mathbb C}\,,
\end{equation}
where $\hat G_{\mathbb C}$ denotes the complexified universal covering of $G$.   (So $e^{2\pi i}$ is non-trivial.)

Let us introduce $
  \varepsilon_0:=
  \begin{pmatrix}
  \epsilon\\
-\bar\epsilon
  \end{pmatrix}
$,
$
  \varepsilon_1:=
  \begin{pmatrix}
  \gamma_{\hat \tau}\epsilon\\
-  \gamma_{\hat \tau}\bar\epsilon
  \end{pmatrix}
$ 
and
$
  \varepsilon_j:=
  \begin{pmatrix}
   \gamma_{j}\epsilon\\
\gamma_{j}\bar\epsilon
  \end{pmatrix}
$  with $j=2,3$,
which satisfy
$  \varepsilon_m^\dagger  \varepsilon^n=2 \delta_m^n$
and $ \varepsilon^m \varepsilon^\dagger_m = 
2  {\bf 1}_{4\times 4}$ for $m, n=0,\ldots, 3$.
We then define 
$$\Lambda_m= \varepsilon^\dagger_m
  \begin{pmatrix}
    \gamma_{\hat\tau}\lambda\\ \gamma_{\hat\tau}\bar\lambda
  \end{pmatrix}
$$
and take
 \begin{equation}
X_0=(X_0^\text{vec};X_0^\text{chi})\equiv(
 \tilde A_j,\tilde \sigma
;\bar\phi,\phi
)\,,
\quad
 X_1=(X_1^\text{vec};X_1^\text{chi})\equiv (
\Lambda_1
,c,\bar c\,;
\epsilon\gamma_{\hat \tau} \psi,\bar\epsilon\gamma_{\hat \tau} \bar\psi
)
 \end{equation}
as superspace coordinates.
Gauge-fixing can be achieved with the localization action  $\hat V=\hat V_\text{vec}+\hat V_\text{chi}$ given as
\begin{equation}\label{vecs1}
  \begin{aligned}
&\hat V_\text{vec} \equiv(Q\lambda)^\dagger\lambda 
+(Q\bar\lambda)^\dagger\bar\lambda
+V_\text{gh}
  \end{aligned}
\end{equation}
and
 \begin{equation}\label{hyps1}
   \begin{aligned}
\hat V_\text{chi}   =&
 (Q\psi)^\dagger\psi 
 +(Q\bar\psi)^\dagger\bar\psi
 \\
 =&-i  (\partial_\tau \bar\phi -i \bar\phi A_\tau^\dagger) \bar\epsilon\psi + \varepsilon_{\tau}{}^{ij} D_i\bar\phi \bar\epsilon\gamma_j\psi
 -i \sigma \bar\phi\bar\epsilon\gamma_{\hat\tau}\psi+i\Delta\bar\phi\bar\epsilon\psi+\bar F\epsilon\gamma_{\hat \tau}\psi
 \\
&
-i (\epsilon \bar\psi) (\partial_\tau +i A_\tau^\dagger \phi) + \varepsilon_{\tau}{}^{ij} D_i\phi\epsilon\gamma_j \bar\psi
 -i  \phi\sigma\epsilon\gamma_{\hat\tau} \bar\psi-i\Delta\phi\epsilon \bar\psi+ F \bar\epsilon\gamma_{\hat \tau} \bar\psi
\,.
\end{aligned}
\end{equation}
Note that a vortex loop introduces a non-hermitian part to $A_\tau$.

The functional $\hat V_{\rm vec}$ contains only first order differential operators acting on $(\lambda, \bar\lambda, c, \bar c)$, 
When we express $\hat V_{\rm vec}$  in terms of $(X^\text{vec}_{0,1}, \hat Q X^\text{vec}_{0,1})$ to read off $D_{10}$, we find second order differentials because $D_\mu^{(0)} c$ appear in $\hat Q \tilde A_j$ and $\hat Q\tilde\sigma$.
Thus the symbol determined by the highest order terms is, strictly speaking, degenerate everywhere.
Instead of separating the first and the second order parts by block-diagonalizing $D_{10}$ as in the $\mathbb S^3_b$ case, for $\mathbb S^1\times \mathbb S^2$ we take an alternative approach, which we believe is more general.
Namely, in order to compute the one-loop determinant around the saddle point (\ref{S1S2saddle}), we consider the Gaussian functional integration of $e^{-t \hat Q \cdot V_u}$ with
\begin{equation}
  V_u\equiv (1-u) \hat V^{(2)}_\text{vec} +u V'  +\hat V^{(2)}_\text{chi}\,.
\end{equation}
Setting to zero the deformation parameter $u$ gives back the original gauge-fixed action.
If the bosonic part of $\hat Q\cdot V'$ is positive definite in directions transverse to the space of saddle point configurations, the path integral is independent of $u$, and can be evaluated at $u=1$.
As $V'$, we take 
\begin{equation}
  V'= (\hat Q\cdot X_1^\text{vec})^\dagger X_1^\text{vec} +(\hat Q^2\cdot X_1^\text{vec})^\dagger \hat Q X_1^\text{vec}
+V_\text{gh} \,.
\end{equation}
This looks almost the same as $\hat V^{(2)}_\text{vec}$, but $D'_{10}$, defined by replacing $\hat V^{(2)}$ in (\ref{hatV2}) with $V'$, has only first order differentials.
Showing only the terms relevant for $D'_{10}$, we have\begin{equation}
  \begin{aligned}
V'=(\hat Q c)^\dagger c+ (\hat Q \Lambda_1)^\dagger \Lambda_1 +\ldots+ \bar c([\sigma^{(0)},\tilde\sigma]-i D^j_{(0)} \tilde A_j) +\ldots
      \end{aligned}
  \end{equation}
Recall that the embedding of $\mathbb S^2$ in $\mathbb R^3$ implies that $T^*\mathbb S^2$ and a trivial real line bundle add up to a trivial rank three bundle.
Thus the combination $(\tilde \sigma,\tilde A_{j})$, $j=\theta,\varphi$, can be expanded in three real scalars, and a convenient orthonormal basis is provided by supersymmetry:
\begin{equation} \label{filed-decompose}
(\tilde \sigma,\tilde A_j)=( \bar\epsilon \epsilon , i \bar\epsilon\gamma_j\epsilon)S+ (i \epsilon\gamma_\tau\epsilon,\epsilon\gamma_j\gamma_\tau\epsilon)T+c.c.\,,
\end{equation}
where the first term is real and the third is the conjugate of the second.
The first section in the basis appears for $Q\cdot c$ in (\ref{Q-ghosts}).
Since only the term $i v^\tau\tilde A_\tau$ is imaginary,
\begin{equation}
  (\hat Q\cdot c)^\dagger \sim 2 (\bar\epsilon\epsilon \tilde \sigma+i v^j\tilde A_j )
\end{equation}
up to $\hat Q$-exact and higher order terms.
Let us set $w^j=\epsilon\gamma^j\gamma_\tau\epsilon$.%
\footnote{%
Explicitly, $w^\theta=i e^{-i\varphi}$, $w^\varphi=e^{-i\varphi}\cot\theta$.
}
We claim that $D'_{10} $ takes the form
\begin{equation}
  X_1^\text{vec}D'_{10} X_0^\text{vec}
=
  \begin{pmatrix}
    2c & \ -i \bar c-4\Lambda_1 & \  i\bar c-4\Lambda_1
  \end{pmatrix}
  \begin{pmatrix}
    1 &0&0
\\
*& w^j D_j+\ldots &\ldots 
\\
*& \ldots & \bar w^j D_j+\ldots
  \end{pmatrix}
  \begin{pmatrix}
    S \\ T\\ \bar T
  \end{pmatrix}\,.
\end{equation}
The ellipses do not involve differentials while $*$'s do.  
The first row in the matrix easily follows from orthonormality, while the rest needs some work.
By rearranging the rows  we can block-diagonalize $D'_{10}$ to decouple $S$ and $c$.
The symbol of the remaining part of $D'_{10}$ is proportional to $|w^j p_j|^2=p_\theta^2+\cot^2\theta p_\varphi^2$.
The symbol is invertible for non-zero momentum $(p_\mu)_{\mu=\tau,\theta,\varphi}$ transverse to $\partial_\tau$ and $\partial_\varphi$, and is transversally elliptic with respect to $H\times K$.

For the purpose of counting zero-modes and determining the index as a distribution, we complexify the complex and treat $T$ and $\bar T$ as independent complex scalars.
For simplicity we suppress $\tau$-dependence.
The index is unchanged if we modify the operator, without changing the leading symbol, to%
\footnote{%
This was explained using K-theory in \cite{pestun}, and the argument with explicit zero-mode solutions is due to~\cite{hosomichi-notes}.   Both considered the four-dimensional $\mathcal N =2$ theory.
See also \cite{MR792703} and \cite{MR0482866}.
}
\begin{equation}\label{Dvec-deformed}
    \begin{pmatrix}
w^j D_j +s e^{-i\varphi}\sin\theta &0
\\
0& \bar w^j D_j  +  is e^{i\varphi}\sin\theta
  \end{pmatrix}\,.
\end{equation}
In the limit $s\rightarrow +\infty$ we get the zero-modes of (\ref{Dvec-deformed}) localized near the north pole $\theta=0$
$$
(T,\bar T)\sim (e^{-i r\varphi} \sin^r \theta, 
0
) e^{-2 s \sin^2 \frac{\theta}{2}}\,,
\qquad
r =0,1,2,\ldots\,,
$$
and those localized near the south pole $\theta=\pi$
$$
(T,\bar T)\sim (0, e^{i r\varphi} \sin^{r }\theta)  e^{-2 s \cos^2 \frac{\theta}{2}}\,, \qquad r =0,1,2,\ldots\,.
$$
The zero-modes of the dual operator also get localized.   
We have  
\begin{equation}
(e^{-i r \varphi} \sin^r\theta,0) e^{-2 s \sin^2\frac{\theta}{2}}\,,
\qquad
r =0,1,2,\ldots
\end{equation}
localized at $\theta =0$, and
\begin{equation}
(0, e^{i r\varphi} \sin^r\theta) e^{-2 s \cos^2\frac{\theta}{2}}\,,
\qquad
r =0,1,2,\ldots
\end{equation}
localized at $\theta =\pi$.
To the index, $e^{i n\varphi}$ in each zero mode contributes $t^{-n}$, where $t\in K$.
We also take into account the $\varphi$ dependence of the basis sections in (\ref{filed-decompose}) as well as the flux contribution to $j_3$ in~(\ref{j3-monopole}).
We apply the reduction formula (\ref{index-reduce}) to obtain 
\begin{equation}\label{index-vec-S1S2}
  \begin{aligned}
  \text{ind}\, D_{10}^\text{vec}&=  
\sum_{n\in\mathbb Z}  \sum_{r=0}^\infty \sum_{\alpha\in\text{adj}}
h^n
\left(
t^{-\alpha(m)/2}(t^{r+1}-t^r)
+
t^{\alpha(m)/2}(t^{-r-1}-t^{-r})
\right)e^{i\alpha(a)}
\\
&=
- \sum_{n\in\mathbb Z} \sum_{\alpha\in\text{adj}}
h^n
\left(
t^{-\alpha(m)/2}
+
t^{\alpha(m)/2}
\right)e^{i\alpha(a)}
\,.
  \end{aligned}
\end{equation}
Since $\epsilon \gamma_\tau\epsilon$ in (\ref{filed-decompose}) vanishes at the poles, the resulting local contributions coincide with those of the complexified de Rham complex with suitable twisting.

The chiral multiplet is simpler.   We obtain from (\ref{hyps1})%
\footnote{We use $\bar\epsilon \gamma_\mu\psi=(\bar\epsilon\gamma_\mu\gamma_{\hat \tau} \epsilon)(\bar\epsilon\psi)+(\bar\epsilon\gamma_\mu\bar\epsilon)(\psi\gamma_{\hat \tau} \epsilon)$, $\epsilon \gamma_\mu\bar\psi=(\epsilon\gamma_\mu\epsilon)(\bar\epsilon\gamma_{\hat \tau}\bar\psi)+(\epsilon\gamma_\mu\gamma_{\hat \tau}\bar\epsilon)(\epsilon\bar\psi)$ and $\bar\epsilon\gamma_{\hat \tau}\epsilon=1$.}
\begin{equation}
\label{D10chirals2}
\begin{aligned}
  X_1^\text{chi} D_{10}^\text{chi} X_0^\text{chi}
=&
\,i \bar w ^j
D_{j}\bar\phi(\psi\gamma_{\hat \tau} \epsilon)
+i w^j D_{j}\phi(\bar\epsilon\gamma_{\hat \tau}\bar\psi)\\ &\,+i \frac{m}{2} \bar\phi(\bar\epsilon\gamma_{\hat \tau} \bar\epsilon)(\psi\gamma_{\hat \tau} \epsilon)  +i  \phi\frac{m}{2}(\epsilon\gamma_{\hat \tau}\epsilon)(\bar\epsilon\gamma_{\hat \tau}\bar\psi)
\end{aligned}
\end{equation}
The operator $D_{10}^\text{chi}$ is thus the ``realification'' of $w^j D_j+\ldots$, where the ellipses  contain no differentials.
The symbol has determinant proportional to $p_\theta^2+p_\varphi^2\cot^2\theta$, and is $H\times K$-transversally elliptic.
By deforming the operator to 
\begin{equation}
w^j D_j+ is e^{-i\varphi}\sin\theta+ \ldots
\end{equation}
and taking $s\rightarrow +\infty$, we find zero-modes localized near $\theta=0$
$$
\phi \sim e^{-i r\varphi} \sin^r \theta e^{-2 s \sin^2 \frac{\theta}{2}}\,,
\qquad
r =0,1,2,\ldots\,,
$$
and the zero-modes of the dual operator, localized near $\theta=\pi$, approximately given by
$$
\epsilon\gamma_{\hat\tau}\psi\sim e^{i r\varphi} \sin^r\theta e^{-2 s \cos^2\frac{\theta}{2}}\,,
\qquad
r =0,1,2,\ldots\,.
$$
Taking into account the R-charges and the gauge group action, the index is $  \text{ind}_g D_{10}^\text{chi}  =   \text{ind}_g D_{10,\mathbb C}^\text{chi} +   \text{ind}_{g^{-1}} D_{10,\mathbb C}^\text{chi}$, where $g=(h,t, e^{ia},f)$ and
\begin{equation}\label{index-chi-S1S2}
  \begin{aligned}
  \text{ind}_g D_{10,\mathbb C}^\text{chi}&=  
\sum_{n\in\mathbb Z}  \sum_{r=0}^\infty 
\sum_{\rho\in R}
h^n
\left(
t^{r-\frac{1}{2}\rho(m)} 
-
t^{-r-1+\frac{1}{2}\rho(m) }
\right)
e^{i\rho(a)}f \,.
  \end{aligned}
\end{equation}
Substitution (\ref{S1S2-group-para}) gives (\ref{S1S2-ind-chi-text}).

\bibliography{refs}
\end{document}